\documentclass[useAMS,usenatbib]{mn2e}
\usepackage{graphicx}
\usepackage{mathptmx}
\usepackage{amssymb}
\usepackage[pdftex,pdfpagemode={UseOutlines},bookmarks,bookmarksopen,colorlinks=false,linkcolor={red},citecolor={blue},urlcolor={green}]{hyperref}

\newcommand{\threetotwo}{$J=3\to 2$}
\newcommand{\twotoone}{$J=2\to 1$}

\newcommand{\thirteenco}{$^{13}$CO}

\newcommand{\ceighteeno}{C$^{18}$O}
\newcommand{\ammonia}{NH$_3$}
\newcommand{\ntwohplus}{N$_2$H$^+$}
\newcommand{\hco}{HCO$^+$}
\newcommand{\htwo}{H$_2$}
\newcommand{\msun}{M$_\odot$}
\newcommand{\mvir}{$M_\rmn{vir}$}
\newcommand{\mjybeam}{mJy\,beam$^{-1}$}

\newcommand{\kms}{km\,s$^{-1}$}
\newcommand{\cmthree}{cm$^{-3}$}

\newcommand{\fturb}{$f_\mathrm{turb}$}

\newcommand{\jcmt}{JCMT}

\newcommand{\scuba}{SCUBA}

\newcommand{\clfind}{{\sc clfind}}
\newcommand{\gclumps}{{\sc gaussclumps}}

\newcommand{\ngc}{NGC\,1333}

\newcommand{\harp}{HARP}

\newcommand{\yso}{YSO}

\newcommand{\fwhm}{FWHM}

\newcommand{\apj}{ApJ}
\newcommand{\apjl}{ApJ}

\newcommand{\apjss}{ApJS}
\newcommand{\aap}{A\&A}

\newcommand{\mnras}{MNRAS}
\newcommand{\aj}{AJ}
\newcommand{\araa}{ARA\&A}

\newcommand{\nat}{Nat}
\defcitealias{hatch07}{HFR07}
\defcitealias{hkirk07}{KJT07}
\defcitealias{paper1}{Paper I}
\defcitealias{paper3}{Paper II}
\defcitealias{jappsen04}{JK04}
\defcitealias{goodman93}{GBF93}
\title[The kinematics of SCUBA clumps in Perseus]{A submillimetre survey of the kinematics of the Perseus molecular
cloud -- III. Clump kinematics}
\author[E.~I.~Curtis and J.~S.~Richer]{Emily~I.~Curtis$^{1,2}$\thanks{E-mail:\href{mailto:e.curtis@mrao.cam.ac.uk}{e.curtis@mrao.cam.ac.uk}} and John~S.~Richer$^{1,2}$\thanks{E-mail:
\href{mailto:jsr10@cam.ac.uk}{jsr10@cam.ac.uk}}\\
$^{1}$Astrophysics Group, Cavendish Laboratory, J. J. Thomson Avenue,
  Cambridge, CB3 0HE\\
$^{2}$Kavli Institute for Cosmology, c/o Institute of Astronomy,
  University of Cambridge, Madingley Road, Cambridge, CB3 0HA}
\begin{document}

\date{Accepted 2010 ...; Received 2010 ...; in original form 2010 ...}

\pagerange{\pageref{firstpage}--\pageref{lastpage}} \pubyear{2010}

\maketitle

\label{firstpage}

\begin{abstract}

\noindent We explore the kinematic properties of dense continuum clumps in
the Perseus molecular cloud, derived from our wide-field
\ceighteeno\ \threetotwo\ data across four regions -- \ngc,
IC348/HH211, L1448 and L1455. Two distinct populations are
  examined, identified using the automated algorithms \clfind\ (85
  clumps) and \gclumps\ (122 clumps) on existing SCUBA
  850\,\micron\ data. These kinematic signatures are compared to the clumps' dust continuum properties. We calculate each clump's non-thermal
linewidth and virial mass from the associated
\ceighteeno\ \threetotwo\ spectrum. The clumps have supersonic
linewidths, $\langle \sigma_\mathrm{NT}/c_\mathrm{s}\rangle= 1.76\pm
0.09$ (\clfind\ population) and $1.71\pm0.05$ (with \gclumps). The linewidth distributions suggest the \ceighteeno\ line probes a
lower-density `envelope' rather than a dense inner core. Similar
  linewidth distributions for protostellar and starless clumps
  implies protostars do not have a significant impact on their
  immediate environment. The proximity to an active young stellar cluster seems to affect the linewidths:
those in \ngc\ are greater than elsewhere. In IC348
the proximity to the old IR cluster has little influence, with the
linewidths being the smallest of all. The virial analysis suggests
that the clumps are bound and close to equipartition, with virial
masses similar to the masses derived from the continuum emission. In
particular, the starless clumps occupy the same parameter space as the
protostars, suggesting they are true stellar precursors and will go on
to form stars. We also search for ordered \ceighteeno\ velocity gradients across the face of
each core. Approximately one third have significant detections, which we
mainly interpret in terms of rotation. However, we note a
correlation between the directions of the identified gradients and outflows across the
protostars, indicating we may not have a purely rotational
signature. The fitted gradients are in the range ${\cal G} = 1$ to
16\,\kms\,pc$^{-1}$, larger than found in previous work, probably as a
result of the higher resolution of our data and/or outflow
contamination. These gradients, if interpreted solely in terms of
rotation, suggest that the rotation is not dynamically significant:
the ratios of clump rotational to gravitational energy are
$\beta_\mathrm{rot} \lesssim 0.02$. Furthermore, derived specific angular
momenta are smaller than observed in previous studies, centred around
$j \sim 10^{-3}$\,km\,s$^{-1}$\,pc, which indicates we have identified
lower levels of rotation, or that the \ceighteeno\ \threetotwo\ line probes
conditions significantly denser and/or colder than $n\sim
10^5$\,\cmthree\ and $T\sim 10$\,K. 

\end{abstract}


\section{Introduction} 

Stars form inside dense cores deep within molecular clouds. The
similarity of the core mass function (CMF) to the initial mass
function of stars (IMF, see \citealt*{motte98,alves07}) has sometimes
been used (e.g.\ \citealt{enoch08}) as evidence against theories of core formation
where no correspondence is anticipated, e.g.\ in the
competitive accretion picture \citep{bonnell01,bate05}. However, the mapping of a given CMF on to
the resultant IMF is complicated by many factors (see
e.g.\ \citealt{hatchell08,scubapaper}), such as the cores' varying multiplicity \citep{goodwin08},
star-forming efficiencies \citep{swift08} and/or mass-dependent lifetimes
\citep*{clark07}. In fact, as shown by \citet{swift08} and \citet{goodwin09}, diverse
evolutionary schemes can map the observed CMFs on to the IMF, implying
the current mass function data are not sufficient to discriminate
between different theoretical models by themselves.   

The kinematics of star-forming cores, probed with spectral-line
observations of dense molecular tracers, are arguably the
best discriminator between different models of core formation
(e.g. \citealt{andre07}; \citealt*{hkirk07}). For
example, cores created by shocks in large-scale flows exhibit large velocity
gradients and are located at local maxima in the line-of-sight velocity
dispersion (e.g.\ \citealt*{ballesteros-paredes03}; \citealt{klessen05}). Alternatively, the magnetically 
controlled scenario has more quiescent velocity fields (e.g.\ \citealp{nakamura05}), well-matched to observations of isolated starless cores
with subsonic levels of turbulence (e.g.\ \citealp{myers83,caselli02}).
Such quiescent cores are in opposition to purely hydrodynamic models of gravoturbulent fragmentation
(see \citealp{maclow04}) which produce a majority of cores with supersonic velocity
dispersions. 

\citet*{offner08a} and \citet*{krumholz05}
maintain that the physical mechanism of star formation depends on the
much debated presence (or absence) of turbulent feedback. In one view,
molecular clouds are short-lived (on timescales of order one
dynamical time), non-equilibrium structures (e.g.\ \citealp{elmegreen00,hartmann01,dib07}) characterized by
transient turbulence, which dissipates quickly. In this scenario, stars could form by
the collapse of discrete cores. Conversely, molecular clouds might form slowly and be
quasi-equilibrium objects (see e.g. \citealp*{shu87}; \citealp{mckee99,krumholz07,nakamura07}). This would require turbulence to be constantly injected
into the clouds, either externally (from e.g.\ supernova blast waves
or H{\sc ii} regions) or internally (from e.g.\ protostellar outflows) and
cores could form through competitive accretion
\citep{bonnell01}. \citet{offner08b} demonstrate that these two
disparate views of star formation give rise to distinguishable
kinematics in dense stellar precursors and thus the study of core
kinematics potentially offers a way to probe the turbulent
state of the surrounding molecular cloud.

This paper is concerned with the kinematical properties of dense clumps in the Perseus molecular cloud (hereafter simply Perseus) and compares
them to theoretical models. We examine the linewidths, virial masses and ordered velocity
gradients of two populations of clumps \citep{scubapaper} identified across
SCUBA dust continuum maps \citep{hatchell05}. These kinematic
signatures are derived for each clump from our
wide-field survey (\citealp*{paper1}, \citetalias{paper1}) of four clusters of
star-forming cores towards NGC~1333, IC348/HH211 (simply IC348 hereafter), L1448 and L1455 in the
\threetotwo\ rotational lines of CO and its common isotopologues \thirteenco\ and
\ceighteeno. In the preceding paper of this series (\citealt{paper3},
\citetalias{paper3}), we explored the evolution of molecular
outflows across these regions. We organize this paper as follows: the remainder of the
introduction explains the naming conventions we have adopted to describe
star-forming clumps/cores. \S \ref{sec:observations} briefly describes the
continuum data and associated clump catalogues alongside our
spectral-line datasets, from which we derive the clump
  kinematics. Various clump kinematic signatures are investigated in
\S \ref{sec:kinematics}, including the clumps' linewidths (\S
\ref{sec:linewidths}) and virial masses (\S \ref{sec:virialmasses}), before we search for ordered
velocity gradients across the face of the clumps in \S \ref{sec:rotation}, which we
interpret with a solid-body-rotation model. Finally, we summarize this
work in \S \ref{sec:summary}. 

\subsection{Nomenclature and clump populations} 

We adhere to the clump hierarchy explained in
\citet{scubapaper}, which follows \citet*{williams00}. Briefly,
molecular clouds contain clumps which in turn harbour cores, the
direct precursors of individual or multiple stars. Clumps without
embedded objects are \emph{starless}, unless they are gravitationally
bound when we refer to them as \emph{prestellar}. 

\section{Observational data}
\label{sec:observations}

\subsection{Dust continuum}

The SCUBA 850\,\micron\ data analysed have a beam size of
14\,arcsec (0.017\,pc at 250\,pc, our adopted distance to Perseus, see
\citetalias{paper1}) and are sampled on a 3\,arcsec grid
with a typical rms noise level, $\sigma_\rmn{rms}=35$\,\mjybeam\ \citep{hatchell05}. 

\subsection{Clump catalogues} \label{sec:clumpcats}

We investigate the kinematics of two populations of dust
  continuum clumps, which we identified in the SCUBA 850\,\micron\ data
in a previous paper \citep{scubapaper}, using the two most popular
automated algorithms: \clfind\ \citep*{williams94} and \gclumps\ \citep{stutzki90}. Our motivation in that study
was to determine which clump properties can be robustly measured and
which depend on the extraction technique. We continue that approach in
this work, analysing the kinematic signatures for each population in
turn. 

In \citet{scubapaper}, we located 85 and 122 clumps using \clfind\ and
\gclumps\ respectively with peak flux densities
$\ge 4\sigma_\mathrm{rms}=140$\,\mjybeam, across the same areas where
we also have \ceighteeno\ \threetotwo\ data
\citepalias{paper1}. The clumps were identified as starless, Class 0
or Class I protostars using the classifications of
\citeauthor{hatch07} (\citeyear{hatch07}, hereafter \citetalias{hatch07}), which are
based on source SEDs incorporating {\it Spitzer} data. We associate a clump with a \citetalias{hatch07}
source, if the separation between the clump and
\citetalias{hatch07} source peak positions
is less than the clump's diameter (see \citealt{scubapaper}). Given the similarities in the extraction method and
resulting population, most of the
\clfind\ clumps are directly on top of the \citetalias{hatch07}
sources and are unambiguously identified. However, three
\gclumps\ sources are large enough to encapsulate two
\citetalias{hatch07} sources and their results are
correspondingly included in the population statistics twice.

\subsection{Spectral-line} 

The \ceighteeno\ \threetotwo\ observations used in this analysis were
undertaken with \harp\ (the Heterodyne Array Receiver Project;
\citealp{harp_paper}) on the James Clerk Maxwell Telescope and have
been described in detail previously \citepalias{paper1}. Striping artefacts resulting from systematic
differences in calibration between \harp\ channels were removed using
a `flatfield' procedure (fully described in \citetalias{paper1}). The final
datasets are sampled on a 3\,arcsec grid, distributed using a Gaussian
gridding kernel with a 9\,arcsec full-width half maximum (FWHM), yielding an equivalent beam
size of 17.7\,arcsec. The median rms spectral noise values, averaged across
each region's map (measured in 0.15\,\kms\ channels), are 0.18, 0.15,
0.15 and 0.14\,K for NGC~1333, IC348, L1448 and L1455 respectively.  

\section{The kinematics of the SCUBA cores}  \label{sec:kinematics}

In molecular clouds, the C$^{18}$O $J=3\to 2$ line is expected to be 
  optically thin and excited in dense regions ($n_\mathrm{crit}\sim
  10^4$\,\cmthree). It should therefore trace material in the
vicinity of or within the dense agglomerations surveyed with \scuba. 
At the peak of every clump in our two catalogues (see \S \ref{sec:clumpcats}), we extracted the C$^{18}$O $J=3\to2$ spectrum and fitted a
Gaussian profile to it using {\sc splat}\footnote{Part of
the Starlink software collection, see
\url{http://starlink.jach.hawaii.edu.}} if the peak of the
  \ceighteeno\ line was $\ge 3\sigma_\mathrm{sp}$, where
  $\sigma_\mathrm{sp}$ is the rms spectral noise. We did not detect \ceighteeno\ at the
  $3\sigma$ level towards 4 (5~per cent) \clfind\ and 6 (5~per
  cent) \gclumps\ positions. Of the detections, 13 (15~per cent)
  \clfind\ and 18 (15~per cent) \gclumps\ spectra have double-peaked
  line profiles. For comparison, \citeauthor{hkirk07} (\citeyear{hkirk07}, hereafter
\citetalias{hkirk07}) made pointed N$_2$H$^+$ $J=1\to 0$ and C$^{18}$O $J=2\to1$
observations on the Institut de Radio Astronomie Millim\'etric (IRAM) 30-m telescope (with beam sizes of 25
and 11\,arcsec respectively) towards 157 candidate
cores in Perseus -- 44 pointings in our fields -- and found 66
(42~per cent) \ceighteeno\ spectra required two component Gaussian fits with three in
one case. Generally, $J=2\to1$ data trace lower density ($n_\mathrm{crit}\sim 10^3$\,\cmthree) and/or colder material than the
$3\to2$ at lower optical depths by a factor of 2--3 in the
optically thin limit. Therefore
the $J=2\to1$ line is unlikely to saturate if the $J=3\to2$ data are
optically thin. This suggests there are separate emitting sources along the line-of-sight with different temperatures and/or densities
which require multiple-component fits for one transition and not the
other. An alternative is simply that at our velocity resolution
  (0.15\,\kms\ compared to \citetalias{hkirk07}'s 0.05\,\kms) many of
  \citetalias{hkirk07}'s separate components are blended into one, which might partly explain
  our larger linewidths (see \S \ref{sec:linewidths}). 

\subsection{Linewidths} \label{sec:linewidths}

The linewidths in molecular clouds are substantially larger than
the sound speed of their constituent gas (e.g.\ \citealp{zuckerman74,solomon87}). Such
supersonic linewidths are thought to arise from chaotic and/or
\emph{turbulent} motions of unknown fundamental origin (see
\citealp{elmegreen04} and references therein). The
velocity dispersion in a region scales with its size
(e.g.\ \citealp{larson81}) akin to the Kolmogorov law
for subsonic incompressible turbulence, although turbulence in the interstellar
medium is actually highly compressible and often supersonic causing
shocks. However, dense star-forming cores
harboured in cloud interiors possess near-thermal linewidths almost devoid of the turbulent movements that cause the
broadening in their outer envelopes
(e.g.\ \citealp{benson89,barranco98,goodman98,pineda10}). The C$^{18}$O $J=3\to 2$
transition is expected to probe a region intermediate in scale between
the dense core and its environment, although it is not the tracer of
choice for young prestellar cores. In the cold ($\la 10$\,K), high-density ($\geq 10^{5}$\,\cmthree) centres of
such objects, the conditions may result in \emph{depletion}, where many
molecules (including CO and its isotopologues) freeze-out on to dust
grains. Such zones are better probed by molecules such as \ntwohplus
and H$_2$D$^+$, but in any case the CO $J=3\to 2$ transitions should probe
higher densities than lower $J$ ones.

\begin{figure*}
\begin{center}
\includegraphics[width=\textwidth]{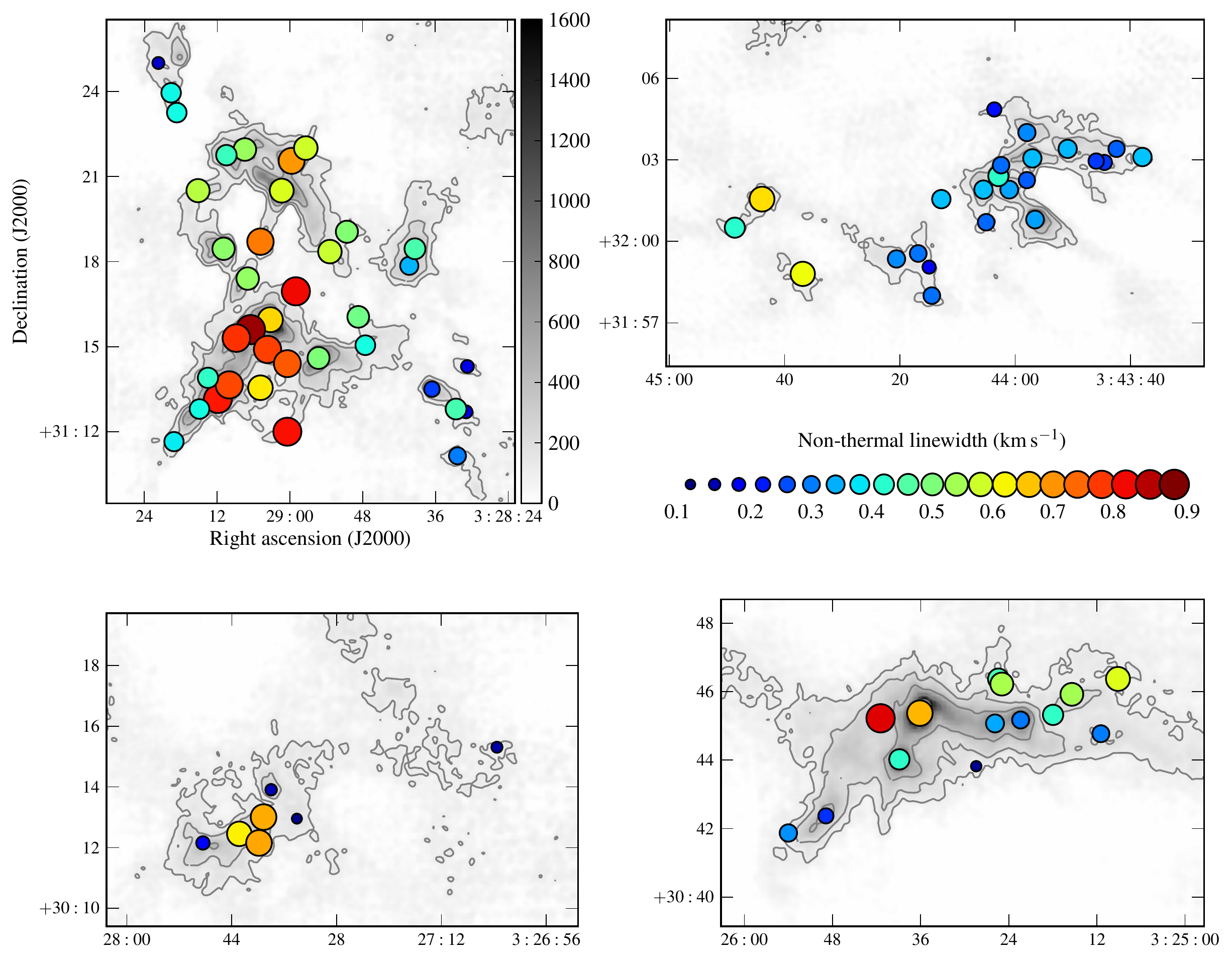}

\caption{Spatial distribution of \ceighteeno\ non-thermal
  linewidth across \ngc\ (upper left), IC348 (upper right), L1448 (lower right)
  and L1455 (lower left). The grey-scale is \scuba\ 850\,\micron\ flux density in
  \mjybeam\ \citep{hatchell05} with contours at 100, 200, 400, 800,
  1600 and 3200\,\mjybeam. Circles are
  positioned at the peak of every SCUBA clump in our catalogue identified
  using \clfind\ \citep{scubapaper} where there is a $3\sigma$
  \ceighteeno\ detection. The circles' size and colour reflect the
  linewidth as shown in the key.}  
\label{fig:linewidths}
\end{center}
\end{figure*}

\begin{figure}
\begin{center}
\includegraphics[width=0.47\textwidth]{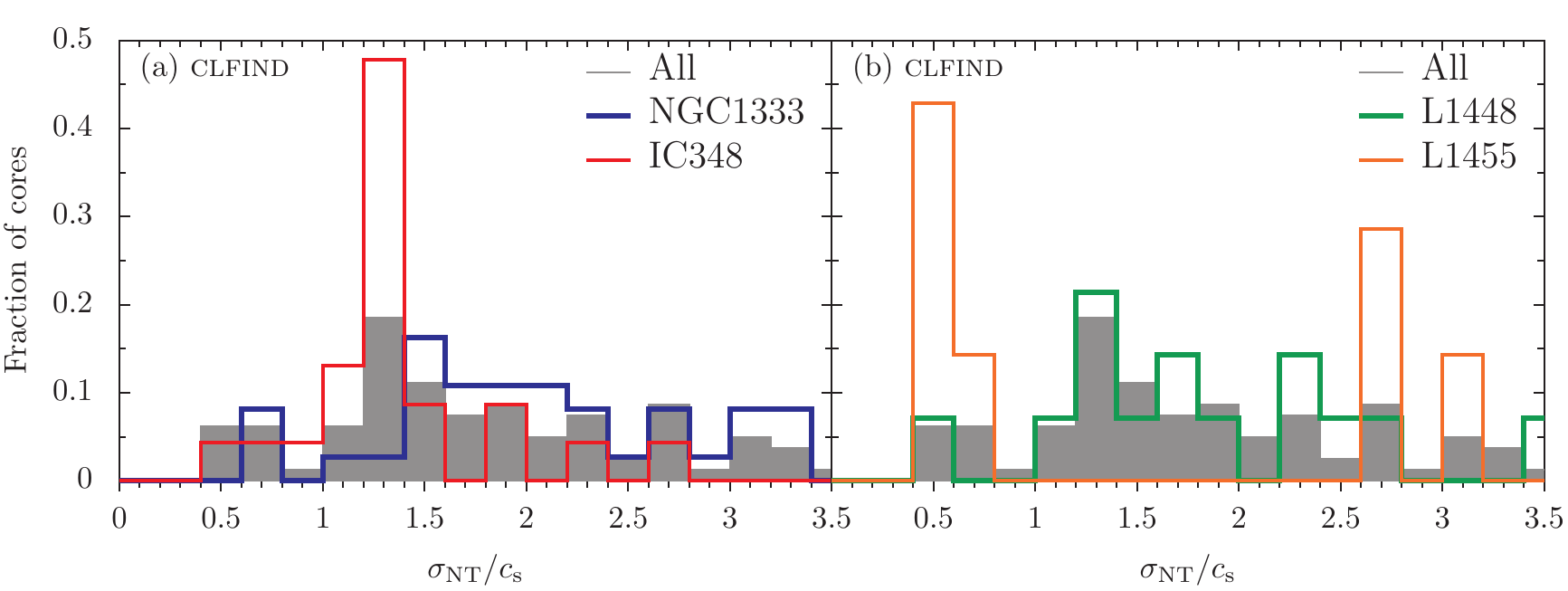}
\includegraphics[width=0.47\textwidth]{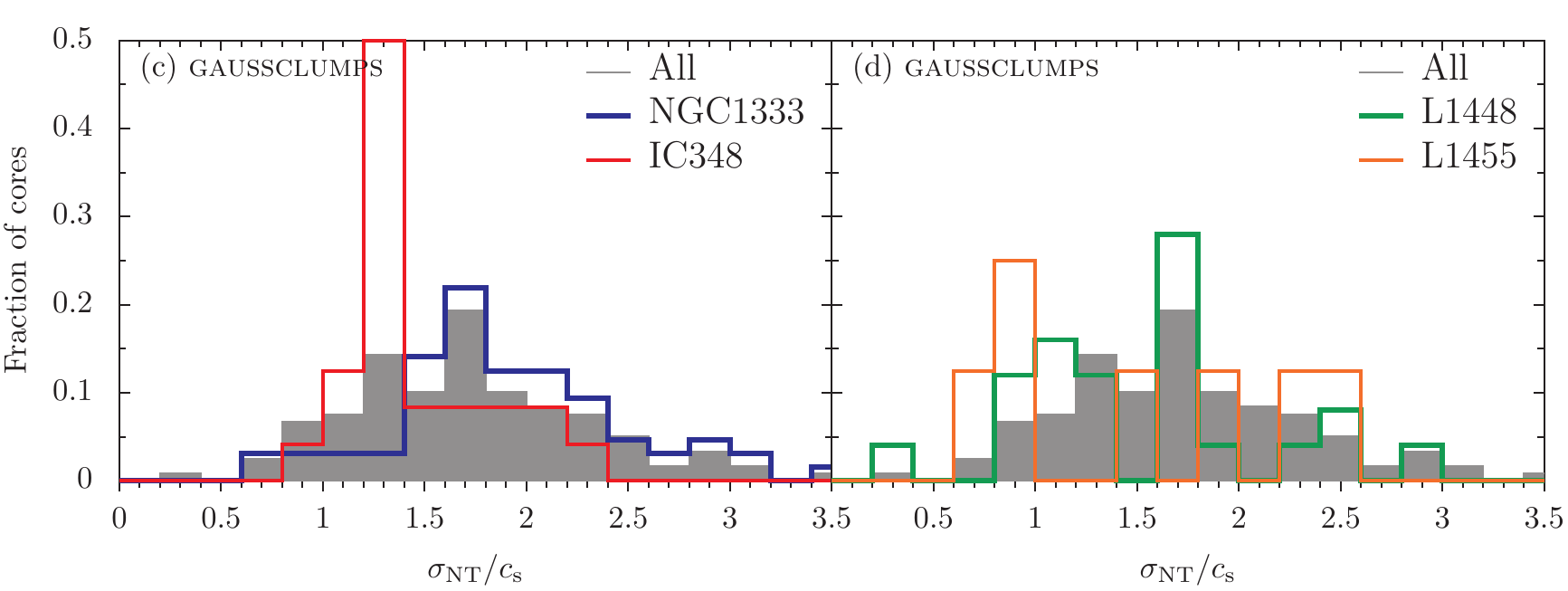}
\caption{Distribution of clump turbulent fractions,
  $f_\mathrm{turb}=\sigma_\mathrm{NT}/c_\mathrm{s}$, measured from the \harp\ C$^{18}$O
  $J=3\to 2$ data (see text), broken down by region for the \clfind\ (upper
  panels) and
  \gclumps\ populations (lower panels).}  
\label{fig:fturb}
\end{center}
\end{figure}

\begin{figure}
\begin{center}
\includegraphics[width=0.47\textwidth]{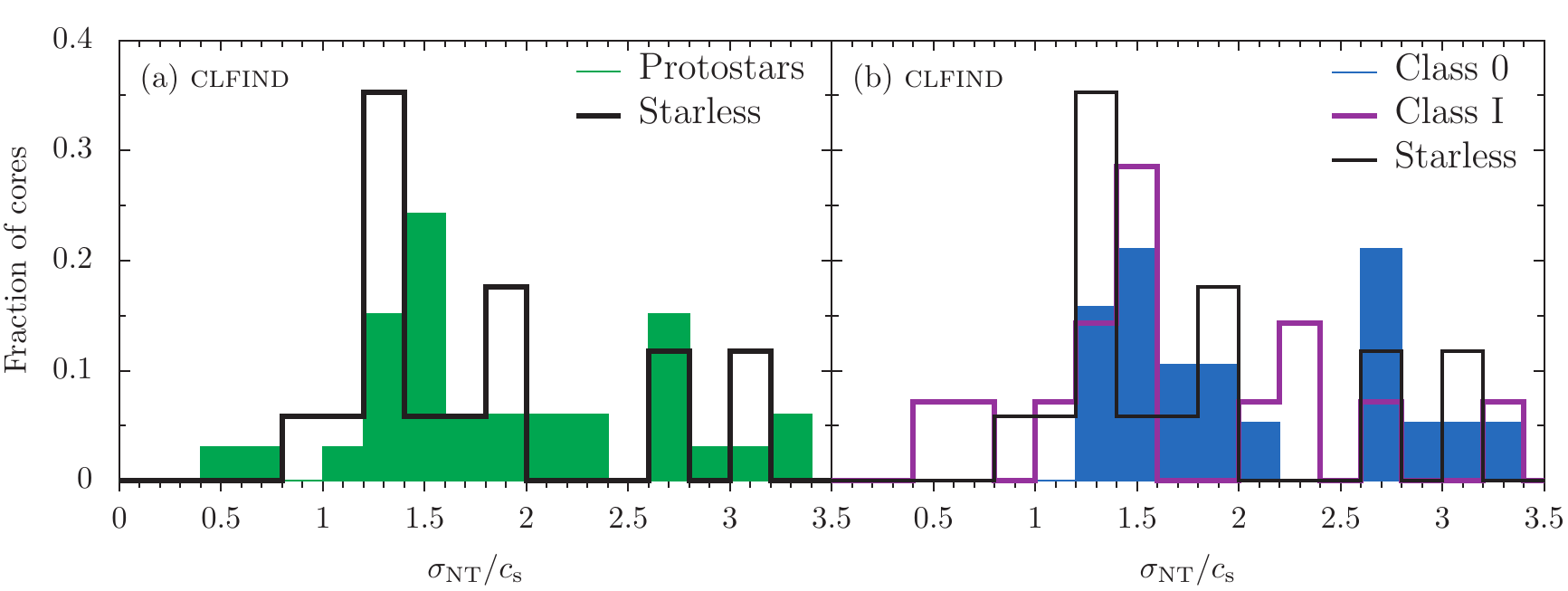}
\includegraphics[width=0.47\textwidth]{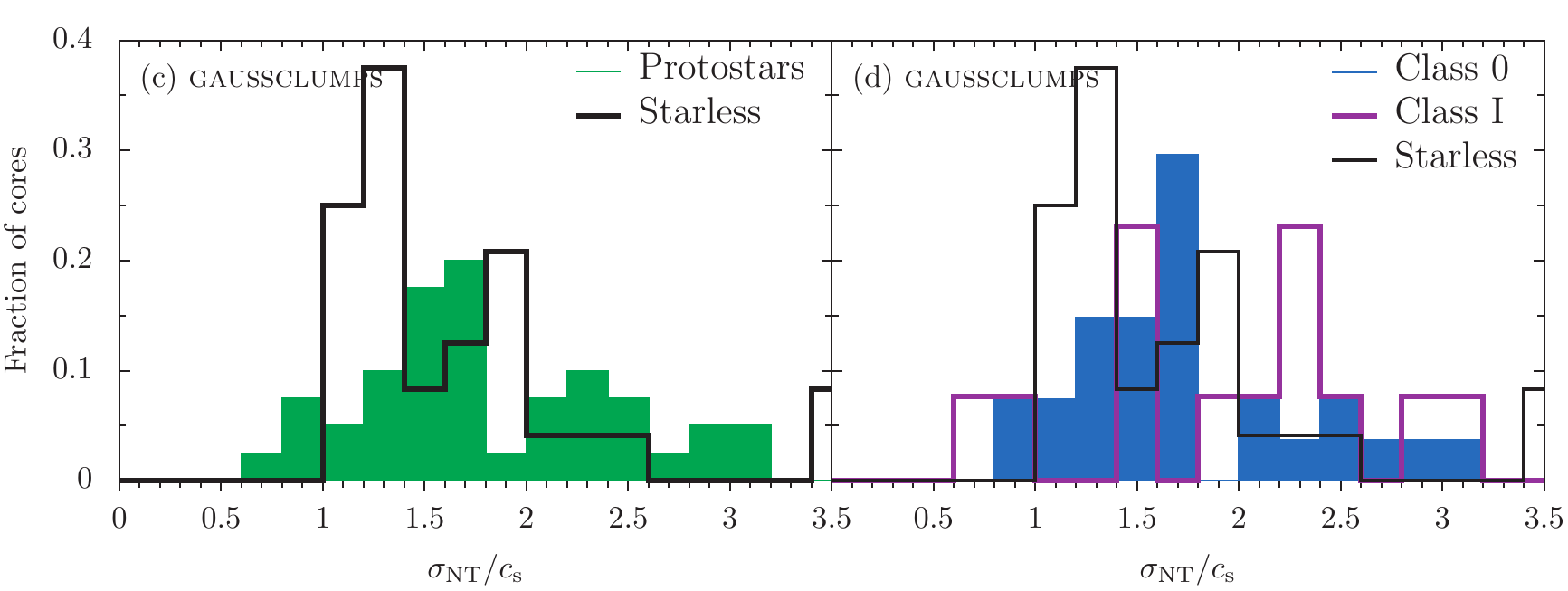}
\caption{Distribution of clump turbulent fractions,
  $f_\mathrm{turb}=\sigma_\mathrm{NT}/c_\mathrm{s}$, measured from the \harp\ C$^{18}$O
  $J=3\to 2$ data (see text), broken down by the different source types for the \clfind\ (upper
  panels) and
  \gclumps\ populations (lower panels).}
\label{fig:fturb_bysource}
\end{center}
\end{figure}

\begin{figure}
\begin{center}
\includegraphics[width=0.47\textwidth]{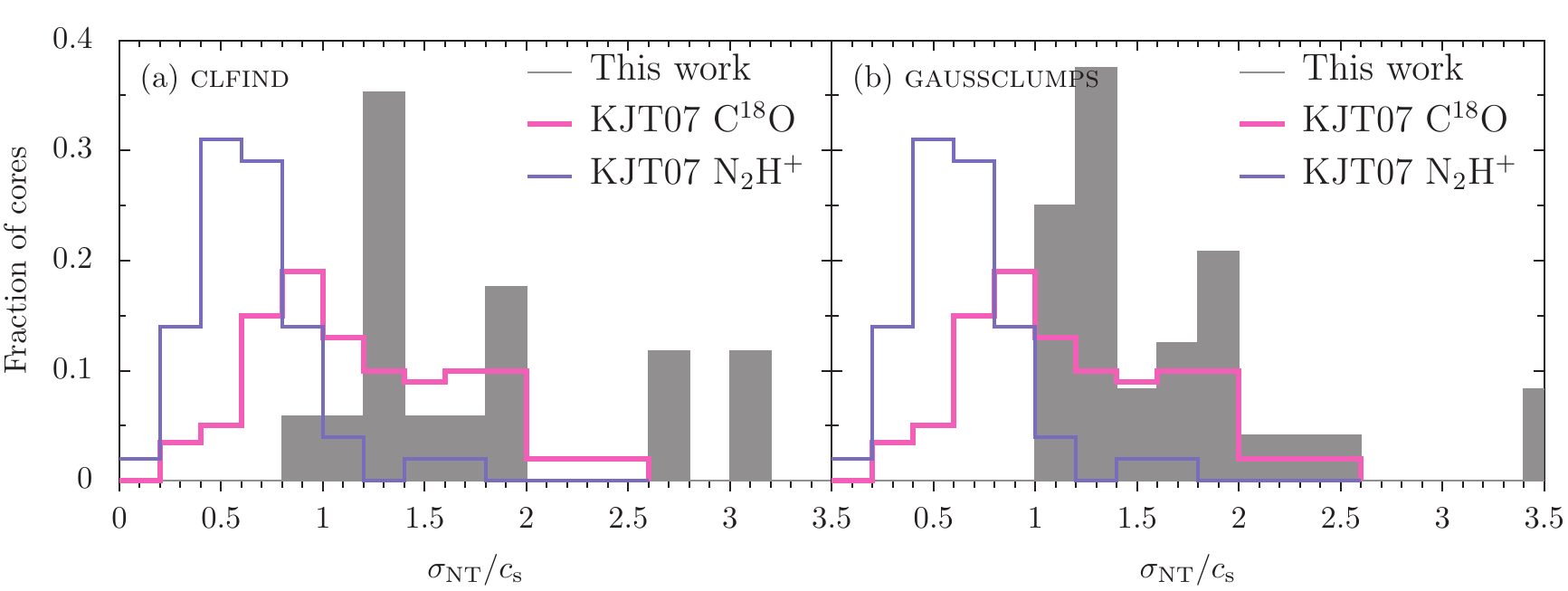}
\includegraphics[width=0.47\textwidth]{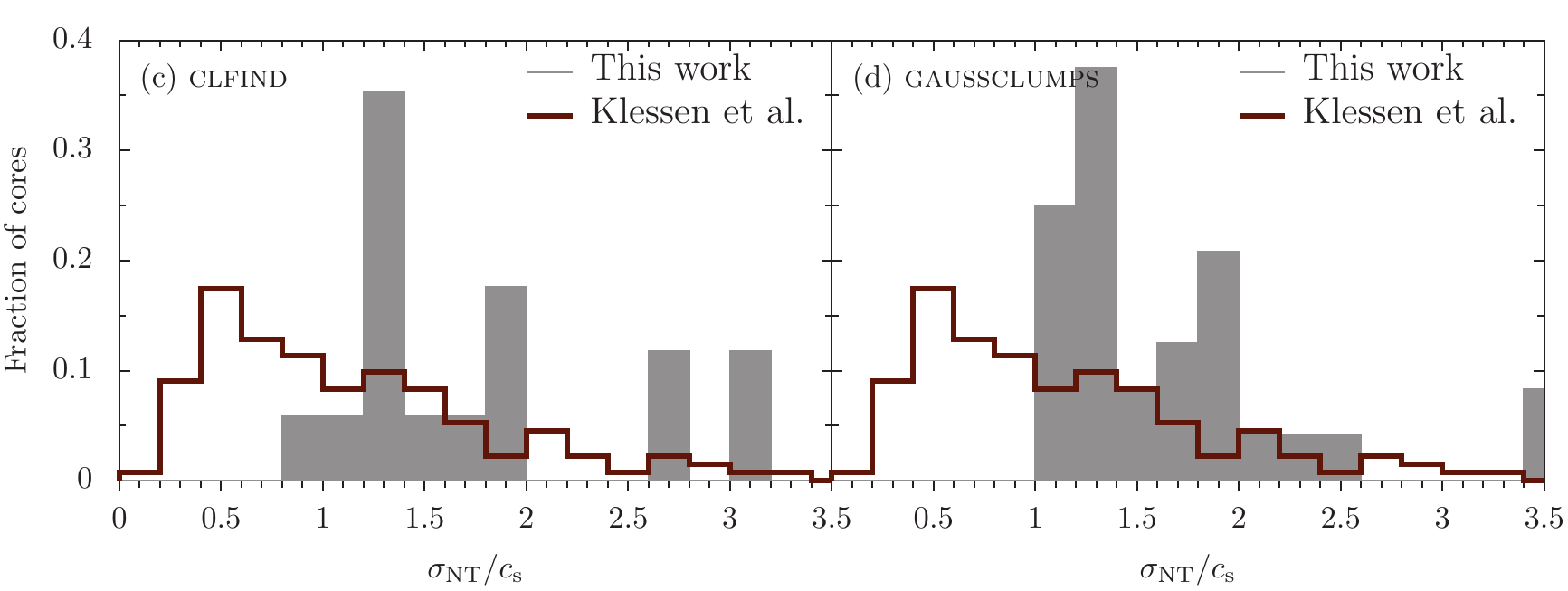}
\caption{Distribution of $f_\mathrm{turb}$, for the starless clump
  populations identified with \clfind\ (left panels) or
  \gclumps\ (right panels), measured from the \harp\ C$^{18}$O
  $J=3\to 2$ data. Overlaid for comparison are the distributions found by
  \citetalias{hkirk07} for 157 candidate cores using C$^{18}$O $J=2\to 1$
  (upper panels, pink) and N$_2$H$^+$ $J=1\to 0$ (upper panels, blue). A prediction for cores from the gravoturbulent
  simulation of \citet{klessen05} is also shown for turbulence driven
  on large scales (lower panels, brown). For that simulation the turbulent
  fraction extends beyond 3.5 to 4.3 (not plotted), though there are small numbers
  of clumps in those bins ($\sim 2$~per cent).}  
\label{fig:fturb_comp_starless}
\end{center}
\end{figure}

\begin{figure}
\begin{center}
\includegraphics[width=0.47\textwidth]{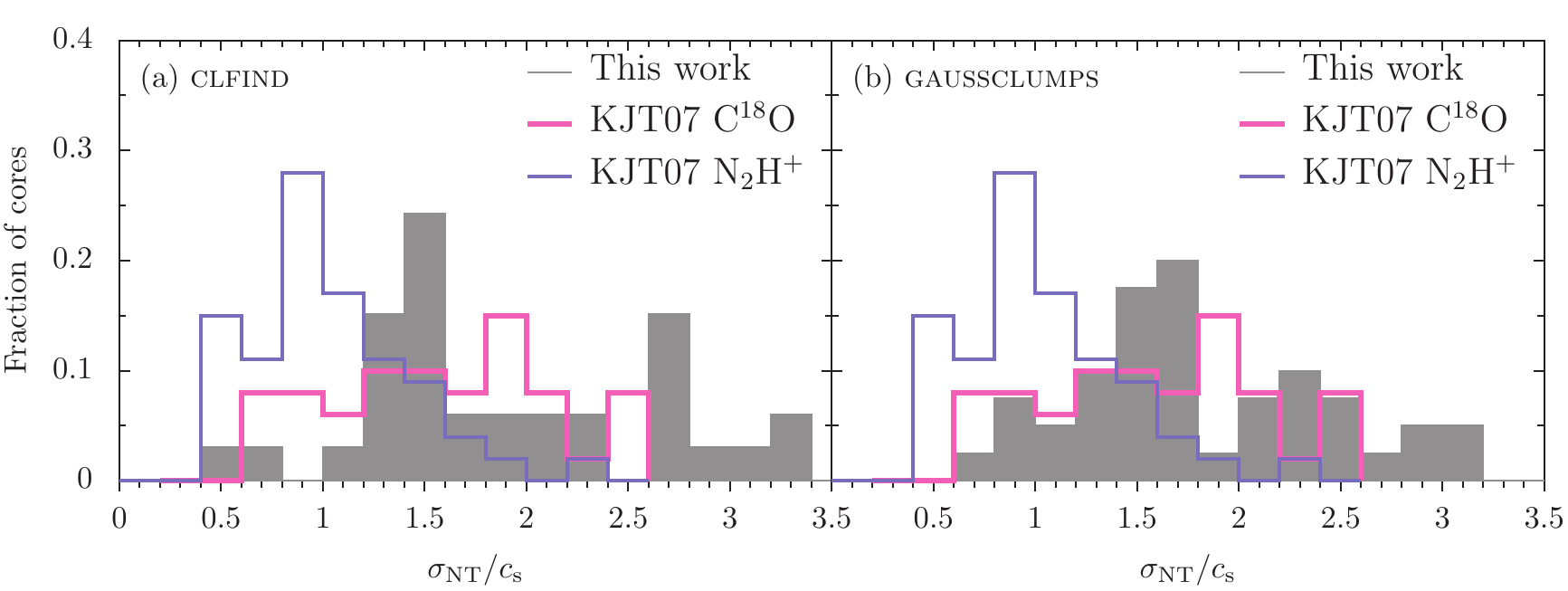}
\includegraphics[width=0.47\textwidth]{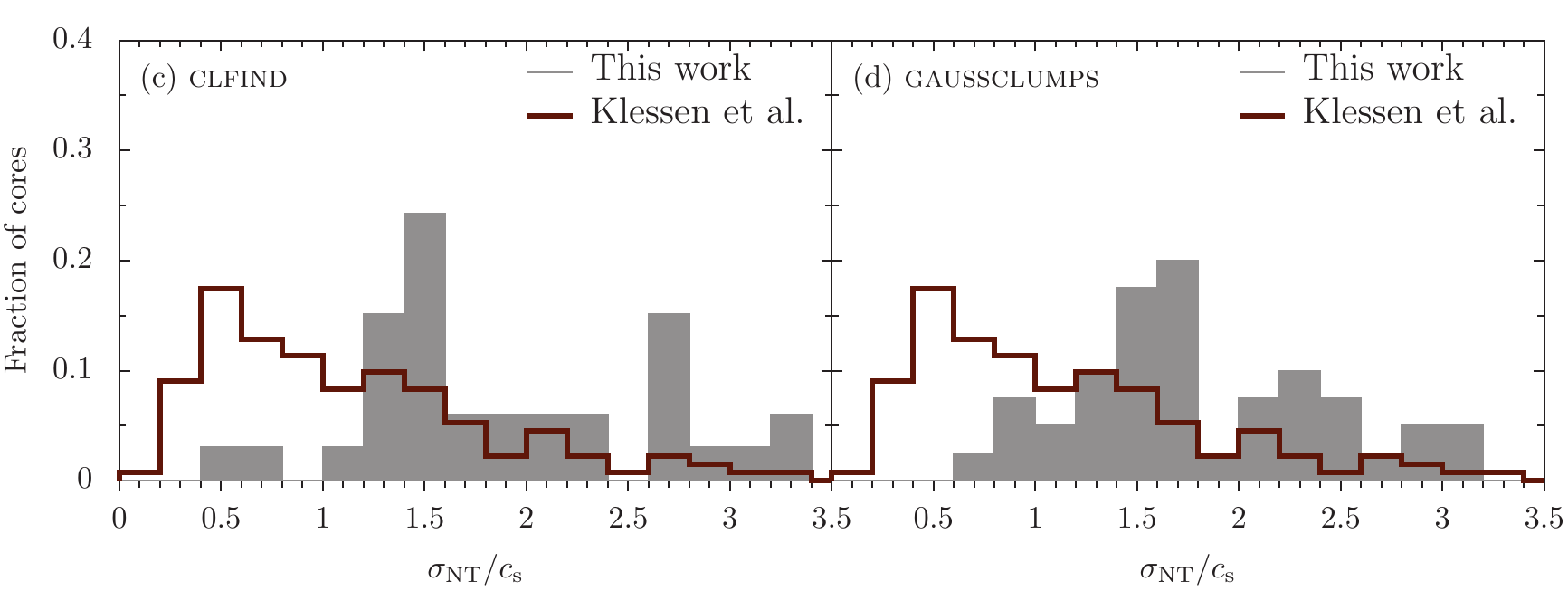}
\caption{Distribution of $f_\mathrm{turb}$, for the protostellar clump
  populations identified with \clfind\ (left panels) or
  \gclumps\ (right panels), measured from the \harp\ C$^{18}$O
  $J=3\to 2$ data. Overlaid for comparison are the distributions found by
  \citetalias{hkirk07} for 157 candidate cores using C$^{18}$O $J=2\to 1$
  (upper panels, pink) and N$_2$H$^+$ $J=1\to 0$ (upper panels, blue). A prediction for cores from the gravoturbulent
  simulation of \citet{klessen05} is also shown for turbulence driven
  on large scales (lower panels, brown). For that simulation the turbulent
  fraction extends beyond 3.5 to 4.3 (not plotted), though there are small numbers
  of clumps in those bins ($\sim 2$~per cent).}  
\label{fig:fturb_comp_proto}
\end{center}
\end{figure}

We take a core's total one-dimensional velocity dispersion,
$\sigma_\mathrm{C^{18}O}$ to be the quadrature sum of
its thermal and non-thermal dispersions, $\sigma_\mathrm{T}$
and $\sigma_\mathrm{NT}$ respectively: \begin{equation}
  \sigma_\mathrm{C^{18}O}^2=\sigma_\mathrm{T}^2 +\sigma_\mathrm{NT}^2.
\end{equation} To find $\sigma_\mathrm{NT}$ (plotted in Fig.\ \ref{fig:linewidths}), we calculate $\sigma_\mathrm{T}$: \begin{equation}
  \sigma_\mathrm{T}^2 =\frac{kT}{m_\mathrm{C^{18}O}m_\mathrm{p}},
  \label{eqn:sigma_t}\end{equation} where $m_\mathrm{p}$ is the mass of a proton and
  $m_\mathrm{C^{18}O}$ the relative molecular mass of C$^{18}$O
  ($m_\mathrm{C^{18}O}=30$). The clump temperature, $T$, is taken from the kinetic temperatures
of ammonia gas \citep{rosolowsky08}, where available, or 10\,K and
15\,K for identified starless
  and protostellar cores respectively where not. For those clumps
    without a \citetalias{hatch07} association we assume a
    temperature of 12\,K. We calculate the
  turbulent fraction, plotted in Figs.\ \ref{fig:fturb} to \ref{fig:fturb_comp_proto},
  $f_\mathrm{turb}=\sigma_\mathrm{NT}/c_\mathrm{s}$, where
  $c_\mathrm{s}=\sqrt{\gamma kT/\bar{m}m_\mathrm{p}}$ is the thermal sound speed in the bulk of the
  gas of mean molecular mass $\bar{m}=2.33m_\mathrm{p}$ (assuming 1 He
  for every 5 \htwo) and $\gamma$ the adiabatic index ($\sim 7/5$ for a
  diatomic molecule). For $T=10$\,K, $c_\mathrm{s}=0.22$\,\kms\ and at
  15\,K $c_\mathrm{s}=0.27$\,\kms. Summaries of \fturb\ for various populations are
  listed in Tabs. \ref{tab:fturbclfind} and \ref{tab:fturbgclumps}. 

A majority of the linewidths are supersonic with means $\langle
f_\mathrm{turb} \rangle=1.76\pm 0.09$\footnote{Errors quoted on
  averages throughout this paper are errors on the mean ($\sigma/\sqrt{N}$) and
  not simple standard deviations ($\sigma$).} and $1.71\pm 0.05$ for
the \clfind\ and \gclumps\ populations respectively. The level of
non-thermal broadening should take into account motions from
both: (a) the source itself i.e.\ infall, rotation and outflows and (b) the
source's natal environment. Young stars can inject significant
energy into their surroundings, increasing the non-thermal motions
within nearby star-forming cores. For example \citet{caselli95} found an
inverse relationship between a core's NH$_3$ linewidth and the distance to
the nearest young stellar cluster in Orion B. There are three such young
clusters in Perseus (see \citealp{hatchell05}): IC348, \ngc\ and the
Per OB2 association. Therefore, of our fields, \ngc\ and IC348 are
very near to young clusters, while L1448 and L1455 are further away (L1455
being the furthest).

\begin{table}
\caption{Summary of the turbulent fractions, \fturb, derived from
  Gaussian fits to the \ceighteeno\ \threetotwo\ line at the peak of
  every SCUBA clump identified with \clfind\ \citep{scubapaper}. The results are broken into either different regions in Perseus or clump classifications (see \S \ref{sec:clumpcats}). The
  quoted errors ($\sigma$) are errors on the mean ($\sigma/\sqrt{N}$) not sample deviations.}
\label{tab:fturbclfind}
\begin{tabular}{lrrrrr}
\hline
Population & Number &  \fturb\ & $\sigma_f$ & \\
& & \\
\hline
All & 81 & 1.76 & 0.09\\
\ngc\ & 37 & 2.05 & 0.12 \\
IC348 & 23 & 1.36 & 0.09 \\
L1448 & 14 & 1.8 & 0.2 \\
L1455 & 7  & 1.5 & 0.4\\
\hline
Starless          & 17 & 1.8 & 0.2\\
Protostars$^\dag$ & 33 & 1.89 & 0.12\\
Class 0           & 19 & 2.04 & 0.16 \\
Class I           & 14 & 1.7 & 0.2\\
\hline
\end{tabular}\\
$^\dag$~The Class 0 and I populations combined.
\end{table}

\begin{table}
\caption{Summary of the turbulent fractions, \fturb, for the
  \gclumps\ clump population as in Tab.\ \ref{tab:fturbclfind}.}
\label{tab:fturbgclumps}
\begin{tabular}{lrrrrr}
\hline
Population & Number &  \fturb\ & $\sigma_f$ & \\
& & \\
\hline
All  & 119 & 1.71 & 0.05 \\
\ngc\ & 64 & 1.91 & 0.07\\
IC348 & 25 & 1.44 & 0.07\\ 
L1448 & 23 & 1.52 & 0.13\\
L1455 & 7  & 1.5 & 0.3\\
\hline
Starless & 24 & 1.61 & 0.11\\
Protostars & 40 & 1.80 & 0.10\\
Class 0 & 27 & 1.73 & 0.11\\
Class I & 13 & 2.0 & 0.2\\
\hline
\end{tabular}\\
\end{table}

The variations in \fturb\ region-to-region (see Fig.\ \ref{fig:fturb}), which we might
  attribute to each region's distance from a young stellar cluster, are
  slightly different for the two populations. For the \clfind\ clumps
  (see Tab.\ \ref{tab:fturbclfind}), \ngc\ and L1448 have
  statistically similar fractions: $2.05\pm0.12$ and $1.8\pm 0.2$
  respectively. We ignore the L1455 populations in these comparisons
  as they are so few in number. IC348, on the other hand, has much smaller
  linewidths overall: $\langle f_\mathrm{turb}
  \rangle=1.36\pm0.09$. The \gclumps\ population (see
  Tab.\ \ref{tab:fturbgclumps}) has clumps in \ngc\ with much
  larger turbulent fractions than the similar \fturb\ means in IC348 and L1448: $1.91\pm 0.07$
  compared to $1.44\pm 0.07$ and $1.52\pm 0.13$. Therefore, for both populations, the IC348
  linewidths are smaller than in \ngc, even though both regions are
  close to stellar clusters. This is possibly (as noted in
  \citetalias{paper1}) because the IC348 IR cluster is old
and no longer actively forming stars, so it affects its
environs much less than the \ngc\ cluster. There is a wide range of
clump \fturb\ in L1448, which could be a result of a significant
contribution to the \ceighteeno\ linewidth from molecular
outflows. Many of the protostellar clumps in L1448 drive particularly
strong flows \citepalias{paper3}.

On dividing the clumps into protostellar and starless subsets (see
Fig.\ \ref{fig:fturb_bysource}), there appears to little difference
between the distributions of turbulent fraction for both source types
and clump populations. At the peak positions of the \clfind\ objects,
we find $\langle f_\mathrm{turb} \rangle=1.8\pm0.2$ for starless
clumps compared to $1.89\pm0.12$ for
protostars. Similarly, $\langle f_\mathrm{turb} \rangle=1.61\pm0.11$
and $1.80\pm0.10$ for starless and protostellar \gclumps\ sources
respectively. Further separation of the protostars into Class 0 and I
clumps does not yield statistical differences between their average
\fturb\ (see Tabs.\ \ref{tab:fturbclfind} and
\ref{tab:fturbgclumps}). Unfortunately, such simple averages do not provide the whole
picture we can see in the distributions (in
Fig.\ \ref{fig:fturb_bysource}). For example, if the small but significant
population of high \fturb\ starless cores were eliminated it would
seem that protostars have marginally higher \fturb\ on
average. However, the large uncertainties on the average \fturb\ do
emphasize the need for larger samples of objects. For the
\clfind\ population (but not the \gclumps\ sources) most of the high \fturb\ protostars are Class 0, and
all of the low \fturb\ Class I, which is perhaps what we might expect
if outflow power decreases from the Class 0 to I stage
(\citealt{bontemps96}; \citetalias{paper3}). Overall, there are few
significant differences between the starless and protostellar
distributions.

We can draw interesting comparisons with the \ceighteeno\ (\twotoone)
and \ntwohplus\ \fturb\ distributions presented by \citetalias{hkirk07}
(see Figs.\ \ref{fig:fturb_comp_starless} and
\ref{fig:fturb_comp_proto}). Our distributions are closer to the C$^{18}$O distributions of \citetalias{hkirk07} than their N$_2$H$^+$
results. The high levels of non-thermal motions, shown by some of our
sources, are simply not present in N$_2$H$^+$ data. A striking
contrast between the \ceighteeno\ distributions is that many of our cores have high
\fturb\ ($\ge 3.5$), which are not seen for either protostars or
starless cores by \citetalias{hkirk07}. In addition, many \citetalias{hkirk07}
starless cores have sub-thermal linewidths (\fturb$<1$). These
differences may be partly caused by the spectral resolution of our
observations ($\Delta v=0.15$\,\kms\ compared to \citetalias{hkirk07}'s
0.05\,\kms), rather than intrinsically different results. Our slightly
poorer resolution means we cannot probe as narrow linewidths and we
possibly blend many of the separate components, noted in
\citetalias{hkirk07}'s \ceighteeno\ spectra, into one. Given the
similarities it seems likely our \ceighteeno\ \threetotwo\ data probe
the same regions as the \twotoone\ observations of
\citetalias{hkirk07}, namely the outer parts of star-forming cores, referred
to by \citetalias{hkirk07} as their {\it envelopes}. This is probably because of \ceighteeno\ freeze-out in the dense, cold interiors of
star-forming cores. Such regions should see an enhancement in the
abundance of say \ntwohplus\ relative to
\ceighteeno. \citetalias{hkirk07} investigated this possibility for
their similar population of candidate cores in Perseus by
measuring the variation in the integrated
\ceighteeno-to-\ntwohplus\ ratio with peak SCUBA flux density (an approximate proxy
for central \htwo\ volume density). They found high
\ceighteeno-to-\ntwohplus\ ratios (i.e.\ little
\ceighteeno\ depletion) occurred mainly in starless cores,
which have the smallest SCUBA fluxes (i.e.\ lowest densities), whereas
high-flux cores (mainly protostars) have low ratios suggesting high
levels of \ceighteeno\ depletion. 

Given that our starless and protostellar C$^{18}$O distributions are
quite similar, we may also conclude (like \citetalias{hkirk07}) that protostars do not affect their
environment significantly. The starless distribution may differ from
the true prestellar one as starless clumps with the largest
linewidths are likely to be \emph{unbound}, and thus the true \emph{prestellar}
turbulent fractions will be smaller. Finally, we also note little
correspondence with the simulations of \citet{klessen05}, whose results for a large-scale
driven (LSD) gravoturbulent model are also plotted in Figs.\ \ref{fig:fturb_comp_starless} and
\ref{fig:fturb_comp_proto}. \citetalias{hkirk07} claim to have a good fit to the
\citeauthor{klessen05} data with their C$^{18}$O starless population, although the model was designed to
match N$_2$H$^+$ observations. 

\subsection{Virial theorem} \label{sec:virialmasses}

If the \ceighteeno\ data trace material that will go on to form
the final star and not just the ambient gas in which a \yso\ is
embedded, the core linewidth can inform us about its \emph{stability}, i.e.\ whether the cores are
gravitationally bound. We can then distinguish bound, starless
i.e.\ prestellar cores from unbound ones that will dissipate without
forming a star. 

The virial theorem is often used to examine this stability and we will
apply it to our two populations of clumps identified in
\scuba\ 850\,\micron\ data. The clump virial mass, $M_\mathrm{vir}$, is
used as an estimate of the internal energy of the clump whilst its dust mass at 850\,\micron, $M_{850}$, is as an estimate of the
potential energy. If we assume that the linewidths only reflect gravity and we have spherical clumps the virial theorem is \begin{equation}
  M_\mathrm{vir} = \frac{a \sigma^2 R}{G}, \end{equation} where
$\sigma$ is the average three-dimensional velocity dispersion, $R$ the
clump radius and $a$ a constant that depends on the form of the
density profile (a derivation can be found in e.g.\ \citealp{binneytremaine,rohlfswilson}). If we assume a power law
density profile, $\rho(r)\propto r^{-n}$, then
(e.g.\ \citealp*{maclaren88}): \begin{equation} a=\frac{5-2n}{3-n}.
\end{equation} Provided the \ceighteeno\ line of \fwhm, $\Delta
v_\mathrm{C^{18}O}$, traces the bulk of the gas with a Gaussian
velocity distribution and the same non-thermal linewidth as \htwo, we can estimate \begin{equation} \sigma^2 = \frac{3}{8\ln 2} \left(
  \Delta v_\mathrm{C^{18}O}^2 + \frac{8\ln 2 kT}{m_\mathrm{p}}\left(
  \frac{1}{\bar{m}}-\frac{1}{m_\mathrm{C^{18}O}} \right) \right),
\end{equation} where $m_\mathrm{p}$ is the proton mass, $\bar{m}=2.33$
to factor in the abundance of He relative to H$_2$ and
$m_\mathrm{C^{18}O}=30$. This yields for a shallow power law clump, $n=1.5$,
  consistent with the shapes found by \citet{enoch08} \begin{eqnarray}
    M_\mathrm{vir} & = & 3.4 \left( \frac{a}{4/3} \right) \left( \left(
    \frac{\Delta v_\mathrm{C^{18}O}}{1\,\mathrm{km\,s^{-1}}}
    \right)^2 +0.22\left(\frac{T}{12\,\mathrm{K}} \right) \right)
    \nonumber \\ & & \qquad \times \left(\frac{R}{\mathrm{0.02\,pc}} \right)\,\mathrm{M_\odot.}
  \end{eqnarray} We expect objects in equipartition to have $M_\mathrm{vir}\sim
  M_{850}$ whilst those that are self-gravitating should have $M_\mathrm{vir}\la
  2M_{850}$. 

We take data on the clumps' dust properties from
\citet{scubapaper}. Each clump's 850\,\micron\ mass ($M_{850}$) assumes the dust is optically thin, has an opacity,
$\kappa = 0.012$\,cm$^{2}$\,g$^{-1}$ and is at a single temperature,
$T_\mathrm{D}$. Again these temperatures are taken as the \ammonia\ kinetic
temperatures \citep{rosolowsky08}, where available, or 10\,K (for
starless cores) and 15\,K (for protostars) where not. The core radius, $R_\mathrm{dec}$, is the geometric mean of the two core
semi-major and -minor axis `sizes' each deconvolved with the beam
size. These `sizes' are the standard deviation of the pixel coordinates about the core centroid, weighted by
the pixel values.

There are considerable uncertainties in any dust and virial
  mass estimates. The errors are very difficult to quantify for individual
  clumps without detailed modelling. Thus, we follow complementary
  studies (e.g.\ \citealt{buckle10,enoch08}) and do not attempt to
  account for the uncertainties in our analysis, except for a discussion of their magnitude, which follows. The dust masses depend
  on the assumed distance to Perseus and the dust properties (temperature and
  opacity). We try to minimize the effects of dust temperature
  by using the \ammonia\ kinetic temperature as an estimate of $T_\mathrm{D}$,
  which should be an accurate measurement at high volume densities
  ($\gtrsim 10^4$\,\cmthree, e.g.\ \citealt*{galli02}), where the gas
  and dust are thermally coupled. Nevertheless our dust temperature
  estimates still do not account for variations in the dust
  temperature across a clump. A range of distances have been used
  in the literature for Perseus (220 to 350\,pc see e.g.\ \citetalias{paper1}) and
  indeed it may not be a contiguous cloud at a single distance. These
  result in an uncertainty of a factor of $\sim 5$ in the dust mass
  estimates. The virial masses depend on the assumed clump
profile, with steeper profiles producing smaller masses, however it only varies by a factor of 1.7 between a constant
density and $1/r^2$ profile. This combined with the uncertainties in
the distance and linewidth produce again around a factor of $\sim
5$ in uncertainty. As we previously noted (\S \ref{sec:linewidths}), it is likely
that the \ceighteeno\ \threetotwo\ line only traces the envelope of
a clump, which could lead to an over-estimate of the non-thermal
\htwo\ linewidth (compared to estimates from say \ammonia\ or
\ntwohplus) and correspondingly the virial mass. Even if our
estimates of the dust and virial masses were infallible, assuming that
a clump with $M_\mathrm{vir}\gg M_{850}$ is unbound may be
misleading; external pressure and/or magnetic fields may contain
such a clump. For instance \citetalias{hkirk07} calculate that cores with external pressures consistent with their previous 
Bonnor-Ebert sphere modelling \citep*{kirk06}, should be considered in
\emph{equipartition} not merely self-gravitating if
$M_\mathrm{vir}\sim 2 M_{850}$. 

\begin{figure}
\begin{center}
\includegraphics[height=0.47\textheight]{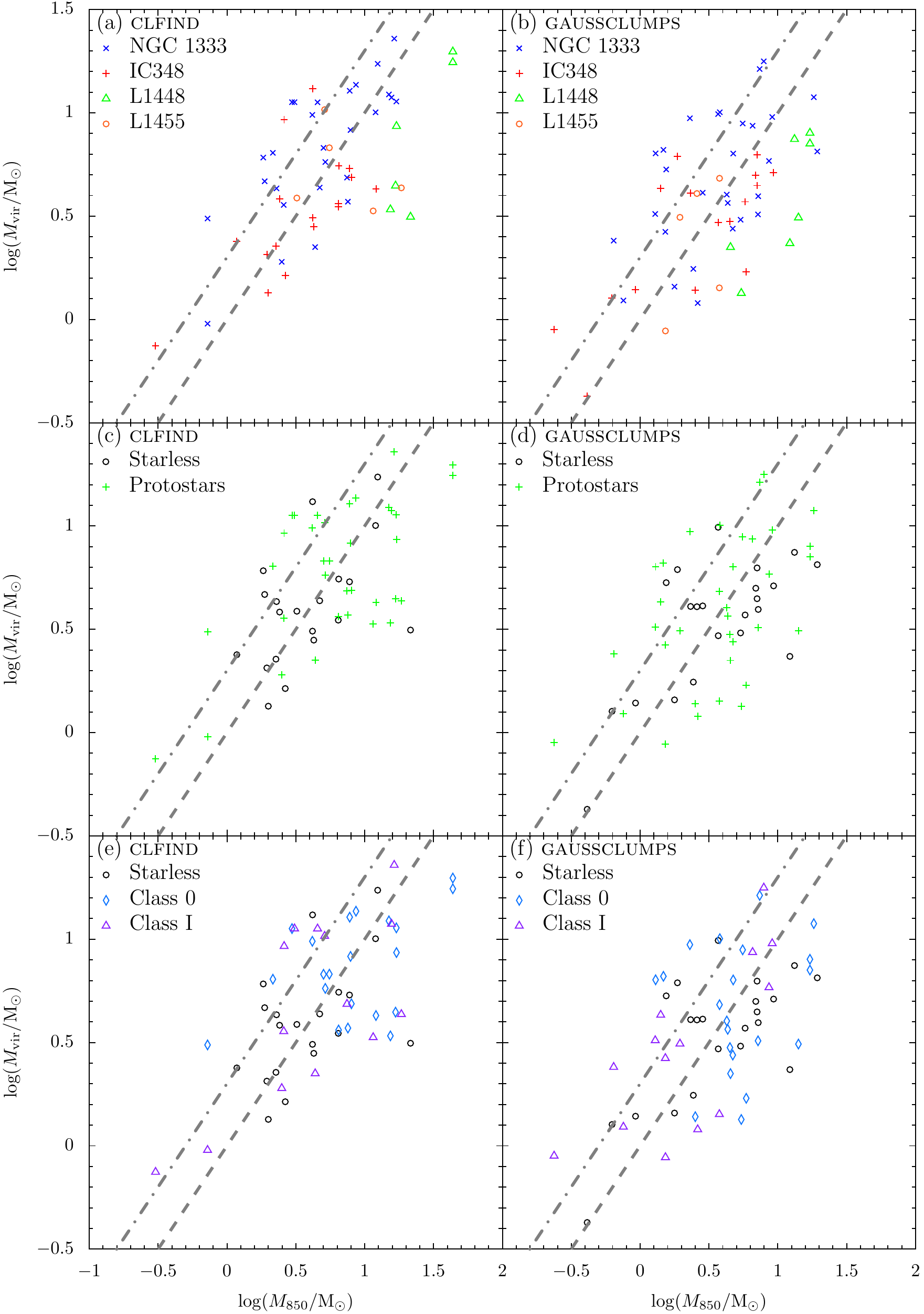}
\caption{Virial mass versus clump
  dust mass for the \clfind\ (left panels, (a), (c) and
  (e)) and \gclumps\ (right panels, (b), (d) and
  (f)) populations, differentiated by region (top panels,
  (a) and (b)) or the classifications of \citetalias{hatch07} (the
  rest). Virial masses are estimated from the C$^{18}$O
  $J=3\to 2$ linewidth at the catalogue positions of
  \citet{scubapaper}. The masses are calculated using the
  temperatures from \citet{rosolowsky08} where possible or 10 and
  15\,K for starless and protostellar cores respectively where
  not. Lines denote where we expect cores to be in equipartition, i.e.\ $M_\mathrm{vir}=M_{850}$ (dashed)
  or the limit where they are self-gravitating,
  $M_\mathrm{vir}=2M_{850}$ (dot-dashed), although the reader should
  note the considerable uncertainties in both masses. Only clumps which have a
    \citetalias{hatch07} source association are plotted.}  
\label{fig:virial}
\end{center}
\end{figure}

The \ceighteeno\ virial masses, which we plot in Fig.\ \ref{fig:virial} against the
850\,\micron\ mass, lie in the range 0.8 to 22.9\,M$_\odot$
for the \clfind\ sources and 0.4 to 17.8\,M$_\odot$
for the \gclumps\ sources, reflecting their smaller radii. Most of the clumps lie scattered near the `equipartition' line, $M_\mathrm{vir}
=M_{850}$. The results are similar for the different algorithms,
regions and evolutionary types (see Tables \ref{tab:mvir_clfind} and
\ref{tab:mvir_gclumps} for summaries of the population statistics). Least-squares straight-line fitting results in poorly-constrained power law exponents of
0.8--0.9 for almost all the correlations. 

Given the uncertainties in both the virial and dust mass
  estimates, it is not possible to draw unequivocal conclusions about
  the stability of individiual clumps. However, we will compare distinct populations. The different regions show similar behaviour with a lot
of scatter. The L1448 population is outlying for both
algorithms with a high $M_{850}/M_\mathrm{vir}$
ratio. L1448 is interesting as it contains large \scuba\ flux
densities and massive outflows but has relatively weaker C$^{18}$O
emission. The \ngc\ population has a smaller
  $M_{850}/M_\mathrm{vir}$ ratio than the others, which we might
interpret as being more unbound and therefore transient, with a larger number of cores
above the $M_\mathrm{vir} = M_{850}$ threshold. This could be a result
of its more perturbed environment. 

Protostellar clumps, by definition should be gravitationally
  bound. If starless clumps occupy a similar $M_\mathrm{vir}-M_{850}$
  sample space, then this suggests that they too are a gravitationally
bound population. In Fig.\ \ref{fig:virial}, the starless and protostellar clumps occupy the
same regions of the plot and their mass ratios are very similar in
Tabs.\ \ref{tab:mvir_clfind} and \ref{tab:mvir_gclumps}. This implies that the starless population
is gravitationally bound as well and truly {\it prestellar} in
nature (as found by \citealt{enoch08}).  
 
The clumps also seem spread around the `equipartition' line
$M_\mathrm{vir} = M_{850}$. Approximate equipartition was also found by \citet{caselli02} in
a sample of 60 starless cores using N$_2$H$^+$ linewidths. Other
studies e.g.\ the C$^{18}$O cores looked at by \citet{tachihara02}
found the virial mass considerably larger than the core mass
estimate. The latter result is successfully explained by the
gravoturbulent models of \citet{klessen05}, who find a large
majority of starless cores with $M_\mathrm{vir}\gg M$ for both large and small
driving-scale turbulence models, with their protostellar sources
non-coincident in parameter space at $M>M_\mathrm{vir}$. Our results
would not favour such a model for Perseus, even in its
clustered environment.

\begin{table}
\caption{Summary of virial mass statistics for the
  \clfind\ \scuba\ clumps identified in \citet{scubapaper}, which have
  a \citetalias{hatch07} classification. All the
  quoted errors ($\sigma$) are errors on the mean ($\sigma/\sqrt{N}$)
  not simple sample deviations.}
\label{tab:mvir_clfind}
\begin{tabular}{lrrrrr}
\hline
Population & No &  \mvir\ & $\sigma_M$ & $M_{850}/M_{vir}$ & $\sigma_\mathrm{ratio}$\\
& & (\msun) & (\msun)\\
\hline
All & 50  & 7.8 & 0.9 & 1.32 & 0.13 \\
\hline
\ngc\ & 24 & 8.9 & 1.2 & 0.96 & 0.15 \\
IC348 & 16 & 4.0 & 0.8 & 1.38 & 0.22 \\
L1448 & 5  & 12.6 & 3.9 & 2.7 & 0.5 \\ 
L1455 & 5  & 7.9 & 1.6 & 1.4 & 0.6\\
\hline
Starless & 17   & 4.6 & 0.8 & 1.4 & 0.2\\
Protostars & 33 & 9.6 & 1.2 & 1.3 & 0.2\\
Class 0 & 19    & 10.1 & 1.4 & 1.3 & 0.2\\
Class I & 14    & 9 & 2 & 1.2 & 0.3\\
\hline
\end{tabular}
\end{table}

\begin{table}
\caption{Summary of virial mass statistics for the
  \gclumps\ \scuba\ clumps identified in \citet{scubapaper}, which have
  a \citetalias{hatch07} classification. All the
  quoted errors ($\sigma$) are errors on the mean ($\sigma/\sqrt{N}$)
  not simple sample deviations.}
\label{tab:mvir_gclumps}
\begin{tabular}{lrrrrr}
\hline
Population & No &  \mvir\ & $\sigma_M$ & $M_{850}/M_{vir}$ & $\sigma_\mathrm{ratio}$\\
& & (\msun) & (\msun)\\
\hline
All   & 64 & 7.1 & 0.7 & 0.9 & 0.1\\
\hline
\ngc\ & 30 & 8.9 & 1.1 & 0.68 & 0.08\\
IC348 & 17 & 5.0 & 0.8 & 0.75 & 0.14\\
L1448 & 12 & 6.5 & 1.5 & 2.2 & 0.4\\
L1455 & 5  & 4.3 & 1.1 & 0.8 & 0.3\\
\hline
Starless   & 24 & 6.0 & 0.6 & 0.87 & 0.16\\
Protostars & 40 & 7.8 & 1.0 & 0.91 & 0.13\\
Class 0    & 27 & 8.4 & 1.1 & 1.06 & 0.18\\
Class I    & 13 & 7 & 2 & 0.67 & 0.14\\
\hline
\end{tabular}
\end{table}

\subsection{Localized velocity gradients} \label{sec:rotation}

Systematic variations in the \ceighteeno\ line centre velocity are
apparent across the face of many of the identified \scuba\ clumps,
e.g.\ the velocity gradually increases along a particular
direction. There are a number of plausible explanations for such
gradients: (i) rotation, (ii) outflows, (iii) motions between smaller unresolved
constituent clumps or (iv) if gravoturbulent
models are true, cores form at the stagnation points in convergent flows
(e.g \citealp{padoan01b}) and velocity gradients may arise from colliding gas
streams. It is difficult to distinguish between each scenario so we
follow the majority of studies and focus on analysing the rotational properties
of the clumps from measured velocity gradients. 
 
A strong motivation to understand the details of rotation in
star-forming cores is provided by the discs out of which most, if not
all, stars are born (e.g \citealp{shu87}). Planets are thought to originate
inside such protoplanetary discs and the details of their
formation crucially depend on various disc parameters, such as surface
density, controlled by the detailed evolution of the angular momentum
of the parent star-forming core
(e.g.\ \citealp{lissauer93,ruden99}). Additionally, there is the
classical `angular momentum problem' of star formation (e.g.\ \citealp{spitzer78}): the angular
momentum of prestellar cores is orders of magnitude larger than that
which can be contained within a single star, even though cores are
observed to be rotating much less than originally predicted
\citep{goodman93,caselli02}\footnote{Furthermore, \citet{dib10} point out
that measurements of core angular momenta from global velocity
gradient fitting tend to overestimate the intrinsic (three-dimensional) angular momenta by a factor
of $\sim8$-10, as complicated fluctuations in the three-dimensional
velocity field are smoothed out.}. A plausible solution is provided by
magnetic braking at the early low-density phases, where the field
lines are strongly coupled to the gas and transfer angular momentum from the contracting core to the
surrounding medium (e.g.\ \citealp{mouschovias87}). Recent
results suggest that gravitational interactions may dominate over magnetic
braking. MHD models of self-gravitating, decaying \citep{gammie03} and
driven \citep{li04} turbulence have angular momenta consistent with
each other and observations but 
\citeauthor{jappsen04} (\citeyear{jappsen04},
hereafter \citetalias{jappsen04}) and
\citet{tilley04} find similar results in purely hydrodynamic
frameworks. Even if magnetic fields do dominate, they cannot indefinitely strip away angular momentum or support a
core against collapse because ambipolar diffusion will
eventually lead to dynamic collapse. Finally, for cores to fragment into multiple systems
some angular momentum must be present and perhaps cores with the largest
angular momenta will go on to form binaries (e.g.\ \citealp{larson03,goodwin07}).

The pre-eminent observational study in this area is the search for solid-body-rotation
in $\sim$40 NH$_3$ cores by
\citeauthor{goodman93} (\citeyear{goodman93}, hereafter \citetalias{goodman93}). In all their objects the
rotational energy is at most a few per cent of the gravitational and
cannot provide support. An evolutionary sequence can be built using
their results with others (e.g.\ \citealp{caselli02}), demonstrating the
specific angular momentum ($j=J/M$) decreases with decreasing scale \citepalias{jappsen04}.  

Rotational signatures become more complicated at higher resolution
and once collapse has started in protostars. \citet{belloche02}
studied a young Class 0 protostar, IRAM 04191+1522, finding two distinct regimes
of collapse: the inner ($r\la 2000-4000$\,AU), rapidly
collapsing and rotating whilst the outer ($4000 \la r\la 11000$\,AU)
has only moderate infall and rotation. The fall in rotational velocity beyond the 4000\,AU boundary and the flat inner profile
suggests that the inner region's angular momentum is conserved whilst it
is dissipated in the outer perhaps by magnetic braking. At the distance
to Perseus the $r\sim4000$\,AU boundary is bigger than the \jcmt\ beam
(32\,arcsec diameter compared to 15\,arcsec) so we may be able to
probe the inner regions, although C$^{18}$O will possibly freeze-out at $\la 5000$\,AU in
starless and young protostellar cores. 

\subsubsection{Velocity gradient fitting}

A clump undergoing solid-body rotation will display a linear
velocity gradient across its face perpendicular to the rotation
axis. we follow \citetalias{goodman93} and fit a linear gradient,
$\mathbf\nabla v_\mathrm{LSR}$, across each
clump using: \begin{equation} v_\mathrm{LSR} = v_0 + a
  \Delta \alpha +b \Delta \delta \label{eqn:velocitygradient} \end{equation} where $v_0$ is the
systemic clump velocity, $\Delta \alpha$ and $\Delta \delta$ the
angular ascension and declination offsets from the clump centre and $a$ and $b$ the projections of the gradient per radian
on to the $\alpha$ and $\delta$ axes. The gradient has a magnitude, ${\cal G}= |\mathbf\nabla
v_\mathrm{LSR}|=(a^2+b^2)^{1/2}/D$, where $D$ is the distance to the
object, at an angle (east of north), $\theta_{\cal G}= \arctan(a/b)$.

We explore the \ceighteeno\ \threetotwo\ velocity field of our two
populations of \scuba\ 850\,\micron\ clumps. At each \scuba\ map
pixel the corresponding spectrum in the \harp\ data has
been fitted with a Gaussian profile to extract its centre velocity and
linewidth. Fits were only performed where the spectral peak was greater than three times the rms noise,
estimated on a line-free portion of the spectrum for every
spatial position. L1455 was omitted, as its weak \ceighteeno\ emission had too few fits. We then used the Levenberg-Marquardt algorithm
implemented in the scientific python module,
SciPy\footnote{\url{www.scipy.org}.}, to perform non-linear
least-squares fitting of Equation \ref{eqn:velocitygradient} to every clump
about its \scuba\ peak. \citetalias{goodman93} performed the same
analysis on sets of randomly generated maps with no gradient and
report that approximately 10~per cent of those random maps were found
to have significant gradients if a ``3 $\sigma$''
criterion is used i.e.\ ${\cal G} \ge 3\sigma_{\cal G}$. Using the
same significance criterion we expect similar levels of reliability.

\subsubsection{A signature of rotation?}

We noted there are four plausible causes of velocity
gradients. In this section, we look at some of the suggestions other than
rotation in more detail. First, some of our clumps may be composed of smaller
constituents. The gradient then might measure the velocity
dispersion of the multiple cores \citepalias{goodman93}. If
  the SCUBA clumps are gravitationally bound, as suggested in \S
  \ref{sec:virialmasses}, such dispersion in a multi-core clump
may signal its rotation about a common centre. \citetalias{hatch07} estimate within
their similar core catalogue, $\ga 10$ per cent
of clumps will break into multiple sources, by comparing the most luminous sources to higher resolution observations. 

Second, the exact region that is assigned to each clump can change the
fitted gradient's magnitude, direction and level of significance. One
of \citetalias{goodman93}'s original sample, TMC-1C, a
starless core in Taurus, has been mapped over a larger area at
higher resolution by \citet{schnee07}. They found a more complicated pattern of local
gradients no longer consistent with solid-body or differential
rotation, which had previously been noted by \citet{caselli02}.  

Finally, if the C$^{18}$O line centre is affected by outflows, then this could be misinterpreted as a sign of
rotation. \citetalias{goodman93} also fitted velocity gradients to
C$^{18}$O $J=1\to 0$ data in a sub-sample of their cores -- both with and without embedded objects --
and found they were in the same direction as the gradients from NH$_3$. However, they did not
formally examine any outflows. Towards starless cores, outflows should
not be a problem. To explore the effect on our gradients, we
compare the direction of the velocity gradient to the orientation of
the outflow from every protostellar source with a significant
gradient in the \clfind\ catalogue. We estimated the outflow
position angle from our CO $J=3\to 2$ datacubes (a full
outflow analysis was presented in \citetalias{paper3}). For some of
the protostellar clumps, the outflows are complicated and we
have ignored such cases, only comparing clumps where a characteristic bipolar
structure is apparent (see Table \ref{table:ouflowvsgradient}). In a number of these cases the velocity gradient is
in the same direction as the outflow, implying we are tracing the
outflow rather than solid-body rotation with the gradient. If we were viewing solid-body rotation of
the core or a disc around it, we would expect the gradient and outflow
to lie perpendicular to one another. 

\begin{table}
\caption{Approximate
  outflow orientation compared to the velocity gradient direction for \clfind\ clumps with clear bipolar
outflows. Classifications are from \citetalias{hatch07}.}
\begin{tabular}{lccccc}
\hline
Sub-region & \clfind\ & Class &
$\theta_{\cal G}$ & Outflow orientation \\
 & ID & & (deg E of N) & (deg E of N) \\ 
\hline
NGC\,1333 & 6 & 0 & 126 & $-113$\\
IC348 & 1 & 0 & $-51$ & $-64$\\
IC348 & 2 & 0 & 50 & 68\\
IC348 & 4 & 0 & 4 & 0\\
L1448 & 1 & 0 & 139 & 124\\
L1448 & 2 & 0 & 78 & 165\\
L1448 & 4 & 0 & 154 & 168\\
\hline
\end{tabular}
\label{table:ouflowvsgradient}
\end{table}  

\begin{figure*}
\begin{center}
\includegraphics[height=5cm]{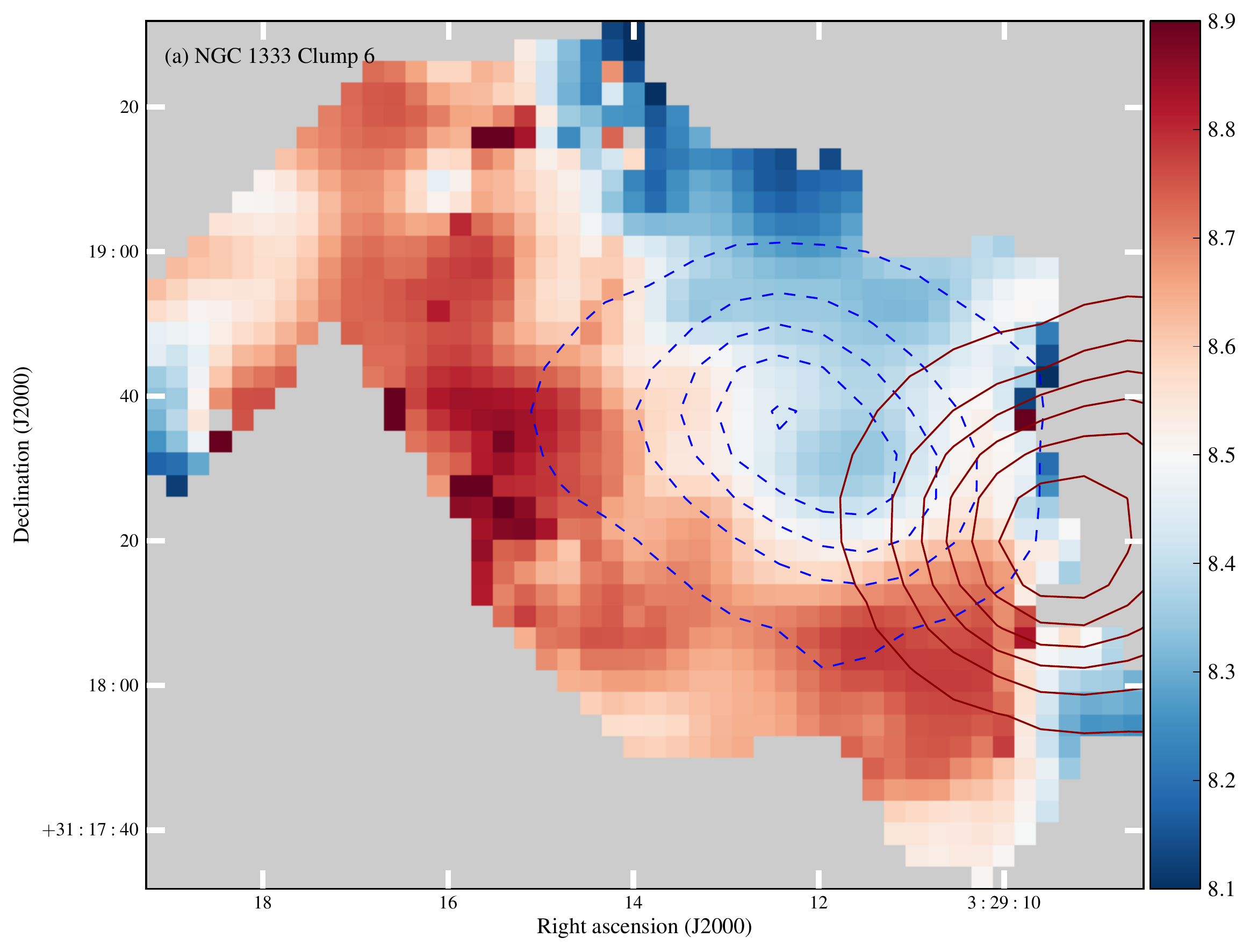}
\includegraphics[height=5cm]{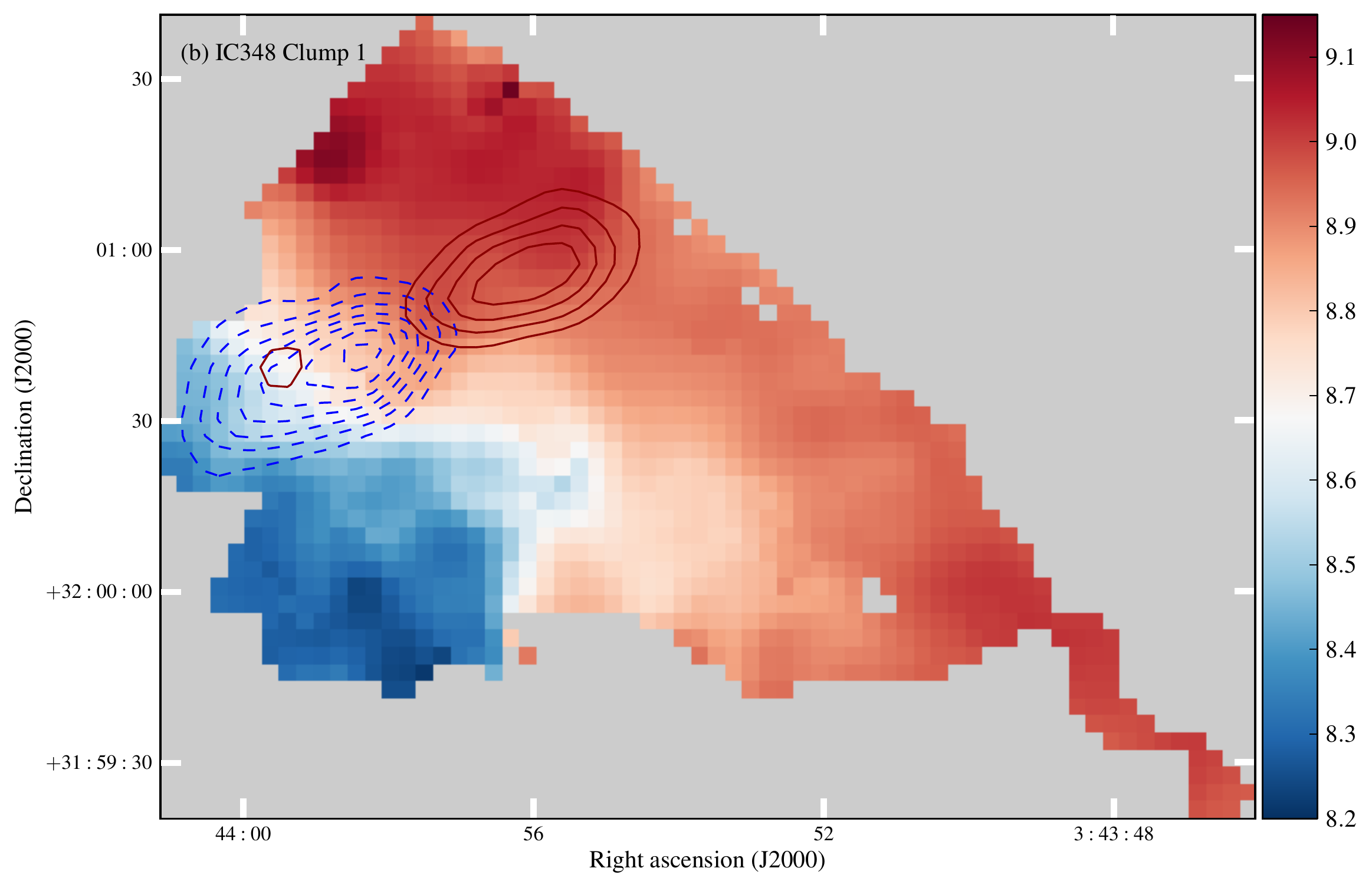}
\includegraphics[height=5cm]{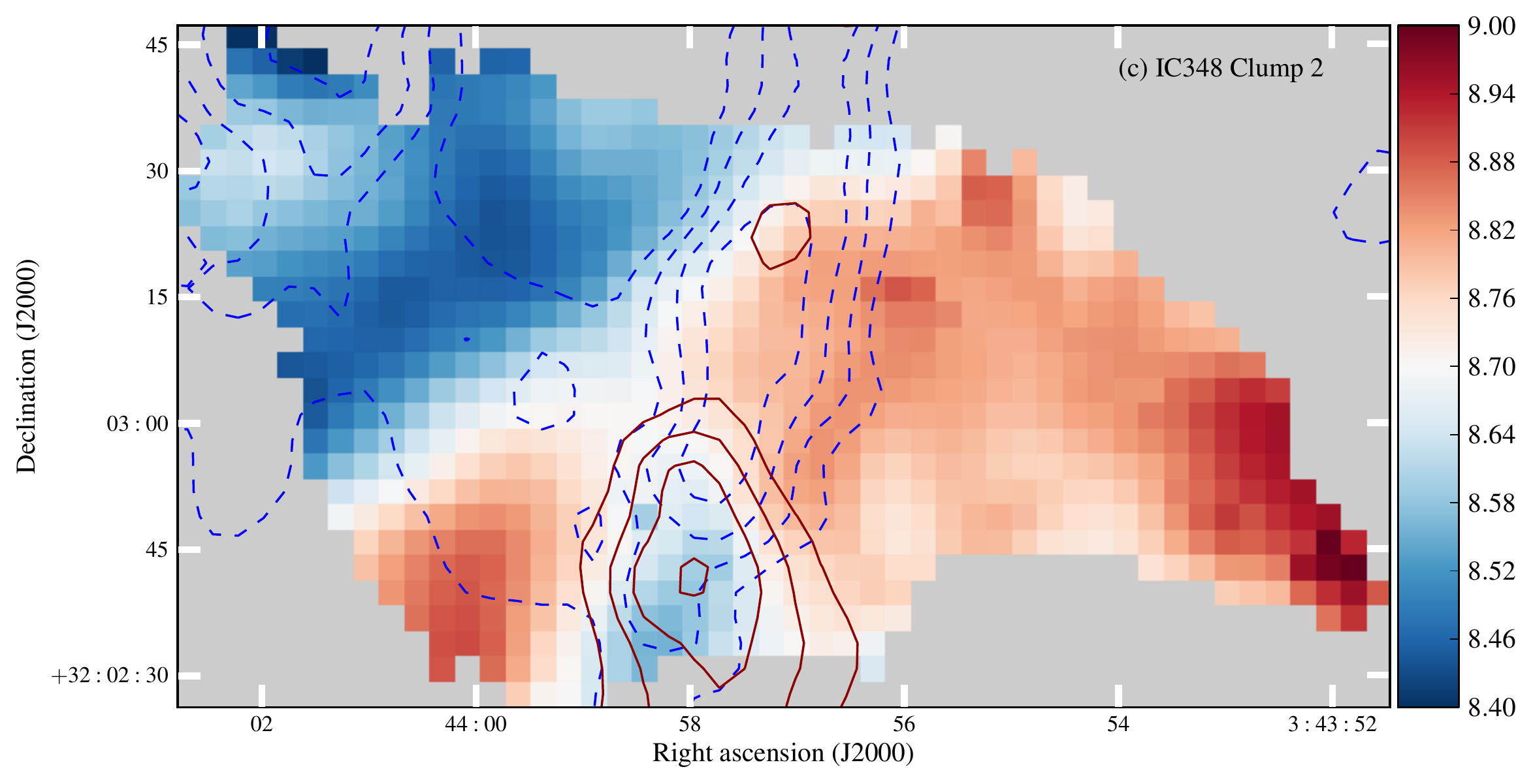}
\includegraphics[height=5cm]{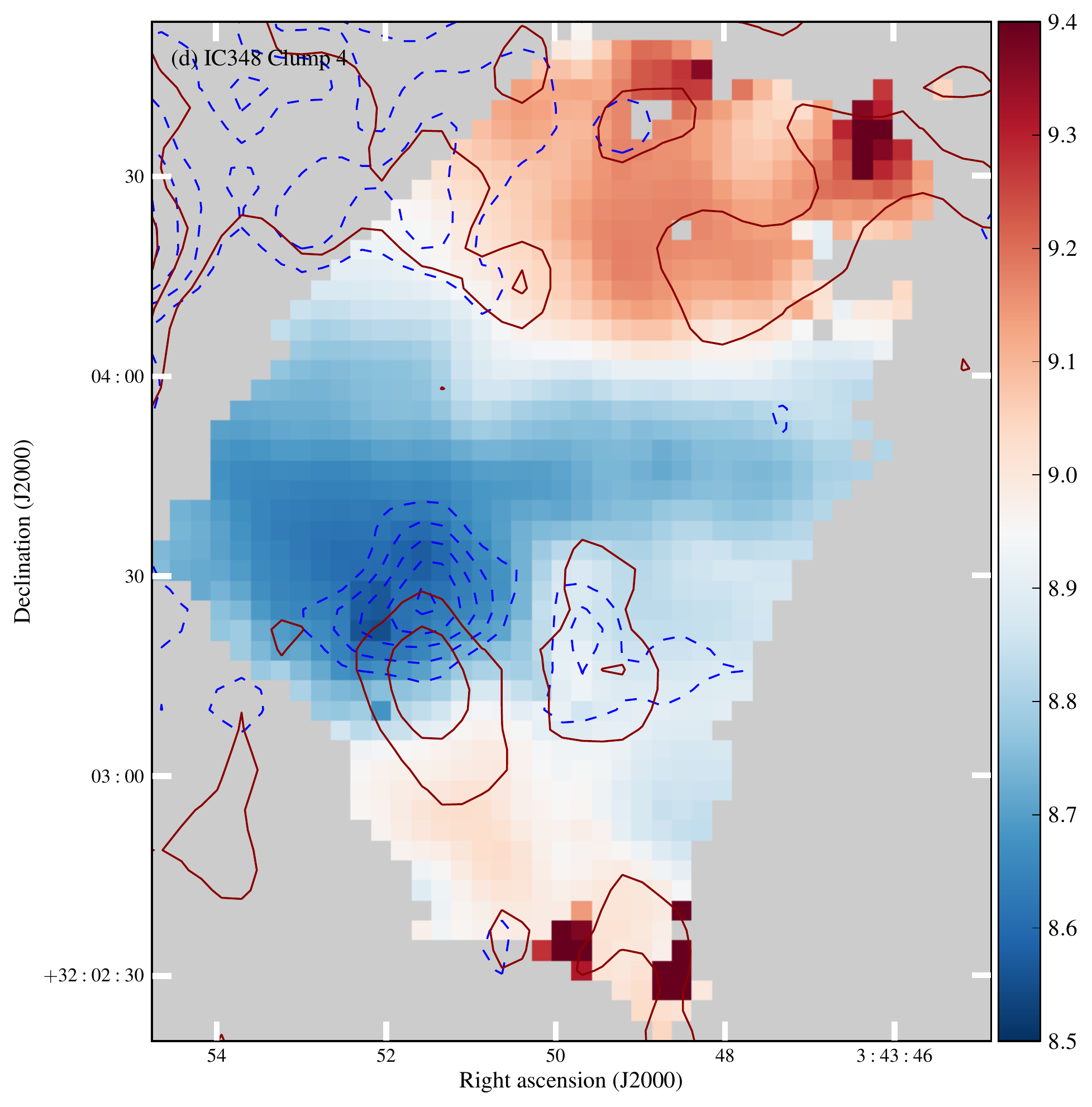}
\includegraphics[height=5cm]{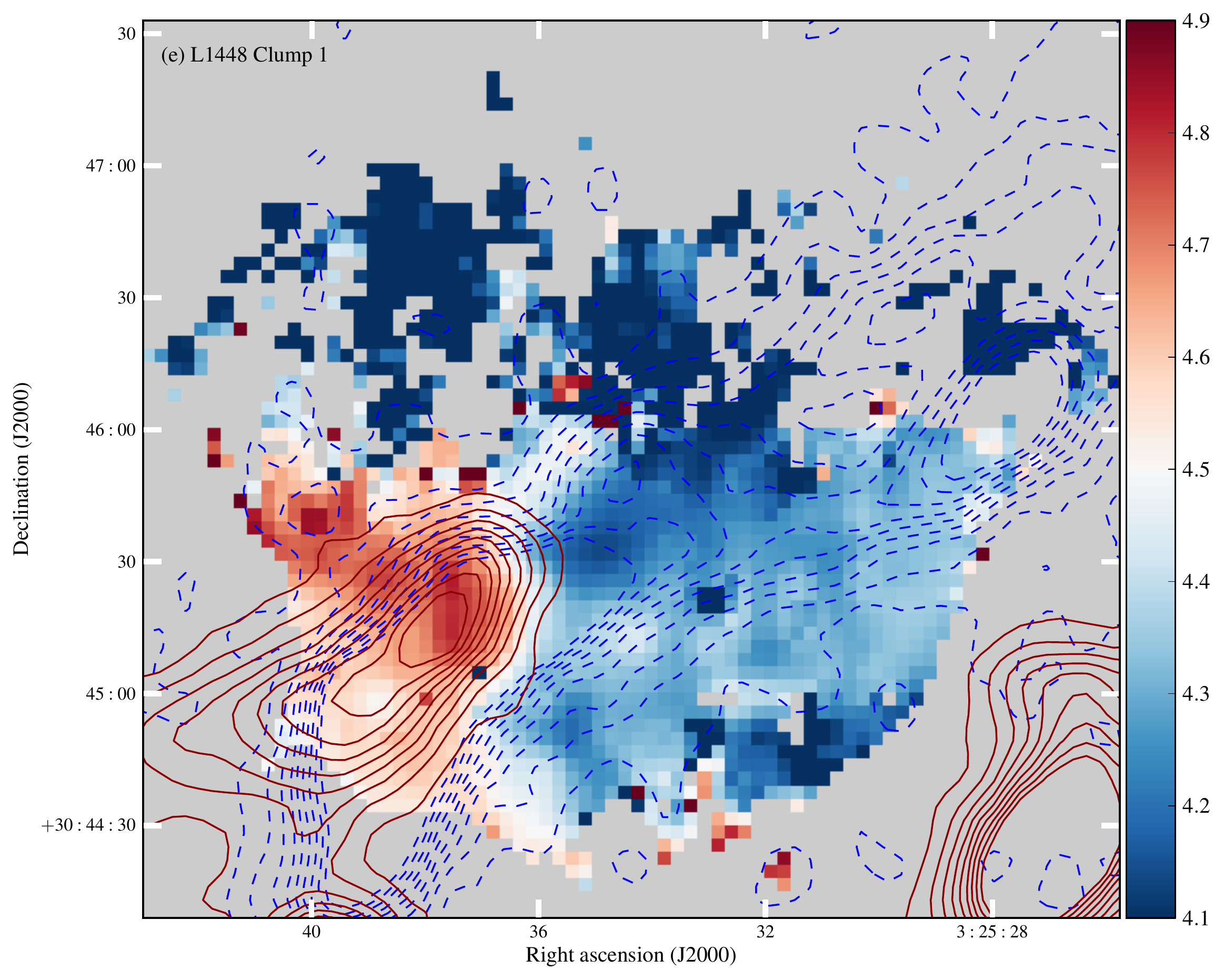}
\includegraphics[height=5cm]{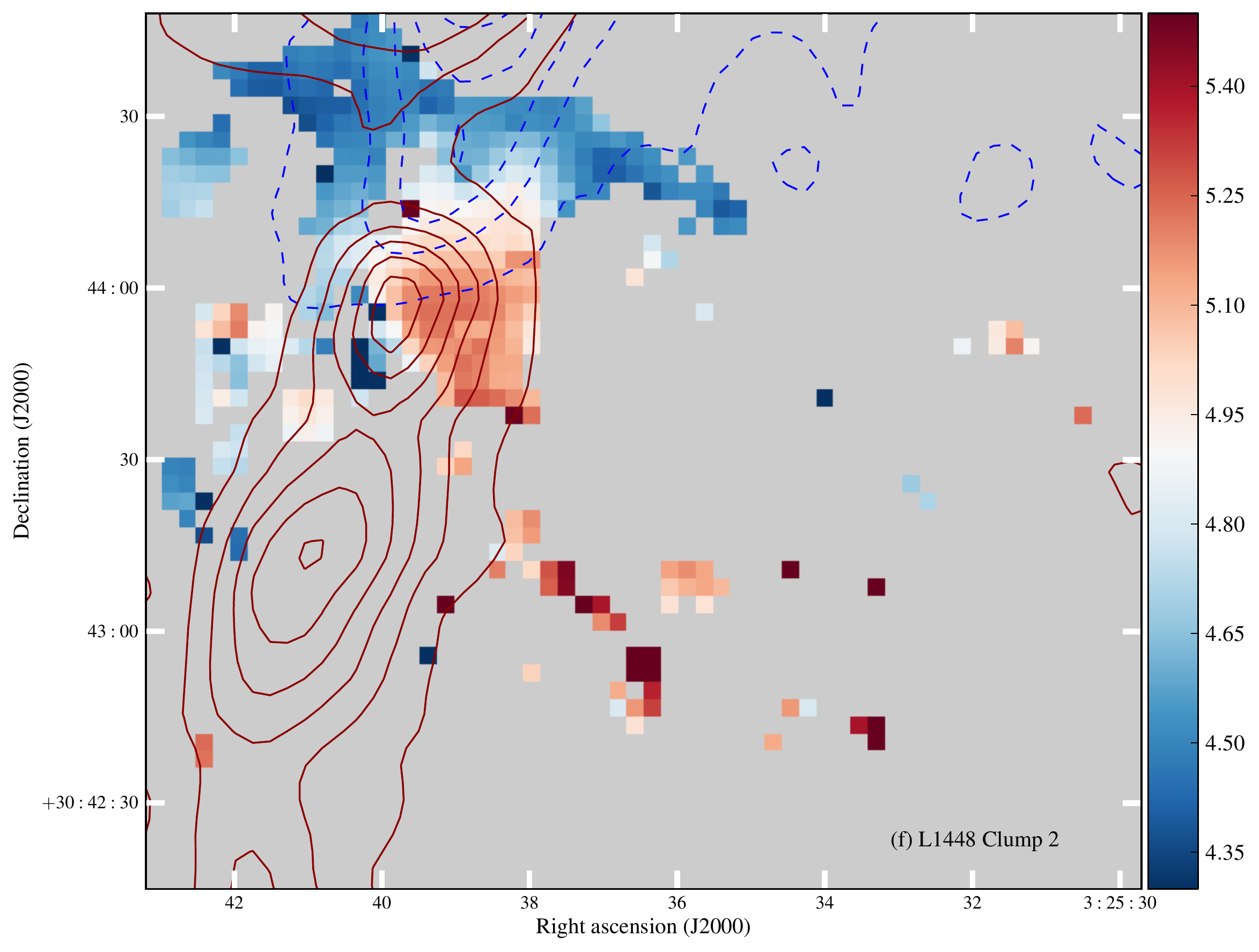}
\includegraphics[height=5cm]{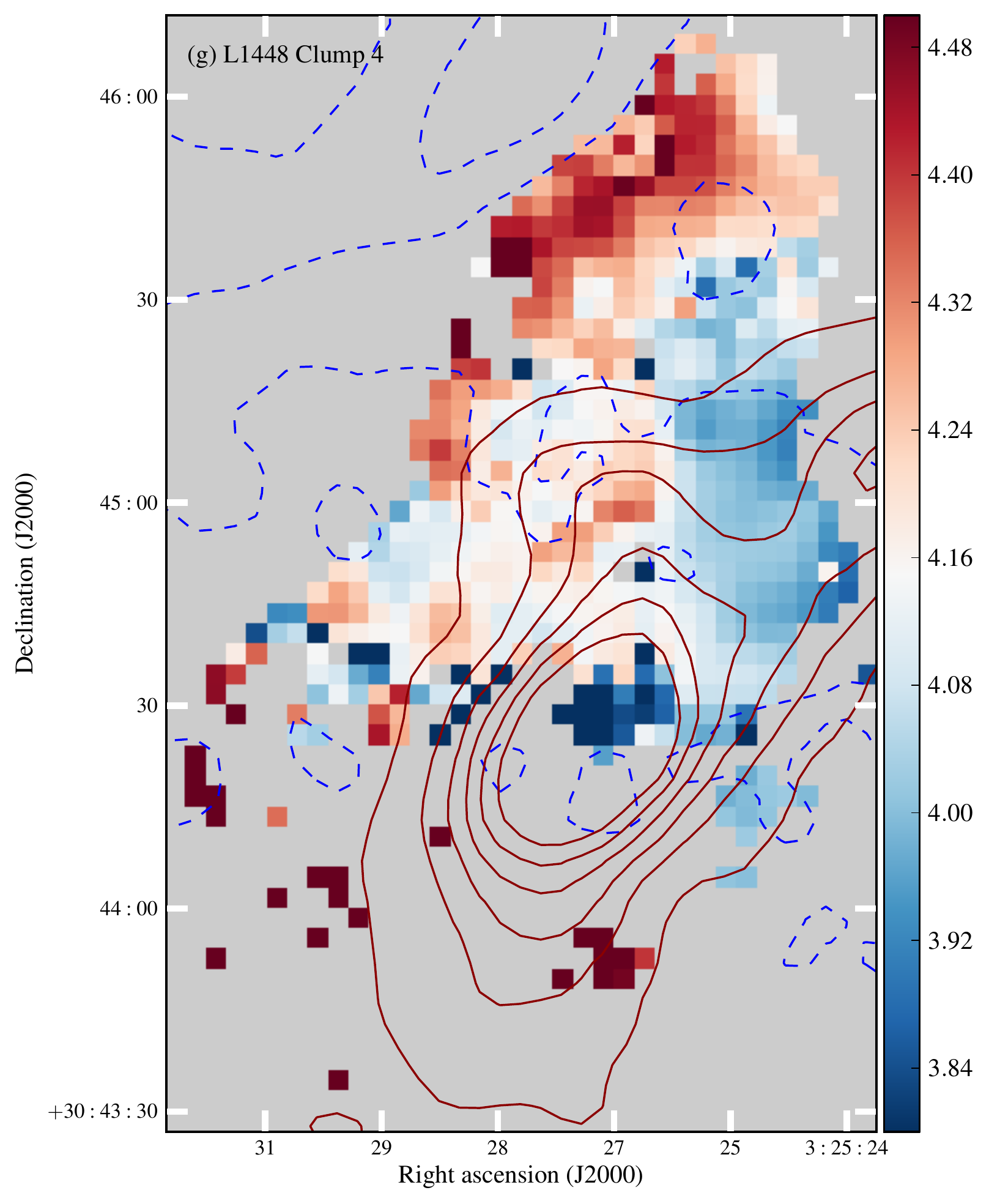}
\caption{Outflow orientation compared to the velocity
gradient direction for clumps in Table \ref{table:ouflowvsgradient}. The colour-scale is the
C$^{18}$O $J=3\to 2$ line centre in km\,s$^{-1}$ for map pixels allocated to each clump. Overlaid are contours of CO $J=3\to$ integrated antenna
temperature $\int T_\mathrm{A}^*\mathrm{d}v$ for the blue- and
red-shifted line wings, integrated from: $-5$ to
3 (blue) and 12 to 18\,km\,s$^{-1}$ (red) for \ngc, $-5$ to
6 (blue) and 11 to 15\,km\,s$^{-1}$ (red) for IC348 and $-5$ to
0 (blue) and 8 to 15\,km\,s$^{-1}$ (red) for L1448. Contours are at
various heights in K\,km\,s$^{-1}$ ($n$ is an integer from zero
to $n_\mathrm{max}$): (a) (2+4$n$), $n_\mathrm{max}=4$
(blue) and (4+4$n$), $n_\mathrm{max}=6$ (red). (b)
(6+2$n$), $n_\mathrm{max}=6$ (blue) and (6+2$n$),
$n_\mathrm{max}=3$ (red). (c) (8+1.5$n$), $n_\mathrm{max}=6$
(blue) and (4+1.5$n$), $n_\mathrm{max}=8$ (red). (d) (5+$n$), $n_\mathrm{max}=4$
(blue) and (2+$n$), $n_\mathrm{max}=2$ (red). (e) (2+$n$), $n_\mathrm{max}=8$
(blue) and (2+2$n$), $n_\mathrm{max}=7$ (red). (f)
(2+4$n$), $n_\mathrm{max}=5$ (both blue and red). (g)
(2+4$n$), $n_\mathrm{max}=5$ (blue and red). }
\label{fig:outflow_vs_orientation_byclump}
\end{center}
\end{figure*}

The data themselves present a
more complicated picture. In Fig.\
\ref{fig:outflow_vs_orientation_byclump}, we overlay the outflows on top
of the \ceighteeno\ centre velocities for the sources in Table
\ref{table:ouflowvsgradient}. Often the higher C$^{18}$O line centres
(i.e.\ red-shifted gas) correspond closely to
the extent of the red-shifted CO outflow lobe and similarly for the
blue-shifted gas. For instance, the complex outflow structure in L1448-1 is closely
followed by the line centre data. Furthermore, the blue-shifted
outflow lobe in NGC\,1333-6 is coincident with blue-shifted
\ceighteeno\ line centres. The outflow itself extends away from
the clump to the west, resulting in orientations that seem very
disparate in Table \ref{table:ouflowvsgradient}. In two cores there
seems to be little outflow-gradient correspondence: IC348-2 and L1448-4. Perhaps it is
more convincing to argue that none of the cores exhibit gradients perpendicular
to their outflows. Therefore, there appears to be a link between the outflow orientation and the
velocity gradient, although how much this influences the gradient found
in every protostar is difficult to quantify. 

Thus, we must view the
interpretation of a linear velocity gradient as unequivocal evidence
of solid-body rotation with some scepticism. However, in the absence of a strong outflow
i.e.\ for starless clumps and assuming we can consider
each clump as a single object, rotation would seem to be a likely
cause of a coherent velocity gradient and the only one we consider in
the following analysis. 

\subsubsection{Fitted gradients}

Table \ref{detectionrates_gradients} lists the rates of detecting
significant velocity gradients towards various core populations,
separated by their different clump-finding algorithms, regions or types of source. They are similar for both algorithms with
approximately a third of clumps displaying significant velocity
gradients. The detection rate for Class 0 sources in the
\clfind\ catalogue is nearly twice as
large as for the other types and Class 0s with
\gclumps. This is probably because Class 0 sources
have typically stronger outflows than Class Is
(e.g.\ \citealp{bontemps96}; \citetalias{paper3}). It is clear in Fig.\
\ref{fig:outflow_vs_orientation_byclump} that many of the outflows
extend beyond their associated clump. However, the clumps found by \gclumps\
are smaller \citep{scubapaper} with a definite elliptical shape,
which does not necessarily follow the orientation of the
outflow. The lower detection rate for Class 0 sources
with \gclumps\ is perhaps another example of how their velocity
gradients are dominated by outflows. If the
smaller \gclumps\ cores trace their shape less
effectively, we would expect a lower detection rate. 

\begin{table}
\caption{Number of
clumps with significant velocity gradients by region, source
type and algorithm. Percentages give the detection rate for the
particular type of source in each region or overall.} 
\begin{tabular}{lrrrr}
\hline
Source Type & \multicolumn{3}{c}{Region} \\
& \ngc\ & IC348 & L1448 & Total\\
\hline
\multicolumn{5}{c}{\clfind\ Population}\\
Starless & 4(67~\%) & 2(17~\%) & 0(0~\%) & 6(32~\%)\\
Class 0 & 7(64~\%) & 3(100~\%) & 3(75~\%) & 13(72~\%)\\
Class I & 4(44~\%) & 1(50~\%) & 0(0~\%) & 5(45~\%)\\
No \citetalias{hatch07} ID & 1(8~\%) & 1(17~\%) & 1(9~\%) & 3(11~\%)\\
All & 16(41~\%) & 7(30~\%) & 4(25~\%) & 27(35~\%)\\
\hline
\multicolumn{5}{c}{\gclumps\ Population}\\
Starless & 4(50~\%) & 6(55~\%) & 0(0~\%) & 10(48~\%)\\
Class 0 & 6(50~\%) & 1(33~\%) & 0(0~\%) & 7(35~\%)\\
Class I & 3(33~\%) & 0(0~\%) & 0(0~\%) & 3(35~\%)\\
No \citetalias{hatch07} ID & 11(31~\%) & 3(38~\%) & 1(6~\%) & 15(24~\%)\\ 
All & 24(37~\%) & 10(42~\%) & 1(4~\%) & 35(31~\%)\\
\hline
\end{tabular}\\
\label{detectionrates_gradients}
\end{table} 

\begin{table*}
\caption{Average rotational properties derived from the
  velocity-gradient fitting for the \clfind\ clump population. All the errors listed ($\sigma$) are errors on the
mean and not sample deviations.}
\label{tab:rotprop_clfind}
\begin{tabular}{llrrllrr}
\hline
Population & Number & ${\cal G}$ & $\sigma_{\cal G}$ &
$\beta_\mathrm{rot}$ & $\sigma_\beta$ & $j$ & $\sigma_j$ \\
           &        & \multicolumn{2}{c}{(km\,s$^{-1}$\,pc$^{-1}$)} & &
& \multicolumn{2}{c}{(km\,s$^{-1}$\,pc)}\\

\hline
All        & 27     & 5.7 & 0.5 & 0.014 & 0.003   & 0.0019 & 0.0002\\
\ngc\      & 16     & 5.7 & 0.7 & 0.0040 & 0.0010 & 0.0017 & 0.0002\\
IC348      & 7      & 5.0 & 0.6 & 0.015 & 0.006   & 0.0017 & 0.0003\\
L1448      & 4      & 7   & 2   & 0.032 & 0.016   & 0.0032 & 0.0009\\
\hline
Starless   & 6      & 5.5 & 1.0 & 0.013 & 0.007 & 0.0021 & 0.0005\\
Protostars & 19     & 5.8 & 0.7 & 0.008 & 0.003 & 0.0019 & 0.0003\\
Class 0    & 14     & 6.1 & 0.8 & 0.010 & 0.004 & 0.0021 & 0.0003\\
Class I    & 5      & 4.9 & 1.5 & 0.003 & 0.001 & 0.0011 & 0.0001\\
\hline
\end{tabular}
\end{table*}

\begin{table*}
\caption{Average rotational properties derived from the
  velocity-gradient fitting for the \gclumps\ clump population. All the errors listed ($\sigma$) are errors on the
mean and not sample deviations.}
\label{tab:rotprop_gclumps}
\begin{tabular}{llrrllll}
\hline
Population & Number & ${\cal G}$ & $\sigma_{\cal G}$ &
$\beta_\mathrm{rot}$ & $\sigma_\beta$ & $j$ & $\sigma_j$ \\
           &        & \multicolumn{2}{c}{(km\,s$^{-1}$\,pc$^{-1}$)} & &
& \multicolumn{2}{c}{(km\,s$^{-1}$\,pc)}\\

\hline
All        & 35 & 6.9 & 0.6 & 0.013 & 0.004 & 0.00120 & 0.00014 \\
\ngc\      & 24 & 7.4 & 0.7 & 0.015 & 0.006 & 0.0012  & 0.0002 \\
IC348      & 10 & 5.6 & 0.8 & 0.007 & 0.001 & 0.0011  & 0.0002\\
\hline
Starless   & 10 & 5.0 & 0.6 & 0.006 & 0.001 & 0.0010  & 0.0002 \\
Protostars & 10 & 7.0 & 1.0 & 0.009 & 0.004 & 0.0012  & 0.0002 \\
Class 0    & 7  & 6.4 & 1.0 & 0.008 & 0.005 & 0.0012  & 0.0002 \\
Class I    & 3  & 8   & 3   & 0.012 & 0.010 & 0.0012  & 0.0003 \\
\hline
\end{tabular}
\end{table*}

\begin{figure}
\begin{center}
\includegraphics[width=0.46\textwidth]{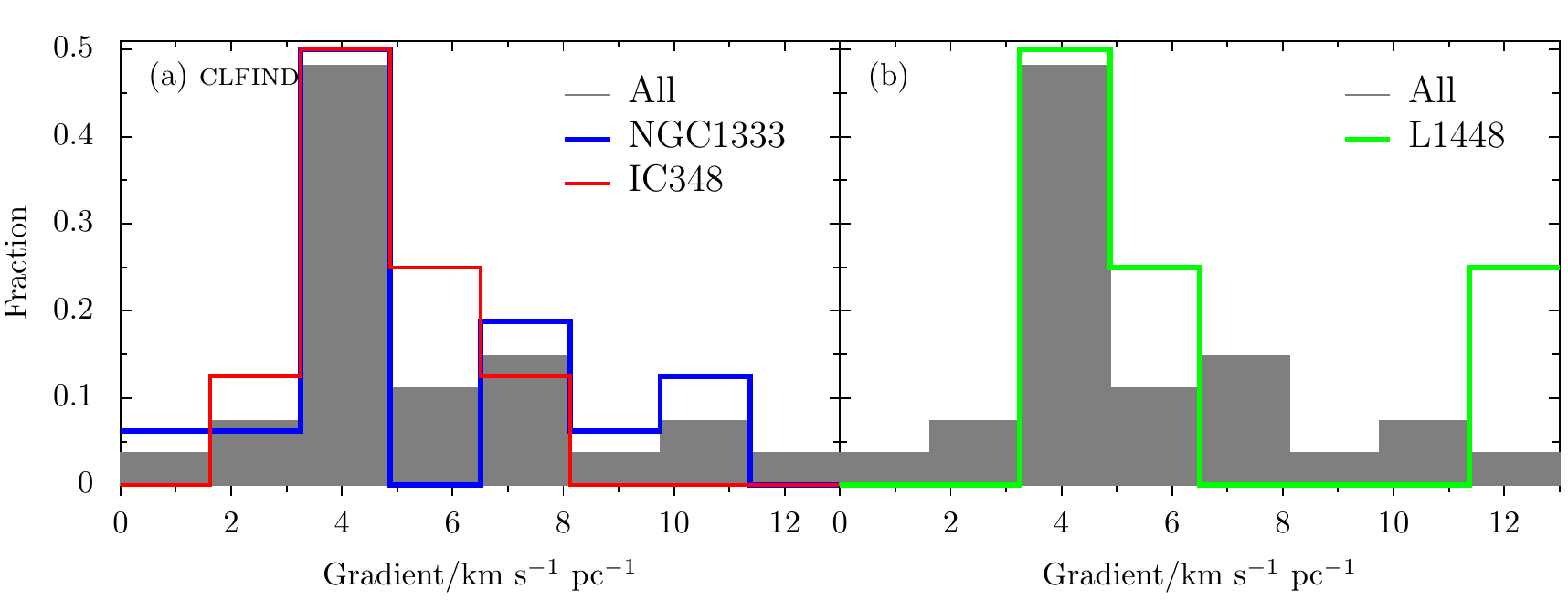}
\includegraphics[width=0.235\textwidth]{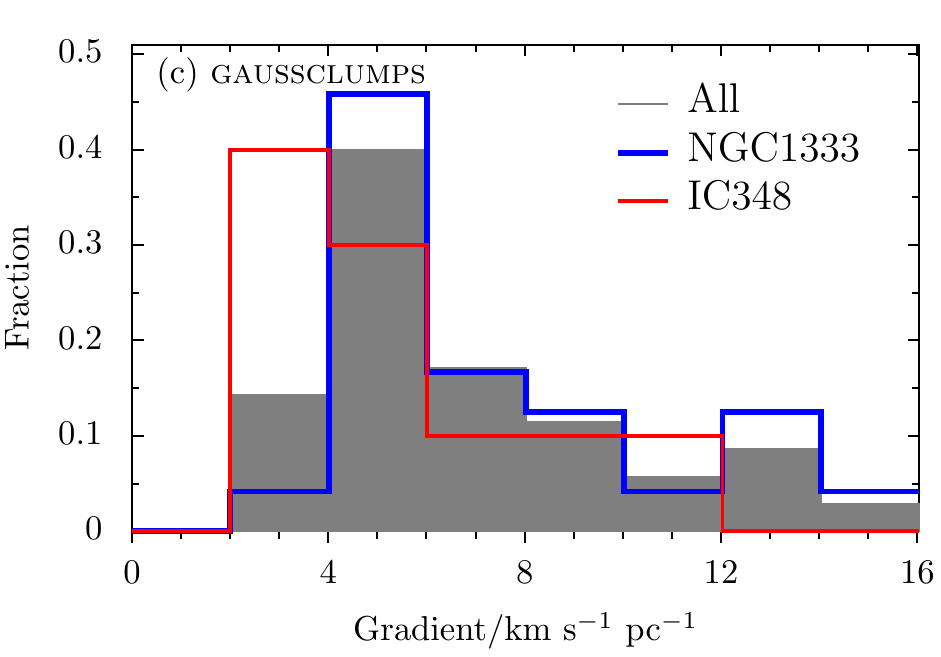}
\caption{Distribution of significant ($\sigma_{\cal G}/{\cal G} \geq
3$) velocity gradients for the \clfind\ (top) and \gclumps\ (bottom) clump
populations. The L1448
distribution is not plotted for \gclumps\ as it has only
one significant gradient. }
\label{fig:histogram_gradients}
\end{center}
\end{figure}

\begin{figure}
\begin{center}
\includegraphics[width=0.47\textwidth]{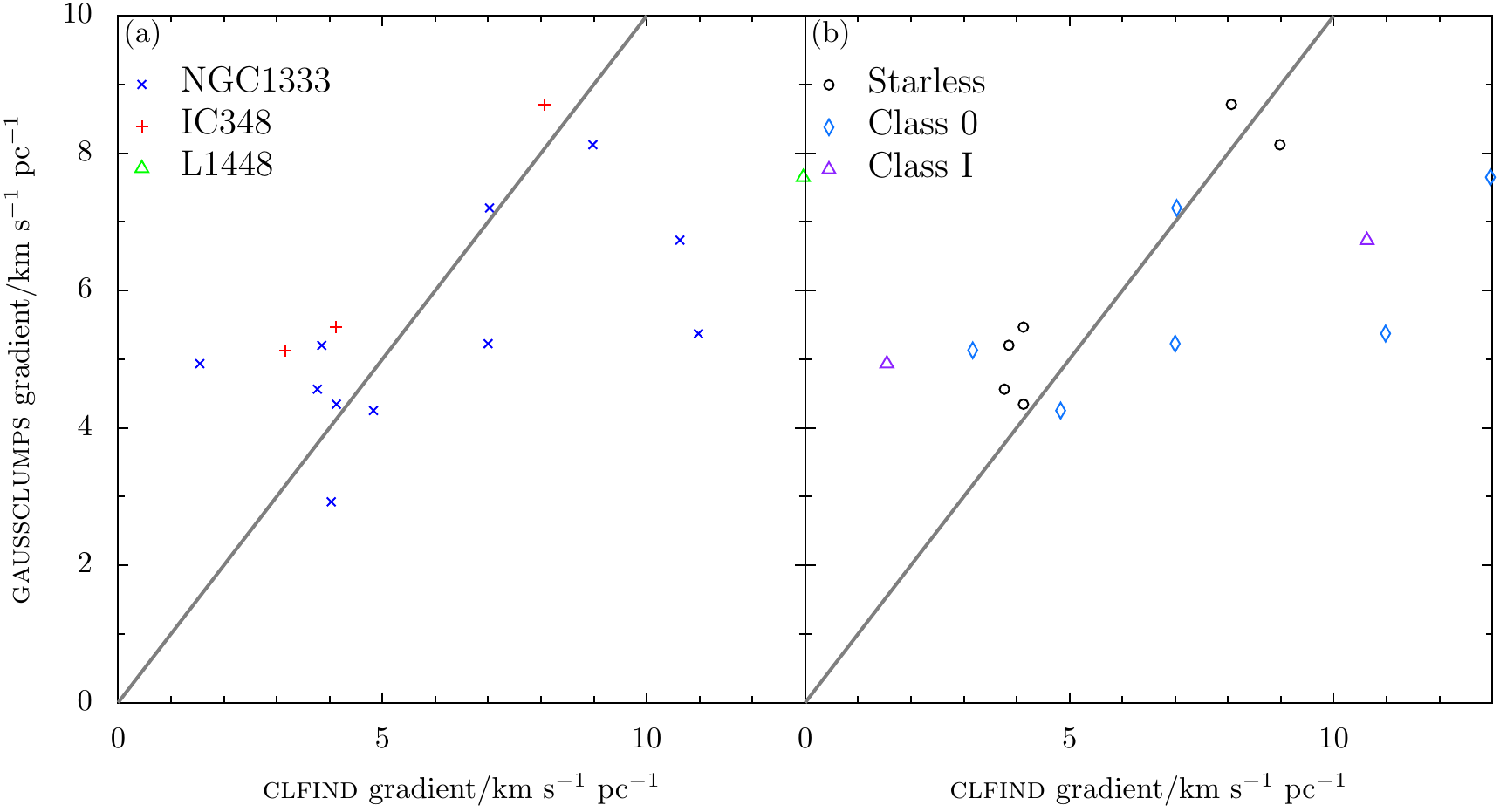}
\caption{Fitted velocity gradient for \gclumps\
versus \clfind\ designations for the 16 clumps which have
significant detections with both algorithms, by region
(left) or by
the classifications of \citetalias{hatch07} (right). The line marks where both
gradients are equal.}
\label{fig:gradients_clfindvsgclumps} 
\end{center}
\end{figure}

The fitted gradients are in the range 1.5 to
13.0\,km\,s$^{-1}$\,pc$^{-1}$ and 2.5 to 16.0\,km\,s$^{-1}$\,pc$^{-1}$ for
\clfind\ and \gclumps\ respectively. We provide a summary of all the
different rotational properties for different categories of clumps in
Tables \ref{tab:rotprop_clfind} and \ref{tab:rotprop_gclumps}. Fig.\ \ref{fig:histogram_gradients}
is a histogram of their magnitude by region. The distributions are
qualitatively similar for the different regions with the usual uncertainty
due to the small sample sizes; Kolmogorov-Smirnov (K-S) tests do not conclusively reject or
confirm the hypothesis that each sample is drawn from the same
population. On average across all the regions, the gradients are
$\langle {\cal G}\rangle = (5.7 \pm 0.5)$\,km\,s$^{-1}$\,pc$^{-1}$ for
\clfind\ and $\langle {\cal G}\rangle = (6.9 \pm
0.6)$\,km\,s$^{-1}$\,pc$^{-1}$ for \gclumps. The
fitted \gclumps\ gradients are larger than those for
\clfind\ but may be misleading as there are only 16
clumps with detections in both. We plot the gradient found with the \clfind\
designation versus that with \gclumps\ for these common sources
in Fig.\ \ref{fig:gradients_clfindvsgclumps}. There is
fairly good agreement on the gradient with the protostellar
sources deviating the most from the line of equal
gradients. This might again be because the C$^{18}$O line centres
are affected by outflowing gas which is not well traced by the
\gclumps\ outlines. 

\begin{figure}
\begin{center}
\includegraphics[width=0.47\textwidth]{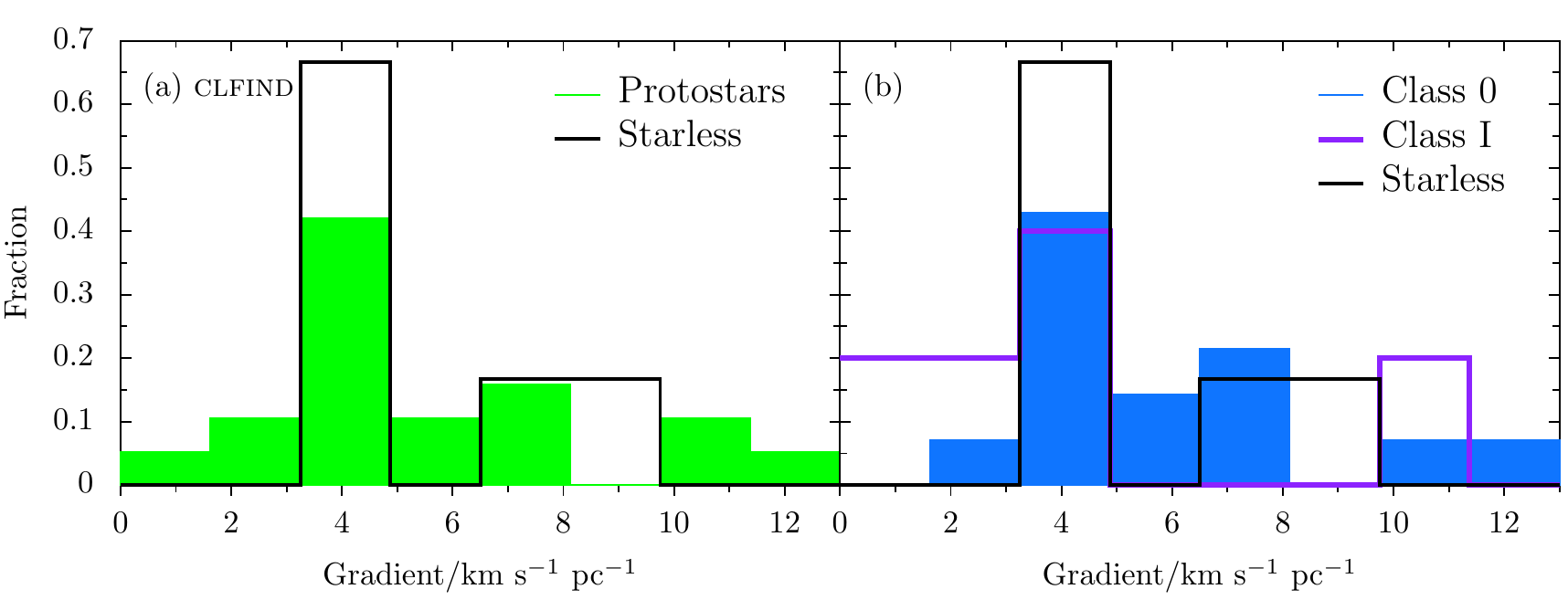}
\includegraphics[width=0.47\textwidth]{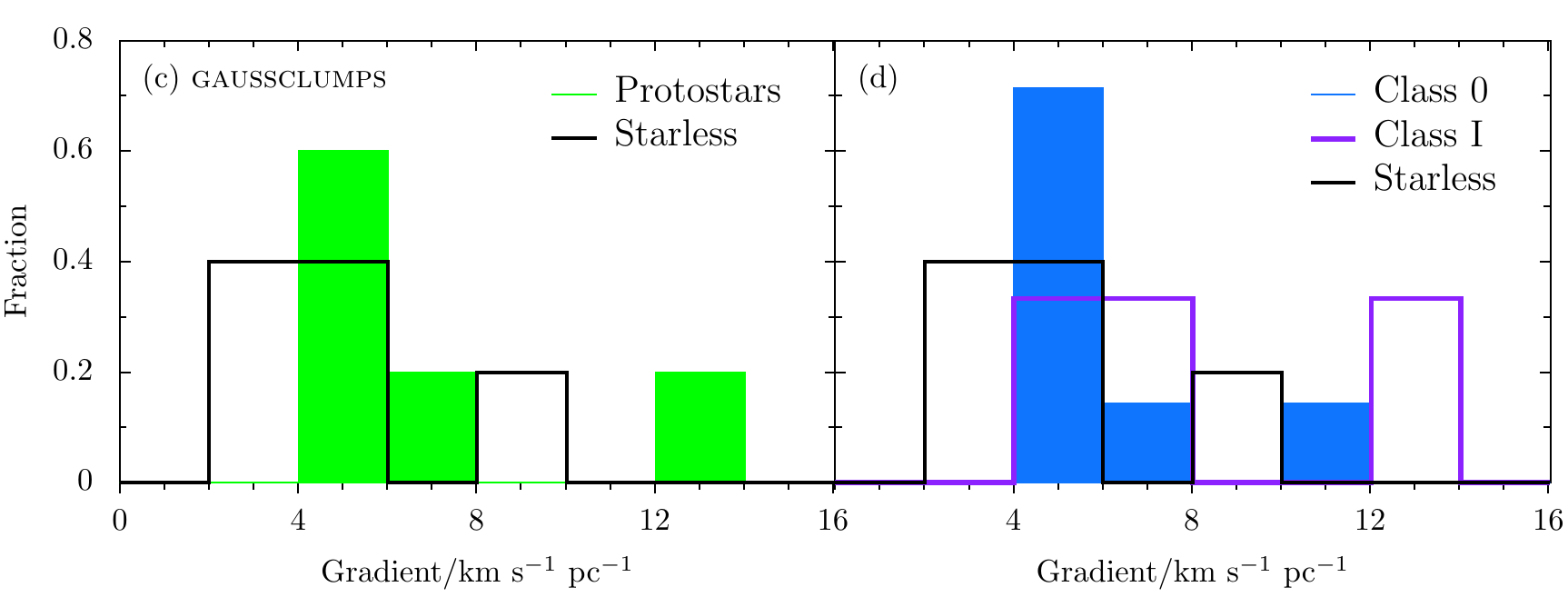}
\caption{Distribution of significant ($\sigma_{\cal G}/{\cal G} \geq
3$) velocity gradients for the classifications of
\citetalias{hatch07} for the \clfind\ (top) and \gclumps\ (bottom) clump
populations.} 
\label{fig:histogram_gradients_bytype}
\end{center}
\end{figure}

There are too few sources of each classification to draw firm
conclusions about any trends with age. It does
seem in the distributions of Fig.\ \ref{fig:histogram_gradients_bytype} that the starless cores have
${\cal G} \la 10$\,km\,s$^{-1}$\,pc$^{-1}$ and larger gradients are protostellar. The protostellar average is
larger than the starless one for the \gclumps\ population: $\langle {\cal G}\rangle = (7.0 \pm 1.0)$ compared to $\langle {\cal G}\rangle = (5.0 \pm
0.6)$\,km\,s$^{-1}$\,pc$^{-1}$. Although they are similar for
\clfind\ $\langle {\cal G}\rangle = (5.8 \pm
0.7)$ compared to $\langle {\cal G}\rangle = (5.5 \pm 1.0)$\,km\,s$^{-1}$\,pc$^{-1}$. This might point
to an outflow contribution to the gradient. 

The gradients we find in Perseus are much greater than those
originally found by \citetalias{goodman93} or \citet{caselli02} in their
N$_2$H$^+$ $J=1\to 0$ survey of 60 low mass cores: 0.3 to 3.9 and
0.5 to 6.0\,km\,s$^{-1}$\,pc$^{-1}$ respectively. Outflows cannot
explain all of this difference because our starless clumps have
significantly larger gradients as well. The discrepancy is
reminiscent of that found by \citetalias{goodman93} towards B361, which was also investigated by \citet{arquilla85} with $^{13}$CO
$J=1\to 0$ data. \citetalias{goodman93} estimated a gradient five times smaller than the
earlier study, 0.7 versus 3.3\,km\,s$^{-1}$\,pc$^{-1}$, which had led \citet{arquilla85} to conclude that rotation was dynamically
significant in dark clouds. \citetalias{goodman93} explain the
difference by emphasizing that tracers
of lower critical densities can produce complicated velocity patterns
dominated by outflows or clump-to-clump motions that mimic solid body
rotation. However, \citet*{olmi05} fitted gradients to a number of
cores in Perseus, finding similar magnitudes for the two common cores also in \citetalias{goodman93} and generally in the same range 0.14 to
2.32\,km\,s$^{-1}$\,pc$^{-1}$. Their study used C$^{18}$O $J=1\to 0$, CS $J=2\to 1$ and
N$_2$H$^+$ $J=1\to 0$ data with their C$^{18}$O fitting not displaying markedly
different gradients. A critical factor will undoubtedly be resolution. The angular resolution of previous surveys is
worse than the 17.7\,arcsec of our maps: $\sim$88, $\sim$54 and $\sim$46\,arcsec
for \citetalias{goodman93}, \citet{caselli02} and \citet{olmi05} respectively. Hence even for their closest sources
in Taurus (at 140\,pc), their best linear resolution (0.037\,pc) is nearly twice as
large as ours (0.021\,pc). A larger
beam will tend to smooth out differences between regions and reduce
the overall gradient magnitude. This effect may be large enough to explain
the differences for cores perhaps three times as distant as
Perseus\footnote{The most distant sources examined by
  \citet{caselli02} are L1031B at 900\,pc and L1389 at 600\,pc.}, in a beam
four times as large as ours. 

\subsubsection{Gradient orientation}

\begin{figure}
\begin{center}
\includegraphics[width=0.47\textwidth]{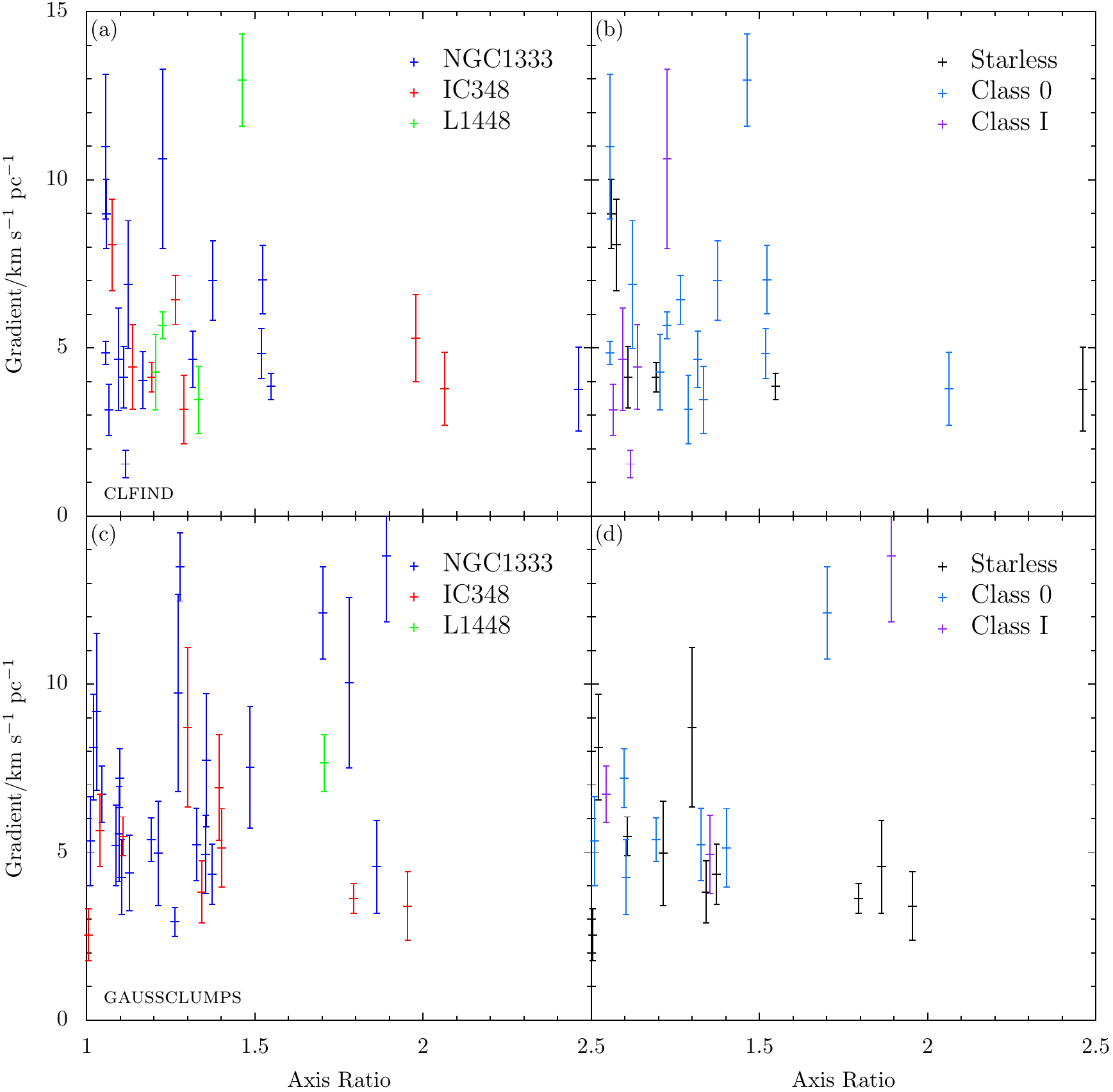}
\caption{Velocity gradient versus axis ratio for
\clfind\ (top panels) and \gclumps\ sources
(bottom panels) broken down by region (left panels) or
by the classifications of \citetalias{hatch07}
(right panels). }
\label{fig:gradvsaxiratio}
\end{center}
\end{figure}

\begin{figure}
\begin{center}
\includegraphics[width=0.47\textwidth]{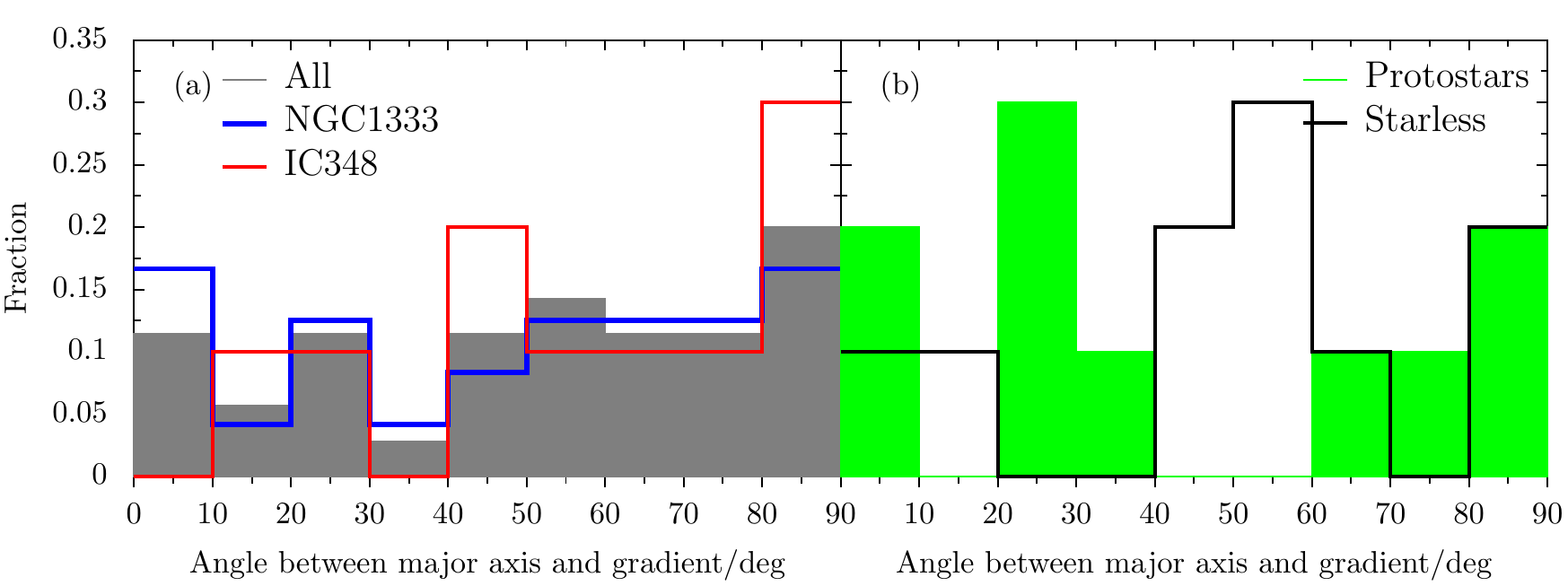}
\caption{Distribution of the angle between the velocity gradient and the
major axis of the \gclumps\ clumps by region (left) or
by the classifications of \citetalias{hatch07}
(right). If rotational
motions are elongating the clumps we would expect the gradient and axis
to be in the same direction making the angle zero.}
\label{fig:angle_gradtomajoraxis}
\end{center}
\end{figure}

Centrifugal stresses due to energetic rotation will flatten cores
along the rotation axis. This tends to produced oblate cores \citepalias{goodman93},
currently the favoured shape (e.g.\ \citealt*{jones01}), which could
also be caused by strong magnetic fields. If rotation dominates, we might expect
the the degree of core elongation to increase with increasing velocity
gradient and a core's major axis to lie parallel to the gradient
i.e.\ perpendicular to the rotation axis. Fig.\ \ref{fig:gradvsaxiratio} shows no relationship between the magnitude of the velocity gradient
and the elongation of a clump, quantified through the axis ratio
(taken from \citealt{scubapaper}). It comes as no surprise then that the angle between the
clump major axis and the velocity gradient in Fig.\
\ref{fig:angle_gradtomajoraxis} (only computed for the \gclumps\ sources), seems to be distributed at
random as well. Therefore, clump rotation is unlikely to be energetically significant. 

\begin{figure}
\begin{center}
\includegraphics[width=0.47\textwidth]{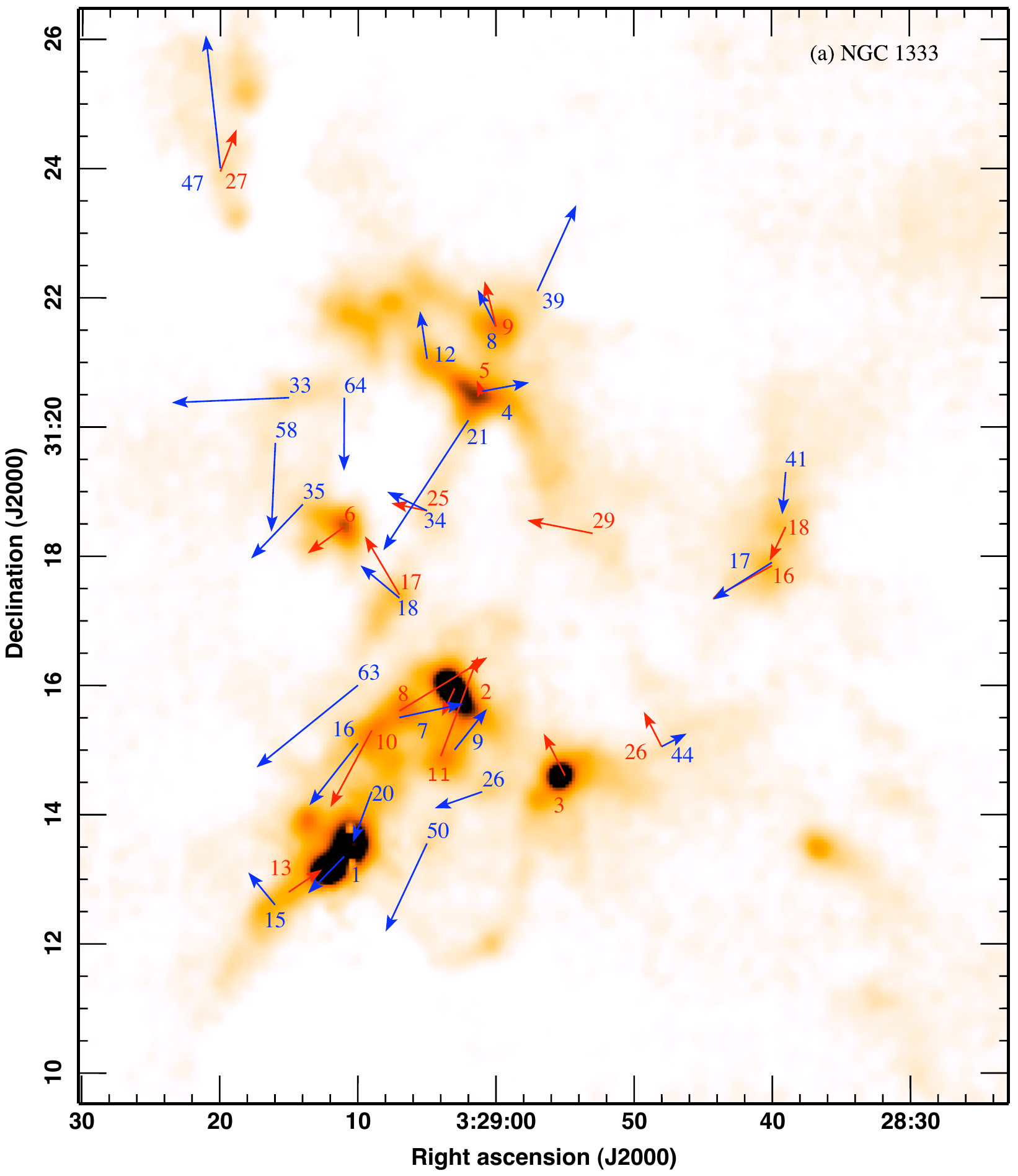}
\includegraphics[width=0.47\textwidth]{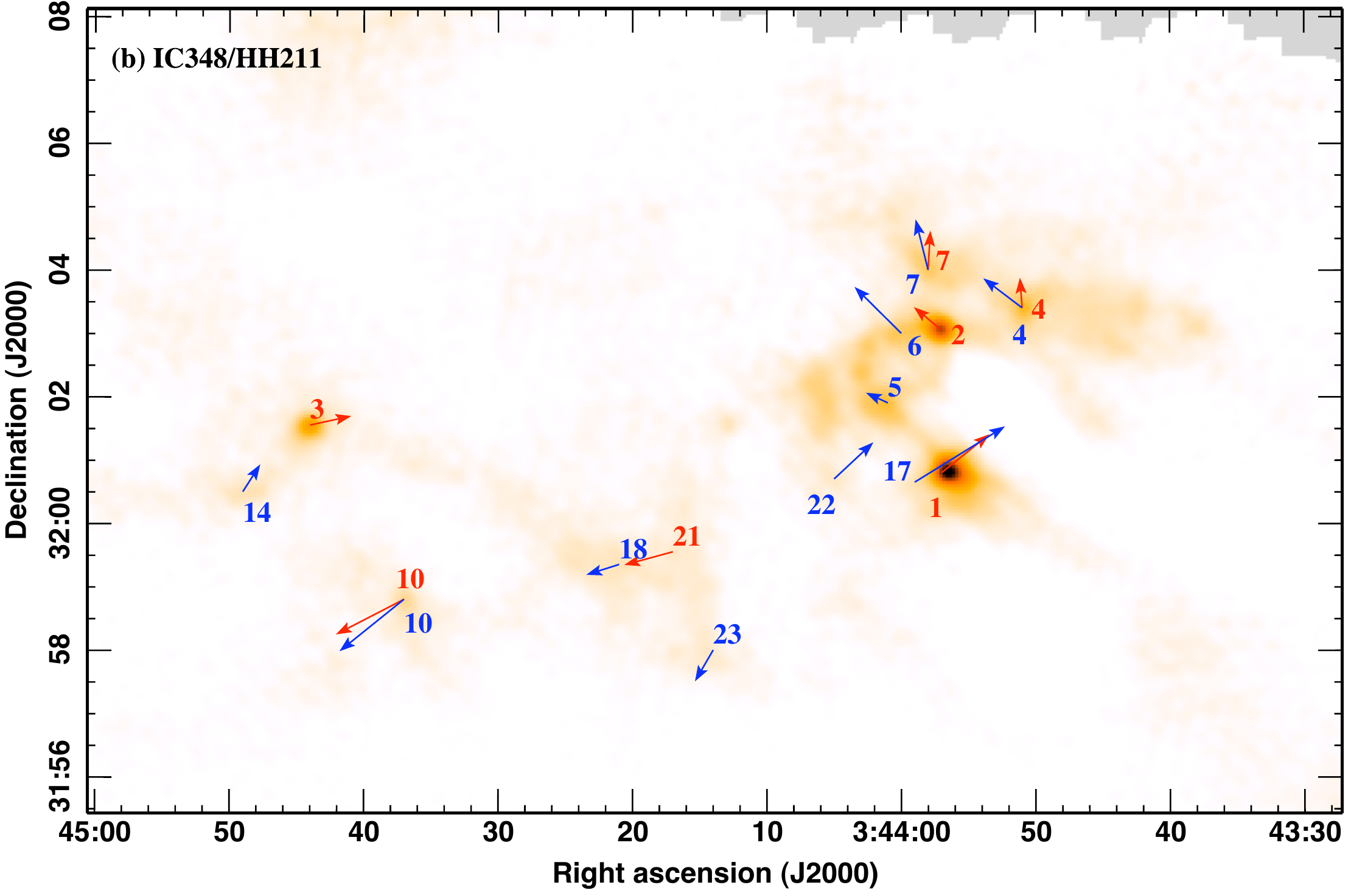}
\includegraphics[width=0.47\textwidth]{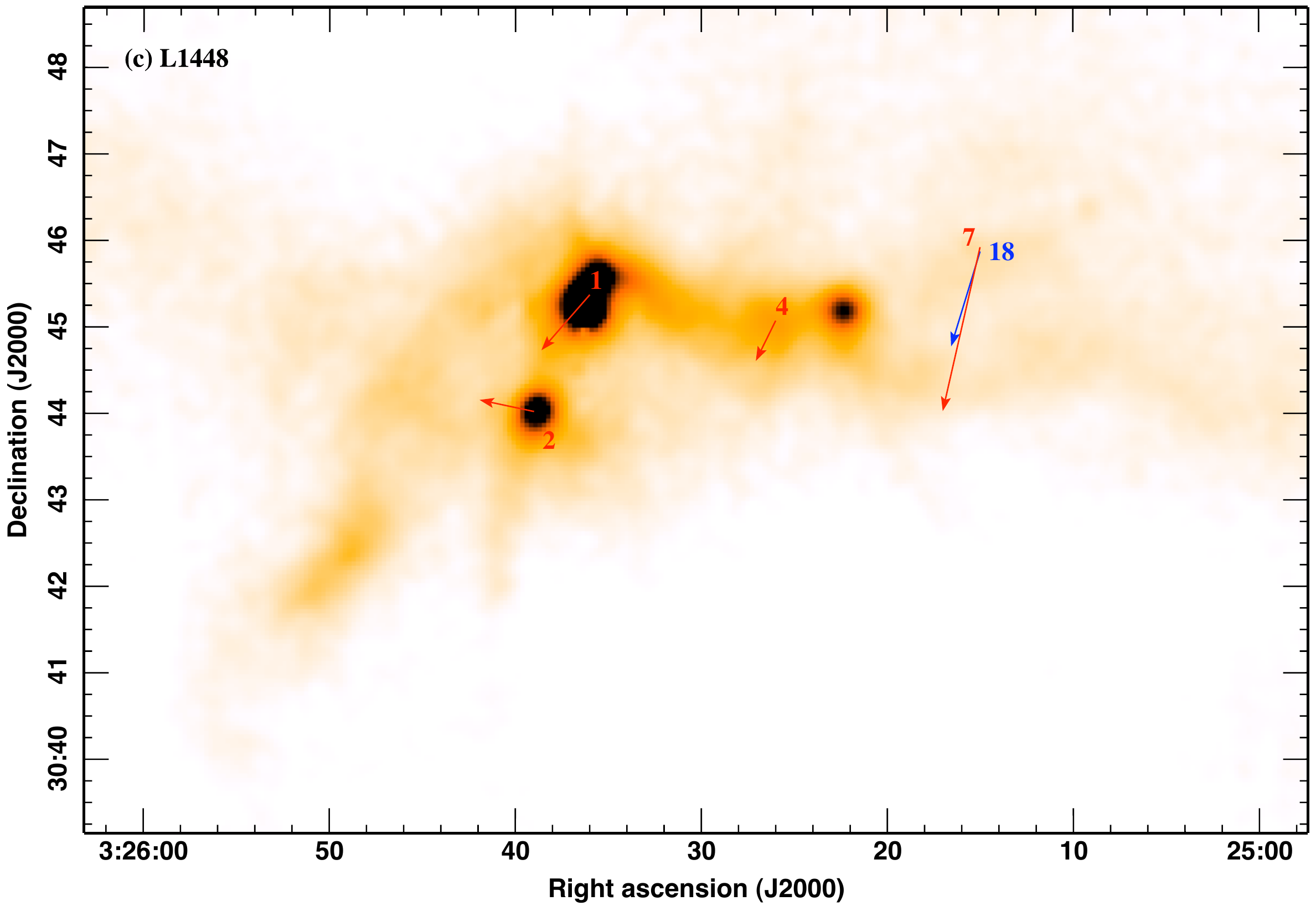}
\caption{Spatial distribution of the
  velocity gradient detections in NGC~1333 (top), IC348/HH211 (middle)
  and L1448 (bottom). Arrows are drawn from the peak position of the
  associated clump for clumps where there are significant
  detections in either the \clfind\ or \gclumps\ catalogues. The
  direction and length of the arrow represents the direction and
  magnitude of the gradient and the detections are labelled with the
  reference from our SCUBA catalogues \citep{scubapaper}. The colour-scale is \scuba\ 850\,\micron\
emission from 0 to 1600\,\mjybeam.}
\label{fig:gradient_orientations}
\end{center}
\end{figure} 

Is there a correlation between the orientation of
the rotation axes of neighbouring clumps or an anti-correlation to keep
the net angular momentum small? In Fig.\
\ref{fig:gradient_orientations}, we overlay the velocity
gradients on dust emission maps of the various
regions. Reassuringly the directions and magnitudes of the gradients
measured using both algorithms' allocations match very well. In small
patches the arrows of neighbouring clumps line up but on large scales there seems
to be little correlation. This is similar to the gravoturbulent models of
\citetalias{jappsen04}, where shortly after collapse starts, the angular momenta in small
regions are well aligned over a moderate correlation length with cores
further away spinning in random directions. Their explanation is that neighbouring cores
are formed from the same reservoir of material and therefore can be
expected to have similar angular momenta. During
subsequent accretion and evolution, the associated correlation
length decreases and neighbours lose their alignment, 
as embedded cores are ejected and turbulence
disrupts the current material or brings in new gas. 

\subsubsection{Dynamic support?}

\begin{figure}
\begin{center}
\includegraphics[width=0.47\textwidth]{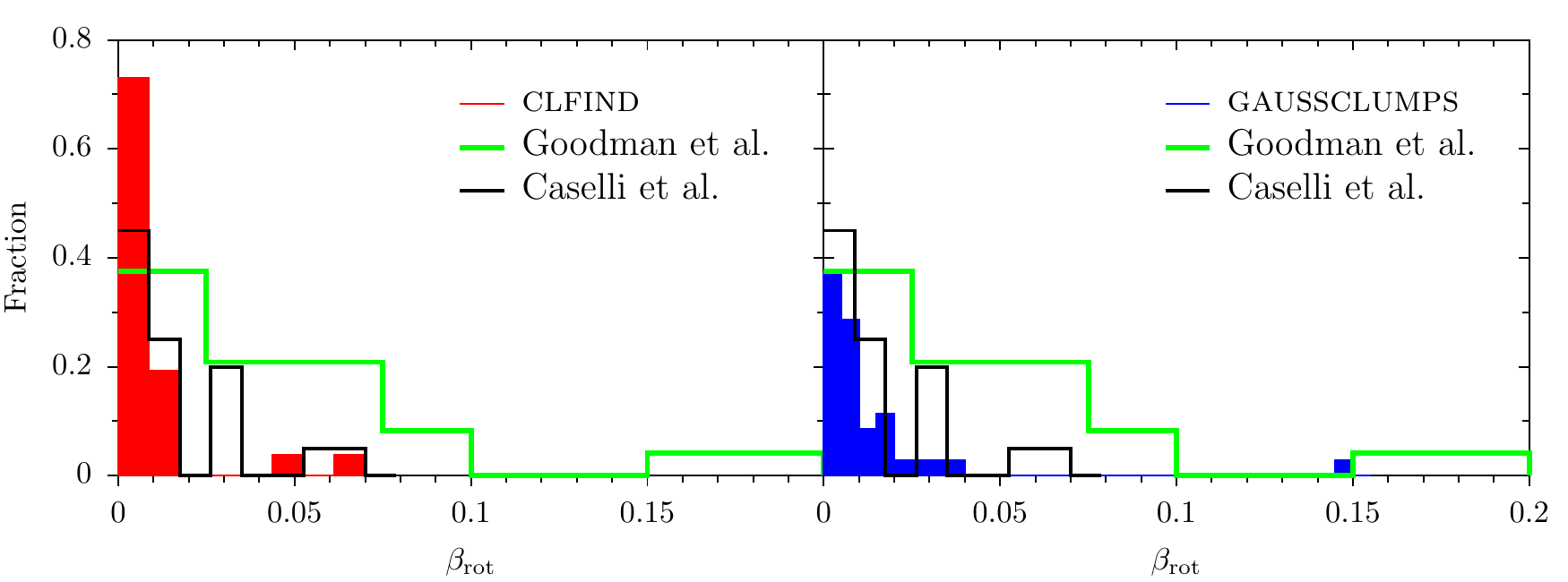}
\caption{Distribution of the ratio of
rotational to gravitational energies, $\beta_\mathrm{rot}$. For comparison the distributions found by
\citetalias{goodman93} (green) and (\citealt{caselli02}, orange) are overlaid.}
\label{fig:beta_comparison}
\end{center}
\end{figure}

To quantify the level of support that the velocity gradients might
provide against gravitational collapse we calculate
$\beta_\mathrm{rot}$, the ratio of rotational to gravitational
energy. \citetalias{goodman93} define this ratio for a uniform density sphere:
\begin{equation} \beta_\mathrm{rot} = \frac{(1/2)I\omega^2}{3GM^2/5R}=\frac{R^3 {\cal G}^2}{3GM\sin
    ^2i} \label{eqn:beta} \end{equation}
where the moment of inertia is $I=pMR^2$ with $p=2/5$ for a
uniform density sphere and the angular momentum is $\omega = {\cal G}/
\sin i$ with $i$ the angle of inclination to the line of sight. Assuming values representative of these clumps and
$\sin i =1$, this becomes:
\begin{equation} \beta_\mathrm{rot} = 6.2 \times 10 ^{-4}
  \left(\frac{R_\mathrm{dec}}{0.02\,\mathrm{pc}}\right)^3
  \left(\frac{M}{\mathrm{M_\odot}}\right)^{-1} \left(\frac{{\cal
  G}}{\mathrm{1\,km\,s^{-1}\,pc^{-1}}}\right)^2. \end{equation}
Across all the clumps the average values of
$\beta_\mathrm{rot}$ are:
$\langle \beta_\mathrm{rot} \rangle = (0.014 \pm 0.003)$ and $(0.013
\pm 0.004)$ for \clfind\ and \gclumps\ respectively. At most the
rotational energy is just six~per cent
for the \clfind\ sample and 14~per cent for
\gclumps\ of the gravitational energy in the clumps. Fig.\ \ref{fig:beta_comparison} shows the distribution of
$\beta_\mathrm{rot}$ for the two clump populations. There is a strong preference for low ratios, with the fraction
of clumps rapidly falling off as $\beta_\mathrm{rot}$
increases. Additionally, both populations closely match each other and
the ratios found by \citet{caselli02}, while the
\citetalias{goodman93} distribution is wider, spreading to higher
values. K-S tests could not confirm the samples
are drawn from the same population as each other or either of the
examples from the literature to any degree of significance. Even
though we found larger velocity gradients than the previous studies,
$\beta_\mathrm{rot}$ is not any higher. This is probably because our
clumps are significantly smaller: on average
the clump radii are 0.03 and 0.02\,pc for \clfind\ and
\gclumps\ respectively while the \citet{caselli02} cores
have an average of 0.06\,pc. As $\beta_\mathrm{rot}$
depends on $R^3$ this will reduce our values by $\sim$9 relative to
theirs for an equally massive source, compensating for our larger
derived gradients. Thus, rotation is not dynamically significant in star-forming cores and
should not support the core against collapse. 

It would be interesting to examine how the level of support varies
region-to-region and as a clump evolves. However, the numbers here are
too small to draw any firm conclusions, with one or two objects
heavily distorting overall averages. This is further compounded
by the outflow ambiguities of the protostellar sources. Nevertheless, there are
no vast differences in the populations by region or age with almost
all the clumps having $\beta_\mathrm{rot} \la
0.02$.   

\subsubsection{Angular momenta}

The core rotation speed can be quantified via the specific angular
momentum, $j=J/M$, i.e.\ the angular momentum per unit mass:
\begin{equation} j=\frac{2}{5}\omega R^2 \label{eqn:j} \end{equation}
where $\omega={\cal G}/\sin i$ is again the angular velocity of the clump, $R$ its radius
and the factor $2/5$ is for a constant density sphere. This yields for
representative values:
\begin{equation} j= 1.6 \times 10^{-4}\left(\frac{\cal
G}{1\,\mathrm{km\,s^{-1}\,pc^{-1}}}\right)\left(\frac{R_\mathrm{dec}}{0.02\,\mathrm{pc}}
\right)^2 \,\mathrm{km\,s^{-1}\,pc}\end{equation} 
The distribution of $j$ has been important, with modellers using the one
derived by \citetalias{goodman93} either as an input to their models
or a target 
(e.g.\ \citealp{burkert00}; \citetalias{jappsen04}). Furthermore, the
distribution of angular momentum across a core and its evolution may prove whether magnetic fields are necessary to strip off angular momentum as
collapse ensues. 

\begin{figure}
\begin{center}
\includegraphics[width=0.47\textwidth]{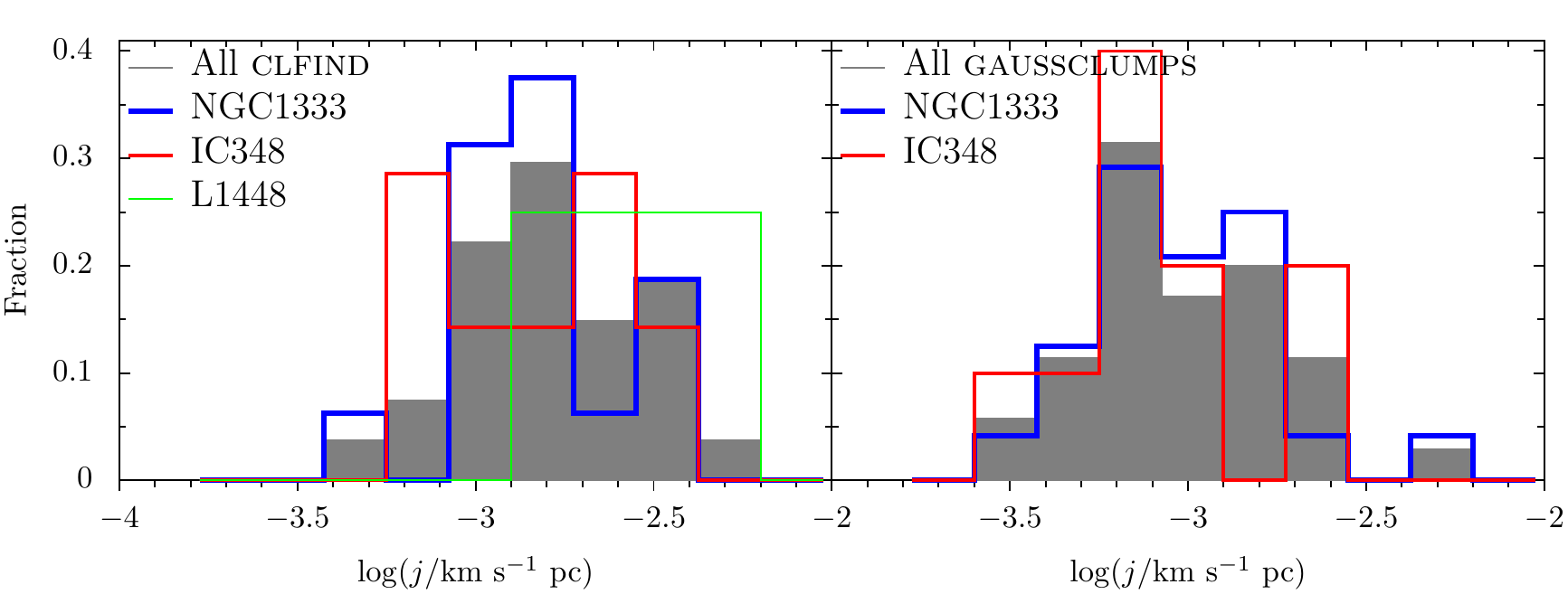}
\caption{Distribution of $j$ by region and algorithm:
\clfind\ on the left and \gclumps\ on the right.}
\label{fig:j_byregion}
\end{center}
\end{figure}

In Fig.\ \ref{fig:j_byregion}, we plot the distribution of $j$ by
region. There is little difference between the different algorithms or
clumps from one region to the next. For \clfind\ the overall average is $\langle j \rangle = (1.9 \pm 0.2) \times 10^{-3}$\,km\,s$^{-1}$\,pc whilst $\langle j \rangle = (1.7 \pm 0.2)
\times 10^{-3}$, $(1.7 \pm 0.3) \times 10^{-3}$ and $(3.2 \pm 0.9)
\times 10^{-3}$ in \ngc, IC348 and L1448 respectively. The
\gclumps\ data are skewed to lower $j$ due to their smaller clump radii but there are still few differences across the regions: $\langle j
\rangle = (1.20 \pm 0.14) \times 10^{-3}$, $(1.2 \pm 0.2) \times
10^{-3}$, $(1.1 \pm 0.2) \times 10^{-3}$\,km\,s$^{-1}$\,pc overall and in \ngc\ and
IC348 respectively. 

\begin{figure}
\begin{center}
\includegraphics[width=0.47\textwidth]{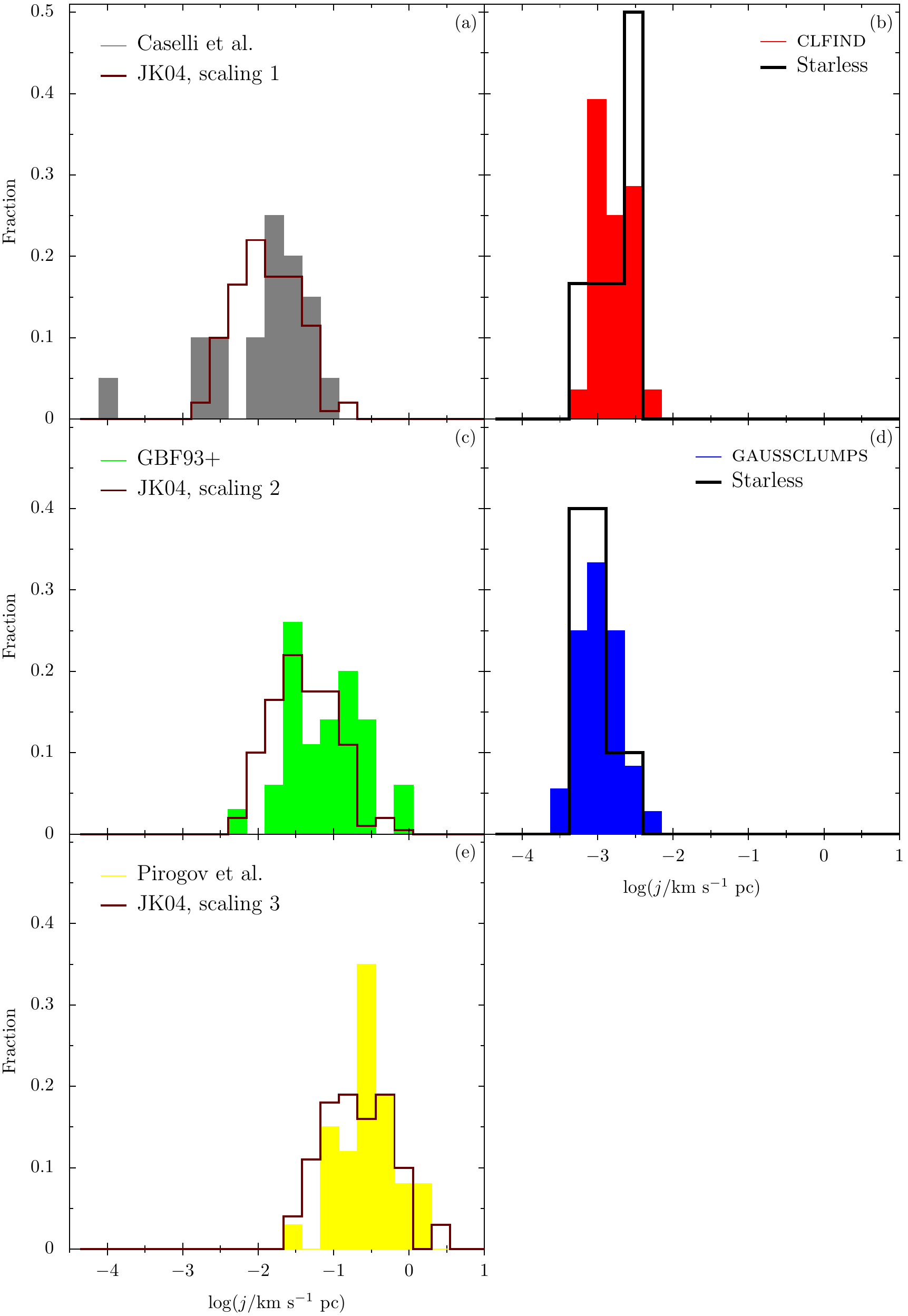}
\caption{Distribution of specific
angular momenta, $j$ for starless cores. (a) N$_2$H$^+$ $J=1\to 0$ cores of
\citeauthor{caselli02} (\citeyear{caselli02}, grey) compared to the \citetalias{jappsen04}
model (brown) scaled by $n=10^5$\,cm$^{-3}$ and $T=10$\,K. (b) All our
\clfind\ detections (red) and just those in starless clumps
(black). (c) NH$_3$ cores of \citetalias{goodman93},
\citet{barranco98} and \citet{jijina99} (green) compared to the
\citetalias{jappsen04} model (brown) scaled by $n=10^4$\,cm$^{-3}$ and
$T=10$\,K. (d) all our \gclumps\ detections (blue) and just those in
starless clumps (black). (e) N$_2$H$^+$ $J=1\to 0$ high-mass cores of
\citet{pirogov03} compared to the \citetalias{jappsen04} model (brown)
scaled by $n=10^4$\,c $^{-3}$ and $T=50$\,K.}
\label{fig:j_modelsvsdata_starless}
\end{center}
\end{figure}

\citetalias{jappsen04} distinguish between starless and
protostellar sources in their gravoturbulent simulations which focus
particularly on rotational properties. They find
wide continuous distributions for $j$, stretching over two orders of
magnitude, as we do. Their prestellar cores have an order of magnitude larger $j$ than the
protostars. Fig.\ \ref{fig:j_modelsvsdata_starless} compares our
distributions of $j$ (overall and for the starless clumps)
to previously observed distributions and \citetalias{jappsen04}. To match a particular set of observations,
\citetalias{jappsen04} scale their code output according to the mean
density, $n$, and temperature, $T$, with $j$ depending on these as
$j \propto T/\sqrt n$. With appropriate scalings their data match three different observational regimes: low-mass cores mapped
in the high-density tracer N$_2$H$^+$ \citep{caselli02}, low-mass cores in the slightly
lower density tracer NH$_3$ (\citetalias{goodman93} with
\citealp{barranco98} and \citealp*{jijina99}) and high-mass cores in the high-density
tracer N$_2$H$^+$ \citep{pirogov03}. Our observed distribution has no
overlap with the high-mass cores of \citet{pirogov03} and
little with \citetalias{goodman93}, having much lower $j$. For the
starless clumps only, we find $\langle j
\rangle = (2.1 \pm 0.5) \times 10^{-3}$ and $(1.0 \pm 0.2) \times 10^{-3}$\,km\,s$^{-1}$\,pc for
\clfind\ and \gclumps\ respectively. The
most similar distribution is that of \citet{caselli02}, which is over twice
as wide with a much larger average, $\langle j
\rangle = 2.2 \times 10^{-2}$\,km\,s$^{-1}$\,pc, but does overlap with
ours at low $j$. For our average to match that of
\citet{caselli02}, presumably our clumps would have to be
considerably denser and/or colder than the values assumed by
\citetalias{jappsen04} of $n=10^5$\,cm$^{-3}$ and $T=10$\,K. In
\S \ref{sec:kinematics} we noted that the critical density is expected to be
$n_\mathrm{crit} = 10^4$\,\cmthree. However, in radiative transfer
models of a free-falling protostellar envelope, we found peak
\htwo\ number densities of greater than a few times $10^5$\,cm$^{-3}$ were required to reproduce the \ceighteeno\ \threetotwo\ line
strengths seen in our data \citep{curtisthesis}. Alternatively or additionally, we are seeing considerably smaller angular momenta then
those measured by previous authors. The narrow width of the
distribution compared to previous work is likely to reflect the
uniform environment of our cores in a single cloud rather
then spread over many different ones as in \citet{caselli02} and
\citetalias{goodman93}. 

On the other hand, the smaller angular momenta observed in our
starless clumps may not indicate lower degrees of rotation in the
clump \emph{core} but may reflect a more slowly rotating outer
\emph{envelope}, making comparisons with tracers such as NH$_3$ inappropriate. \citet{redman04} examined the asymmetric \hco\ line profiles
across L1689B, a prestellar core in Ophiuchus, which they modelled as a rapidly-rotating inner core contained within a static
envelope. Again, the key diagnostics are the densities and temperatures
traced by the \ceighteeno\ \threetotwo\ line, which can only be properly
answered with radiative transfer modelling. We would expect to probe high
densities, $n_\mathrm{crit}\sim 10^4$\,cm$^{-3}$, but earlier we saw the similar non-thermal linewidths for the \threetotwo\ line
compared to the \twotoone, suggesting we might not be detecting
very different material. A further consideration in the low
temperature environments of starless cores is CO
depletion. Estimating the degree of depletion in these cores is
difficult without similar observations in ``late-depleting''
molecules, such as \ntwohplus. However, models
(e.g.\ \citealp*{walmsley04}) suggest that CO freezes-out 
in the central 7000\,AU of a prestellar core (this is really a
temperature and density dependence). Furthermore, we noted in \S
\ref{sec:linewidths} that there is some existing evidence for
considerable \ceighteeno\ depletion in candidate star-forming cores in
Perseus \citepalias{hkirk07} Thus, our \ceighteeno\ signal in evolved
starless clumps and young protostars at least should have little contribution from
the dense inner core and this reasoning supports the idea we are
probing a more slowly rotating envelope. 

\begin{figure*}
\begin{center}
\includegraphics[width=1.0\textwidth]{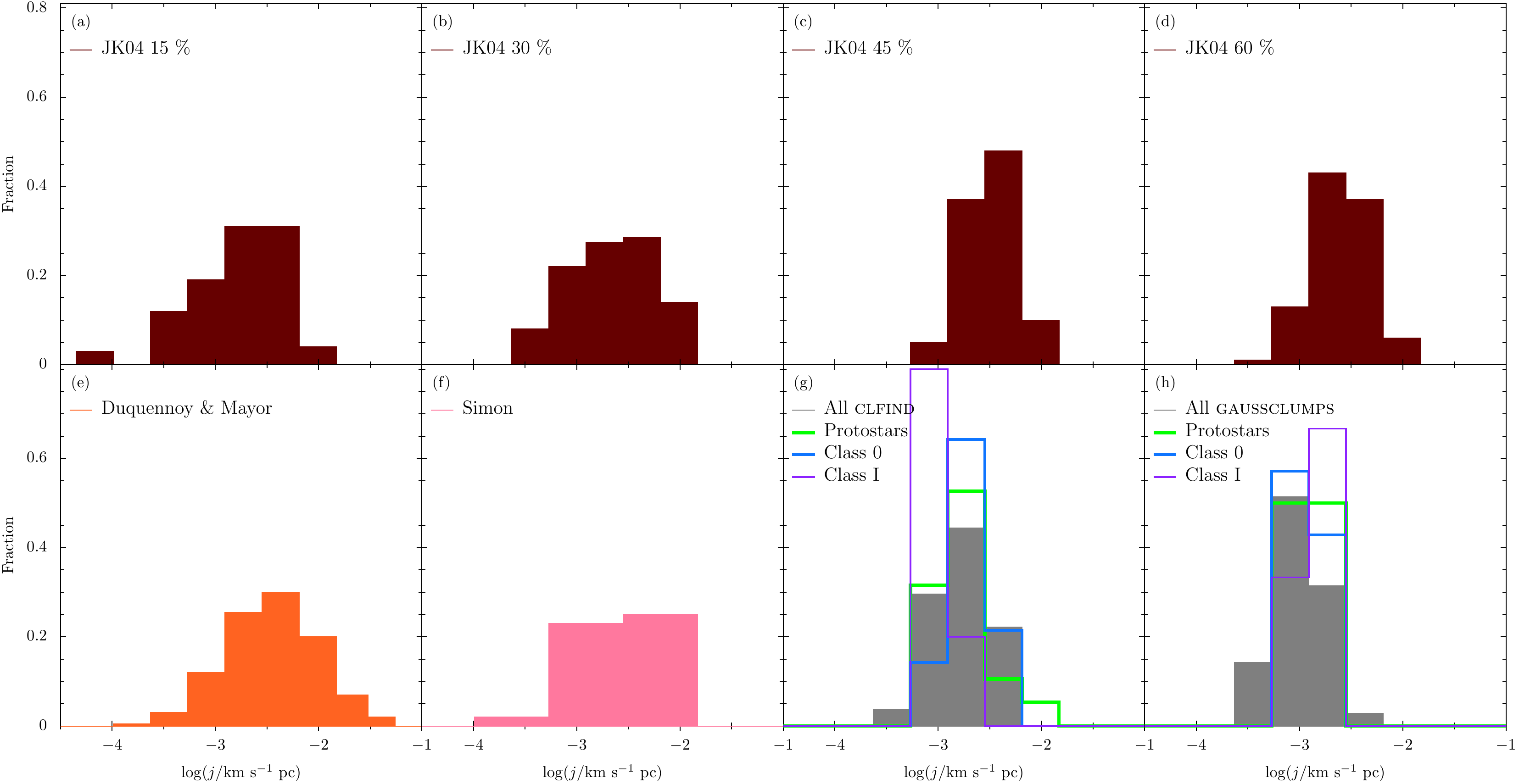}
\caption{Distribution of the specific
angular momenta, $j$, for protostars. (a)
to (d) show \citetalias{jappsen04}'s model M6k2 with different star
formation efficiencies: (a) 15, (b) 30, (c) 45 and (d)
60~per cent. (e) Binaries among nearby G
stars of \citet{duquennoy91}. (f) Binaries in Taurus
\citep{simon92}. (g) and (h) show the distributions from paper for the \clfind\ and \gclumps\ designations
respectively, decomposed according to the designations of \citetalias{hatch07}.}
\label{fig:j_modelsvsdata_proto}
\end{center}
\end{figure*}

In \citetalias{jappsen04}, protostars have a distribution of $j$
that narrows with time, staying around the same average. Their
distribution is also similar to that observed for binaries (see
Fig.\ \ref{fig:j_modelsvsdata_proto}). Our distributions for all
the sources are as
narrow as those for the protostars alone in \citetalias{jappsen04}'s
models. For \clfind\ there is a trend of
decreasing $j$ with protostellar age: $\langle j \rangle = (2.1 \pm
0.3) \times 10^{-3}$ and $(1.1 \pm
0.1) \times 10^{-3}$\,km\,s$^{-1}$\,pc for Class 0 and I clumps, which
is not reproduced with \gclumps: $\langle j \rangle = (1.2 \pm
0.2) \times 10^{-3}$ (Class 0) and $(1.2 \pm 0.3) \times
     10^{-3}$\,km\,s$^{-1}$\,pc (Class I). Of course in these sources there is
     potential confusion of the rotational signatures with outflows
     which is perhaps why the expected decrease in $j$ from
     starless to protostellar clumps is not observed. 

\subsubsection{Scalings with clump size}

\begin{figure}
\begin{center}
\includegraphics[width=0.47\textwidth]{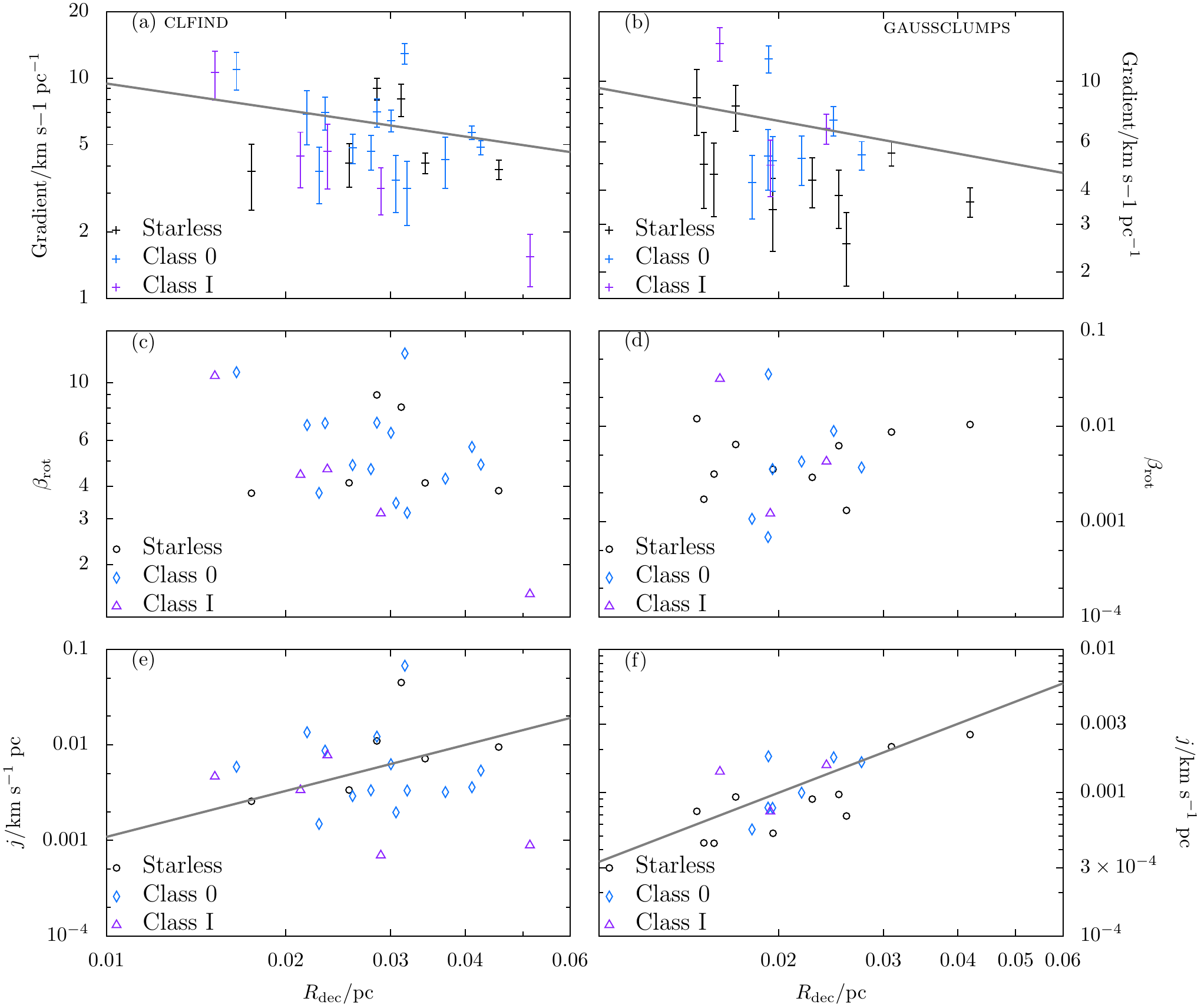}
\caption{Rotational
parameters as functions of clump size for \clfind\ (left) and \gclumps\ (right). ${\cal G}$
versus clump radius; $\beta_\mathrm{rot}$ versus $R_\mathrm{dec}$ (middle panels);
$j$ versus $R_\mathrm{dec}$ (bottom panels). Lines mark the \citetalias{goodman93} relations:
${\cal G} \propto R^{-0.4}$ and $j \propto R^{1.6}$.}
\label{fig:rotvsr}
\end{center}
\end{figure}

\citetalias{goodman93} found that many of the derived rotation
parameters scale with radius, including ${\cal G}$ ($\propto
R^{-0.4}$) and $j$ ($\propto R^{1.6}$)
but not $\beta_\mathrm{rot}$. We plot the
same parameters as a function of clump size in Fig.\
\ref{fig:rotvsr}. Most of the relations found by
\citetalias{goodman93} hold for these clumps, although there is
considerable scatter. \citetalias{goodman93} show
that the relation for the gradient is implied from a
linewidth-size relation (e.g.\ \citealp{larson81}), $\Delta v = R^{0.6}$, for
a core in virial equilibrium. we find $\beta_\mathrm{rot}$ is independent
of $R$ as well, implying $M \propto R^3$ for solid body rotation
($\omega =$ constant) or for differential rotation ($\omega \propto
R^{-1}$) that $M \propto R$. The \scuba\ clumps follow $M \propto
R^2$ \citep{scubapaper}, so we can infer
a more complicated form of rotation than pure solid body or
differential to ensure $\beta_\mathrm{rot}$ is independent of size. 

The scaling of $j$ with size is of particular
interest. \citet{ohashi97} looked at rotation in protostars in Taurus using C$^{18}$O $J=1\to
0$ interferometric spectra. They found at distances $\la
3500$\,AU, that $j$ was at a constant value of $\sim$10$^{-3}$\,km\,s$^{-1}$\,pc. These findings are consistent with 
magnetically-controlled collapse elucidated by \citet{belloche02},
where in a dense inner core, the angular momentum is locked at a minimum
value. This region could be interpreted as a magnetically supercritical core, decoupling from
its subcritical environment, where $j$ scales with $R$. In our data a horizontal line is almost as convincing as the slope of \citetalias{goodman93}, indeed the spread of
clump sizes is not really wide enough to
constrain the trend unambiguously. Our results are therefore
consistent with this magnetically-controlled scenario: there is a lot of spread
about $j\sim$10$^{-3}$\,km\,s$^{-1}$\,pc but no $j$ much below this
value. How the rotational velocity changes across a clump is less clear and is
a key discriminator between different theories. Such measurements
would require high-resolution, interferometric observations of many cores in different stages of
collapse. 

\section{Summary} \label{sec:summary}

We have analysed the kinematic signatures of two populations of
  dust continuum clumps identified in \scuba\ 850\,\micron\ data using
  either \clfind\ or \gclumps\ \citep{scubapaper}, across four regions
(\ngc, IC348/HH211, L1448 and L1455) of active star formation in the
Perseus molecular cloud. The kinematics were derived from
our large-scale (600\,arcmin$^2$) \ceighteeno\ \threetotwo\ survey \citepalias{paper1}. First, the non-thermal contribution to the C$^{18}$O linewidth was
calculated and compared to the results of \citetalias{hkirk07} and the
gravoturbulent models of \citet{klessen05}:
\begin{enumerate}
\item  Most clumps have supersonic non-thermal linewidths, $\langle
  \sigma_\mathrm{NT}/c_\mathrm{s}\rangle= 1.76\pm 0.09$
  (\clfind\ population) or $1.71\pm 0.05$ (\gclumps), with overall distributions similar to the C$^{18}$O $J=2\to 1$ data of \citetalias{hkirk07} towards dense core
candidates in Perseus. This implies we are also probing the `envelopes' of the
star-forming clumps rather than deep within their interiors. 
\item There is little difference in the level of the non-thermal
  linewidth contribution between protostars and starless clumps,
  implying that protostars do no affect their environment
  significantly. Clumps in \ngc\ and IC348 are in the
vicinity of young stellar clusters, explaining the wider
linewidths in \ngc\ ($\langle\sigma_\mathrm{NT}/c_\mathrm{s}\rangle=1.91\pm 0.07$) than L1448 ($1.52\pm0.13$) for the
\gclumps\ objects (although their linewidths are similar for
\clfind\ population: $2.05\pm0.12$ and $1.8\pm0.2$ respectively). IC348 however, has the
narrowest lines of all ($1.36\pm0.09$ with \clfind\ and with
$1.44\pm0.07$ \gclumps), possibly related to its statistically younger 
population \citep{hatchell05}, containing more starless clumps.
\end{enumerate}

Second, the C$^{18}$O linewidths allowed an estimate of each clump's virial mass: 
\begin{enumerate}
\item[(iii)] Most clumps in Perseus appear bound and close to
  equipartition. The ratio of the \scuba\ to virial mass for clumps
  based on flux allocation using \clfind\ is $\langle
  M_{850}/M_\mathrm{vir} \rangle = 1.32 \pm 0.13$, compared to $0.9\pm
  0.1$ for the \gclumps\ decomposition. 
\item[(iv)] Starless clumps occupy a similar section of
  $M_{850}$-$M_\mathrm{vir}$ parameter space to the protostars
  suggesting that they are gravitationally bound and therefore truly
  prestellar. This has implications for our previous work
  investigating the starless CMF in Perseus \citep{scubapaper}. 
\end{enumerate}

These results are in contrast to the models of \citet{klessen05} where
a majority of starless core have $M_\mathrm{vir}\gg M$ and occupy a
different parameter space to the protostars. 

Finally, we fitted a linear velocity gradient to the C$^{18}$O line
centres, across the face of each \scuba\ clump, again identified
either with \clfind\ or \gclumps. Significant detections were found in approximately a third
of the clumps, which we interpreted in terms of solid-body rotation.
\begin{enumerate}
\item[(v)] A correlation between CO $J=3\to 2$ outflows and
the C$^{18}$O line centres is observed, implying any gradients in
outflow sources, i.e.\ protostars, may reflect outflows rather than rotation. 
\item[(vi)] The fitted gradients, in the range $\sim 1$ to
16\,km\,s$^{-1}$\,pc$^{-1}$, are larger than those
previously observed. This probably results from the higher angular
resolution of our observations and/or outflow contamination.
\item[(vii)] There is no correlation between the gradient direction and the
clump orientation nor between the gradient and clump axis
ratio, implying that the rotation is not energetically
significant. Furthermore, the ratio of the clump rotational to gravitational energy is
typically $\la 0.02$, demonstrating that rotation is not dynamically
important as well, with a distribution very similar to that of
the N$_2$H$^+$ cores of \citet{caselli02}. 
\item[(viii)] The distribution of specific angular momentum is narrower
and centred around lower values, $j \sim 10^{-3}$\,km\,s$^{-1}$\,pc, than previous 
studies. Interpreting the results of \citetalias{jappsen04}
this suggests a denser and/or colder environment for the Perseus
clumps than seen in the \citet{caselli02} sample (taken to have $n \sim
10^{5}$\,cm$^{-3}$ and $T \sim 10$\,K). This would seem unlikely given
the large linewidths we observed towards the clumps, although radiative
transfer modelling does suggest we are probing moderately high
densities -- peak \htwo\ number densities of a few $\times 10^5$\,\cmthree\ for a free-falling
protostellar envelope \citep{curtisthesis}. Thus, we are probably
seeing lower levels of rotation in our clumps. 
\item[(ix)] There are no strong trends in the rotational parameters with radius. 
\end{enumerate}   

The somewhat inconclusive findings of this paper motivate further work on the star-forming
cores in Perseus. First, the construction of accurate radiative
transfer models to determine precisely the conditions traced by the
CO \threetotwo\ transitions. Second, higher-resolution or
higher-tracer-density observations. $j$ found in our clumps, suggestively hovers around the background value,
$10^{-3}$\,\kms\,pc$^{-1}$, found by \citet{ohashi97} at radii
$\lesssim 0.03$\,pc in Taurus's protostellar cores. To see how exactly
$j$ varies with radius we need to probe across the clumps themselves, using
higher-density tracers or higher-resolution observations.  

\section{Acknowledgments}

EIC thanks the Science and Technology Facilities Council (STFC) for studentship support while carrying out this
work. We thank Jane Buckle and Gary Fuller for reading carefully an
early version of this paper. We are grateful to the referee, for
comments and suggestions which improved the clarity of this paper. The JCMT is operated by The Joint Astronomy
Centre (JAC) on behalf of the STFC of
the United Kingdom, the Netherlands Organisation for Scientific
Research and the National Research Council (NRC) of Canada. We have also
made extensive use of the SIMBAD data base, operated at CDS,
Strasbourg, France. We acknowledge the data analysis
facilities provided by the Starlink Project which is maintained by JAC
with support from STFC. This research used the facilities of the Canadian
Astronomy Data Centre operated by the NRC with the support of the Canadian Space Agency. 

\renewcommand{\bibname}{References}

\label{lastpage}

\appendix

\section{Derived core dynamical and rotational properties}

\begin{figure*}
\begin{center}
\begin{tabular}{ccc}
\includegraphics[height=0.18\textheight]{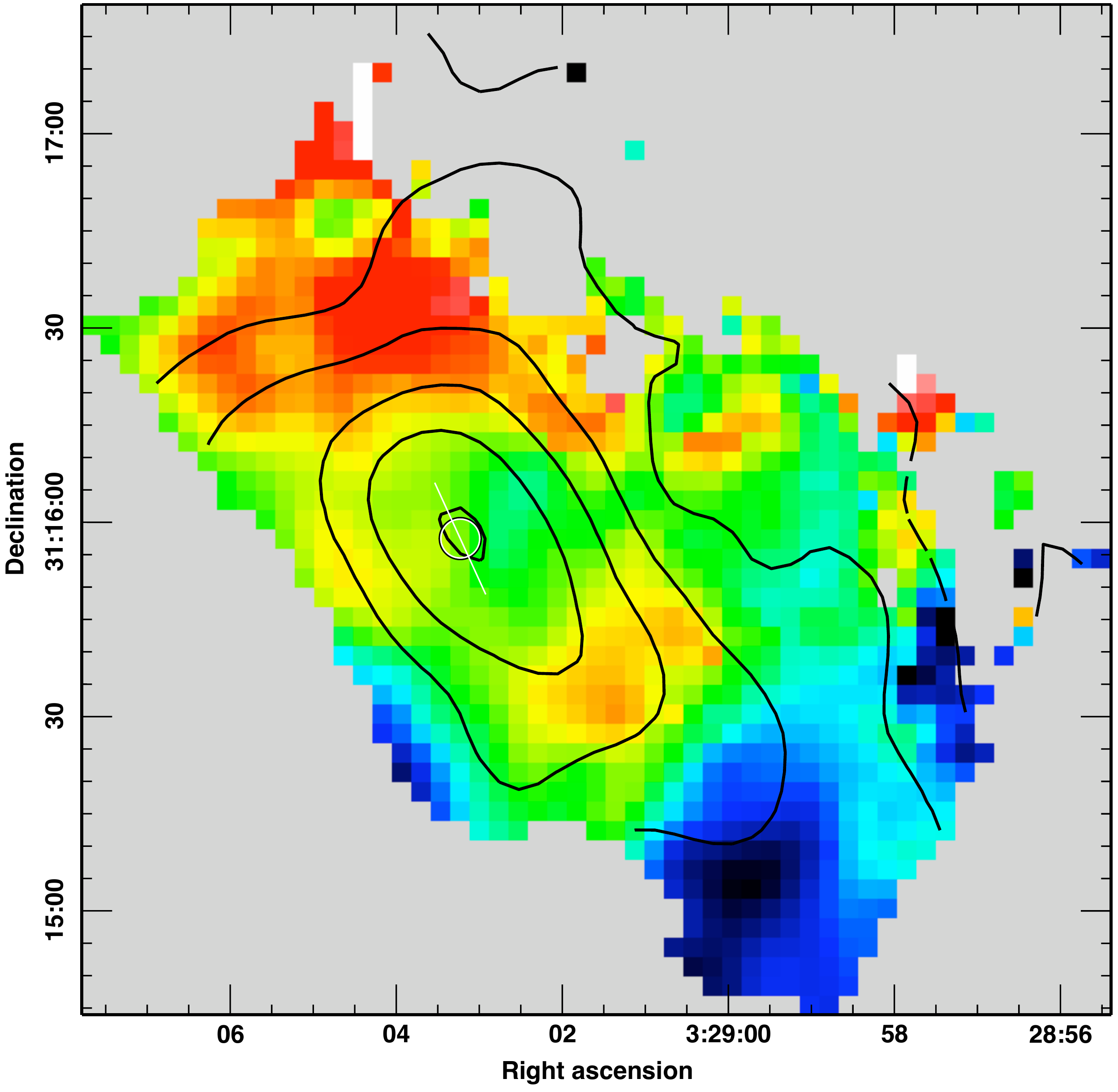}
&
\includegraphics[height=0.18\textheight]{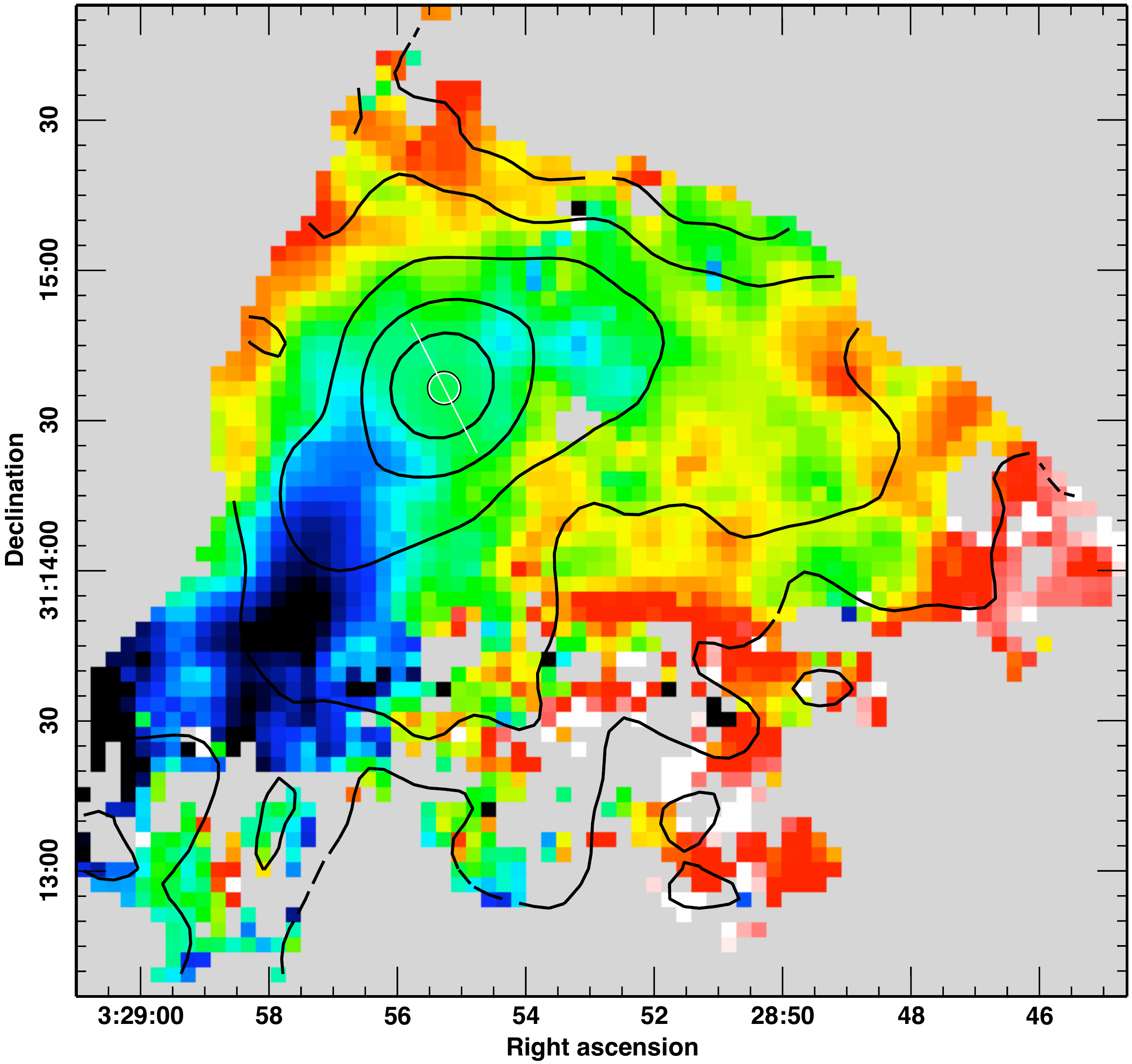}
&
\includegraphics[height=0.18\textheight]{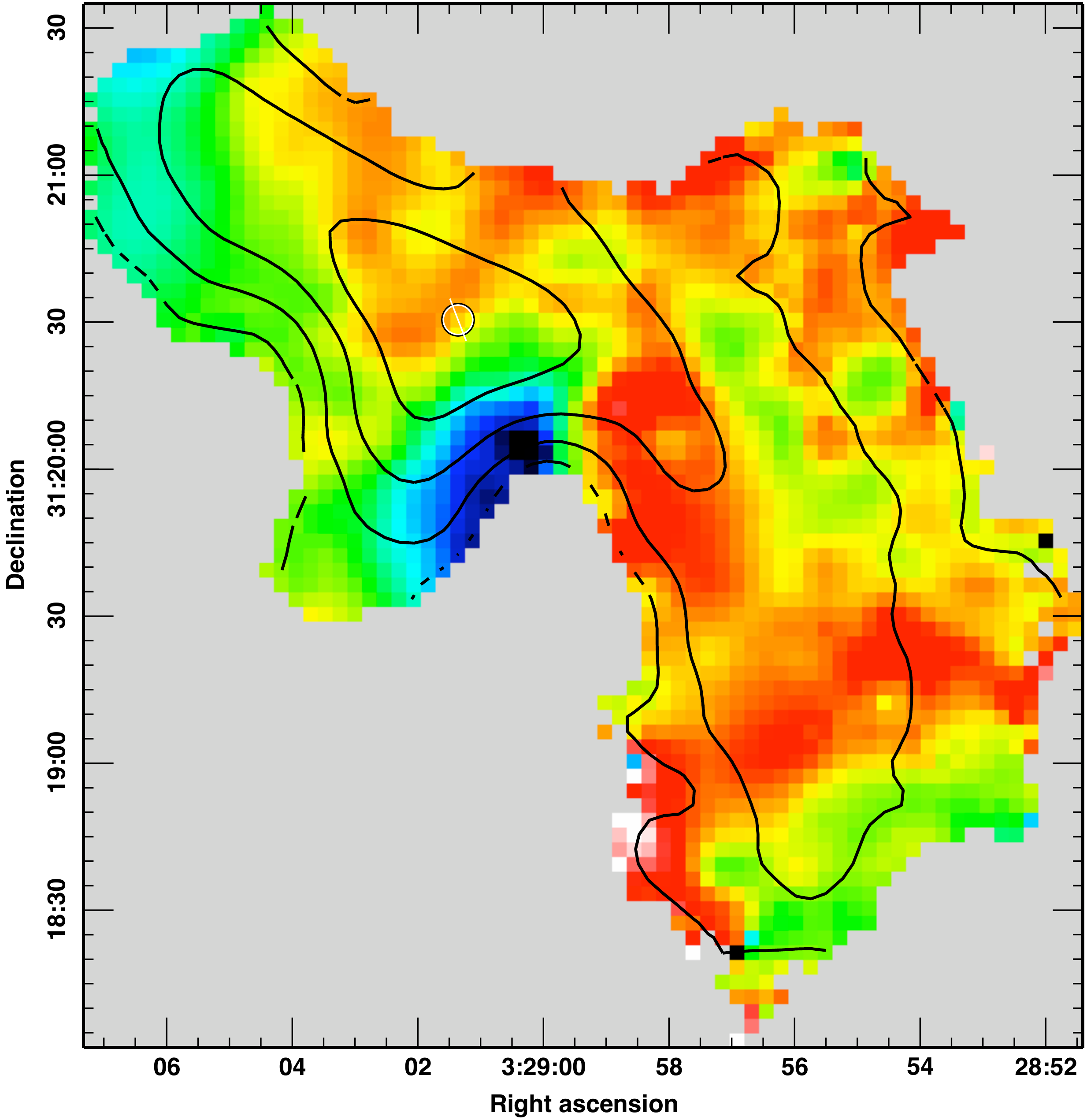}
\\
\ngc\ 2 & \ngc\ 3 & \ngc\ 5 \\
\end{tabular}
\end{center}
\caption{SCUBA 850\,\micron\ clumps identified with \clfind\
containing significant velocity gradients, $\sigma_{\cal G}/{\cal
  G}\geq 3$ in \ngc, IC348 and L1448. The colour-scale denotes
the C$^{18}$O $J=3\to 2$ line centre velocity, linearly scaled from
the clump minimum (blue/black) to maximum
(red/white). Contours of 850~$\mu$m flux density are overlaid
(black) at 100, 200, 400, 800, 1600 and 3200\,mJy\,beam$^{-1}$. A
white/black circle marks the position of the clump peak emission with
a white line positioned to show the direction of the fitted velocity
gradient. The full version of this figure is available as Supporting
  Information to the online version of this article.} \label{fig:velocitygradientmaps_clfind}

\end{figure*}

\begin{figure*}
\begin{center}
\begin{tabular}{ccc}
\includegraphics[height=0.18\textheight]{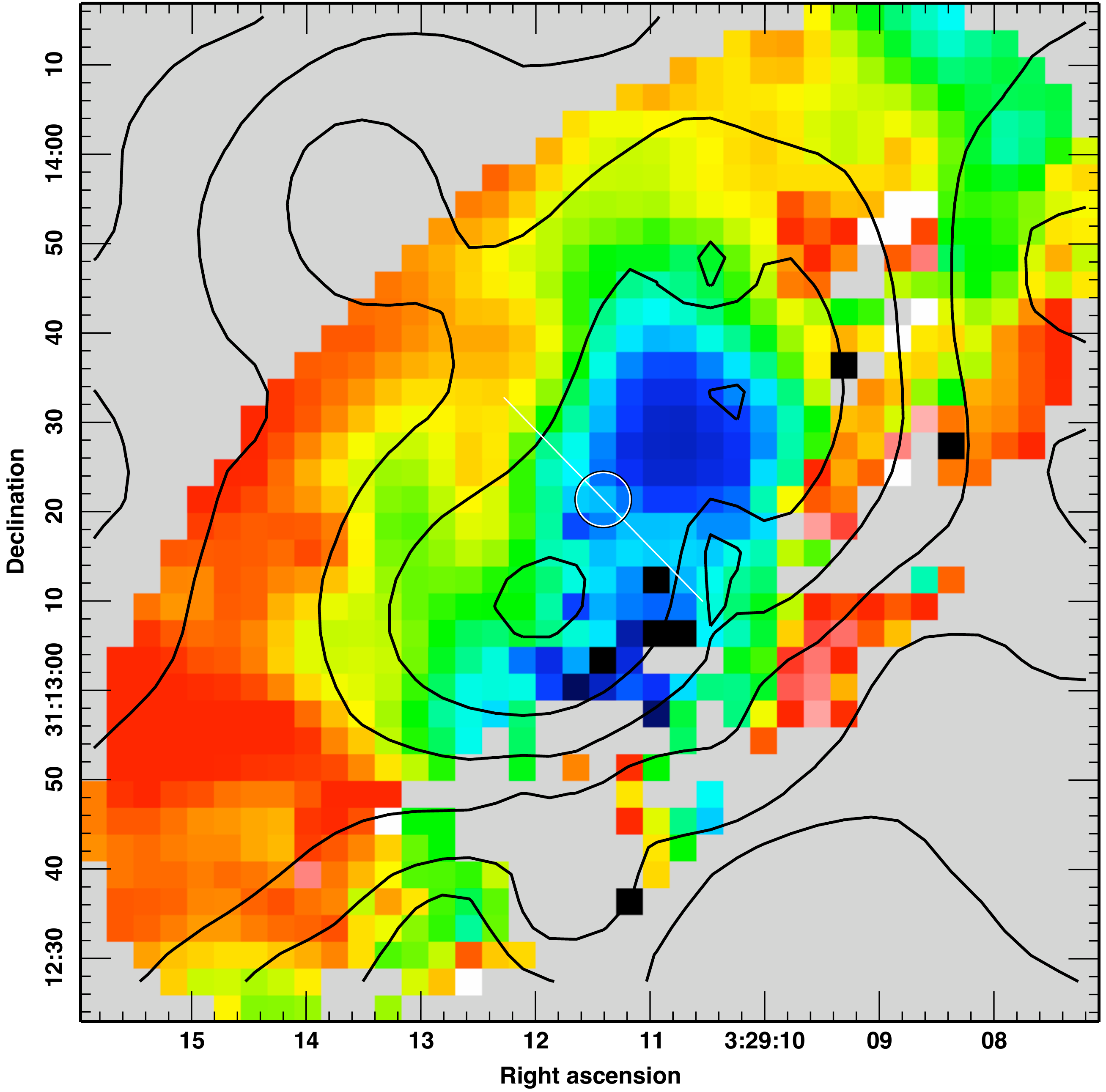}&
\includegraphics[height=0.18\textheight]{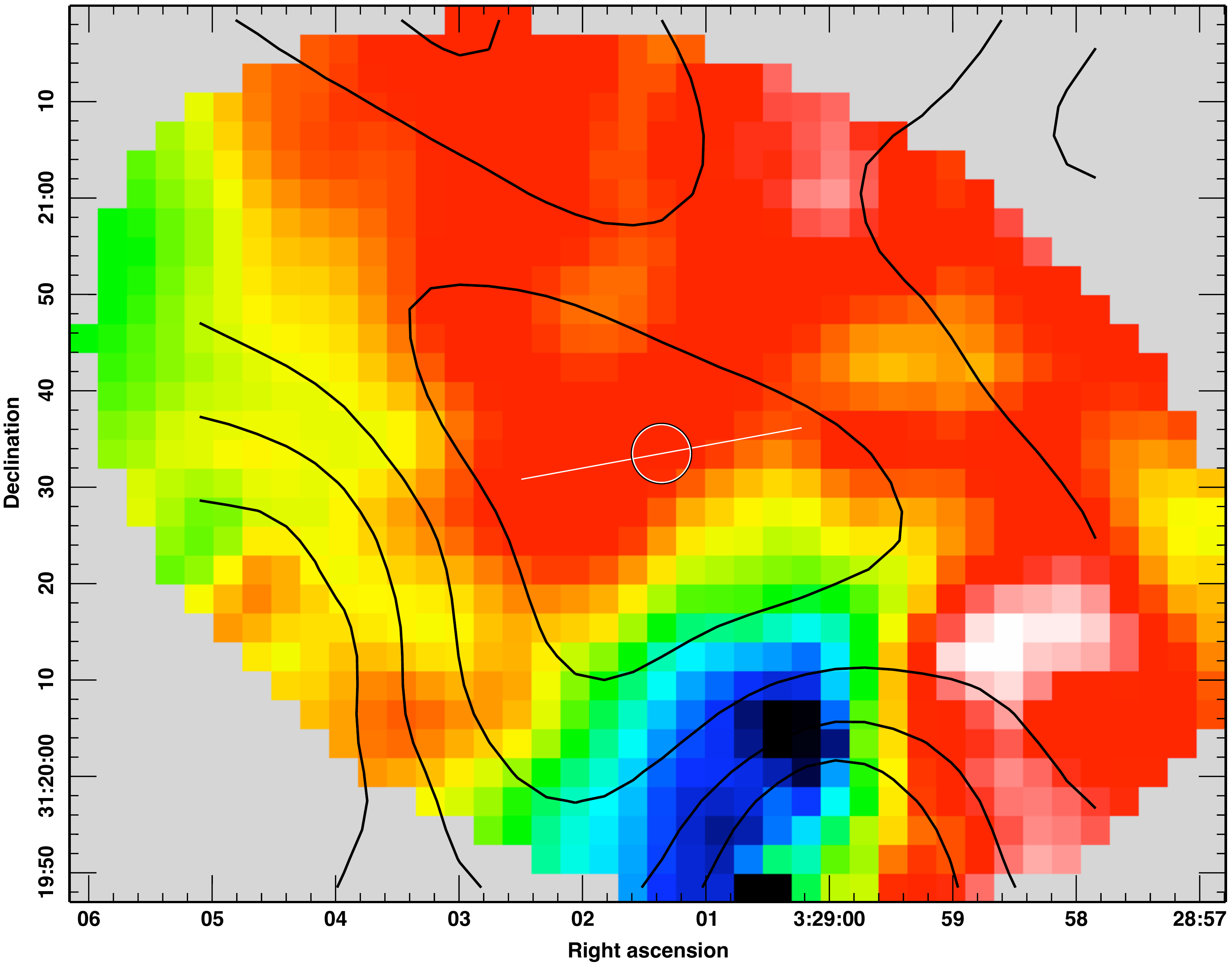}&
\includegraphics[height=0.18\textheight]{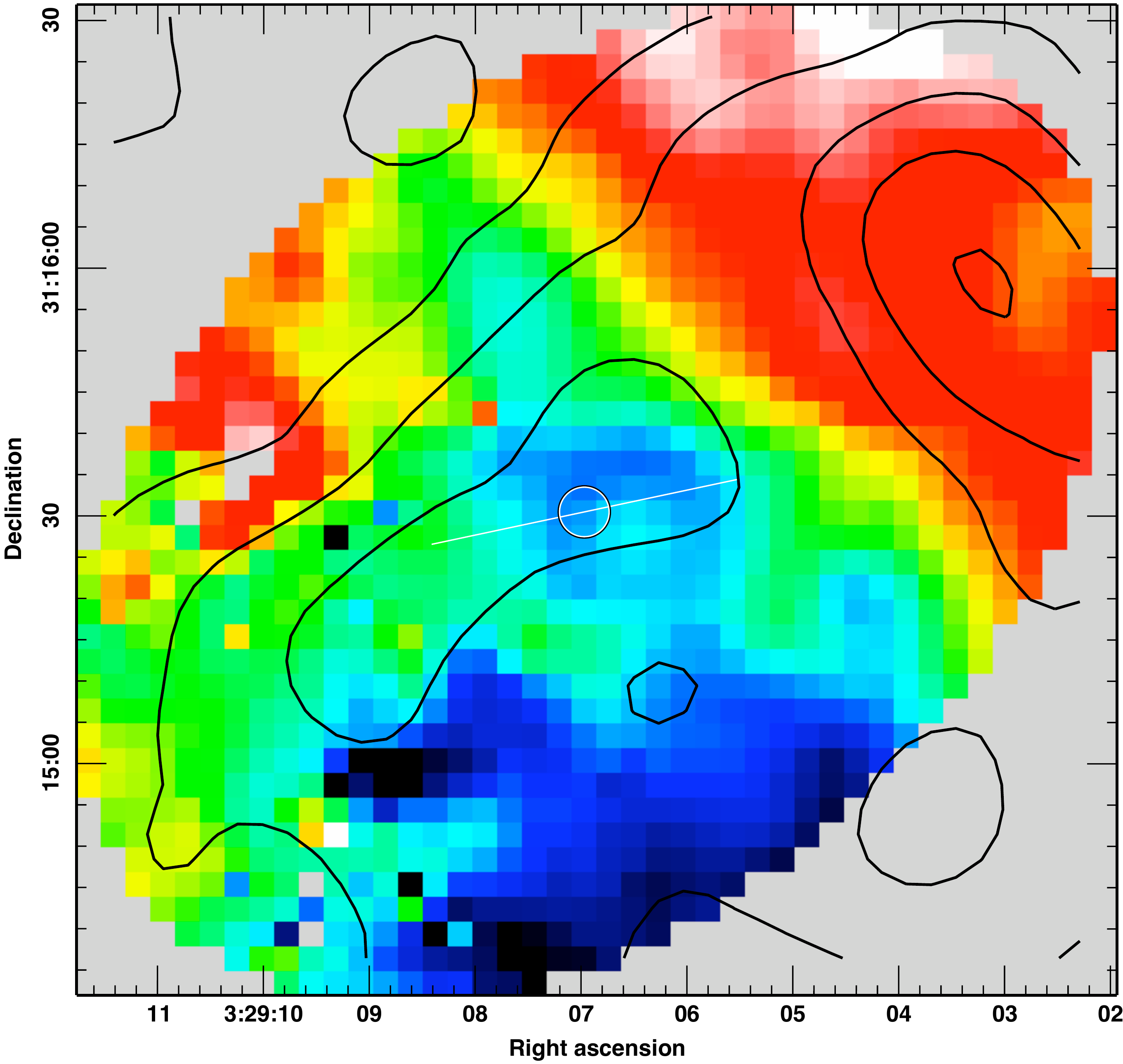}\\
\ngc\ 1 &  \ngc\ 4 & \ngc\ 7\\
\end{tabular}
\end{center}
\caption{SCUBA 850\,\micron\ clumps identified with \gclumps\ containing significant velocity gradients, $\sigma_{\cal G}/{\cal
  G}\geq 3$ in NGC1333, IC348 and L1448. The colour-scale denotes
the C$^{18}$O $J=3\to 2$ line centre velocity, linearly scaled from
the clump minimum (blue/black) to maximum
(red/white). Contours of 850~$\mu$m flux density are overlaid
(black) at 100, 200, 400, 800, 1600 and 3200\,mJy\,beam$^{-1}$. A
white/black circle marks the position of the clump peak emission with
a white line positioned to show the direction of the fitted velocity
gradient. The full version of this figure is available as Supporting
  Information to the online version of this article.}
\label{fig:velocitygradientmaps_gclumps}
\end{figure*}

We list linewidths, virial masses and rotational parameters alongside
other properties for every clump examined in Tables \ref{table:clfind_virial}, \ref{table:gclumps_virial},
\ref{table:rotationalproperties_clfind} and \ref{table:rotationalproperties_gclumps}. In Figs. \ref{fig:velocitygradientmaps_clfind} and \ref{fig:velocitygradientmaps_gclumps}, we plot the
\ceighteeno\ \threetotwo\ line centre velocity and best-fitting velocity
gradient for every SCUBA clump with a significant detection. 

\begin{table*}
\caption{Dynamical properties of \scuba\ clumps found with
  \clfind. The full version of this table is available as Supporting
  Information to the online version of this article.}
\begin{tabular}{c c c c c c c c}
\hline
Sub-region~$^{a}$  & Clump~$^{b}$  &
Hatchell~$^{c}$  & $T_\mathrm{NH_3}$~$^{d}$ & $v_\mathrm{c}(\mathrm{C^{18}O})$~$^{e}$
& $\Delta v_\mathrm{C^{18}O}$~$^{f}$ & $\sigma_\mathrm{NT}$~$^{g}$ & $M_\mathrm{vir}$~$^{h}$  \\
 &  ID  & ID & (K)  & (\kms)  & (\kms) &
(\kms) & (M$_\odot$)  \\
\hline
NGC\,1333 & 1 & 42 & 13.5 & 7.36 & 1.91 & 0.81 & 11.3 \\
NGC\,1333 & 2 & 43 & 16.3 & 8.24 & 1.53 & 0.64 & 11.9 \\
NGC\,1333 & 3 & 44 & 16.5 & 7.62 & 1.19 & 0.50 & 12.3 \\
NGC\,1333 & 4 & 41 & 15.0 & 7.48 & 1.81 & 0.77 & 13.7 \\
NGC\,1333 & 5 & 45 & 16.4 & 7.79 & 1.40 & 0.59 & 22.9 \\
... \\
\hline
\end{tabular}\\
\begin{flushleft}
$^a$~Name of sub-region map. \\
$^b$~Clump number from Curtis \& Richer (2010). \\
$^c$~H07 identifier. \\
$^{d}$~Rosolowsky et al. (2008) NH$_3$ kinetic temperature where they
exist or 10.0, 15.0 and 12.0\,K for starless clumps, protostars and
clumps with no H07 identification where not respectively. \\
$^{e,f}$~Measured C$^{18}$O $J=3\to2$ line centre velocity,
$v_\mathrm{c}$ and FWHM, $\Delta v_\mathrm{C^{18}O}$ from a Gaussian
fit to the line profile at the clump peak. \\
$^{g}$~Non-thermal contribution to the linewidth estimated using the
  temperatures of column (d). \\
$^{h}$~Virial masses calculated using the temperatures of column (d). \\
\end{flushleft}
\label{table:clfind_virial}
\end{table*}

\begin{table*}
\caption{Dynamical properties of \scuba\ clumps found with \gclumps. The full version of this table is available as Supporting
  Information to the online version of this article.}
\begin{tabular}{c c c c c c c c}
\hline
Sub-region~$^{a}$  & Clump~$^{b}$  &
Hatchell~$^{c}$  & $T_\mathrm{NH_3}$~$^{d}$ & $v_\mathrm{c}(\mathrm{C^{18}O})$~$^{e}$
& $\Delta v_\mathrm{C^{18}O}$~$^{f}$ & $\sigma_\mathrm{NT}$~$^{g}$ & $M_\mathrm{vir}$~$^{h}$  \\
 &  ID  & ID & (K)  & (\kms)  & (\kms) &
(\kms) & (M$_\odot$)  \\
\hline
NGC\,1333 & 1 & 41,42 & 15.0 & 6.77 & 1.04 & 0.44 & 11.9\\
NGC\,1333 & 2 & 43 & 16.3 & 8.18 & 1.51 & 0.64 & 5.8\\
NGC\,1333 & 3 & -- & 16.5 & 7.62 & 1.19 & 0.50 & \\
NGC\,1333 & 4 & 45 & 16.4 & 7.82 & 1.35 & 0.57 & 8.7 \\
NGC\,1333 & 5 & 46 & 14.3 & 8.40 & 1.27 & 0.54 & 4.0 \\
...\\
\hline
\end{tabular}\\
\begin{flushleft}
$^a$~Name of sub-region map. \\
$^b$~Clump number from Curtis \& Richer (2010). \\
$^c$~H07 identifier. \\
$^{d}$~Rosolowsky et al. (2008) NH$_3$ kinetic temperature where they
exist or 10.0, 15.0 and 12.0\,K for starless clumps, protostars and
clumps with no H07 identification where not respectively. \\
$^{e,f}$~Measured C$^{18}$O $J=3\to2$ line centre velocity,
$v_\mathrm{c}$ and FWHM, $\Delta v_\mathrm{C^{18}O}$ from a Gaussian
fit to the line profile at the clump peak. \\
$^{g}$~Non-thermal contribution to the linewidth estimated using the
  temperatures of column (d). \\
$^{h}$~Virial masses calculated using the temperatures of column
  (d). \\
\end{flushleft}
\label{table:gclumps_virial}
\end{table*}

\begin{table*}
\caption{Rotational properties of \scuba\ clumps located with
  \clfind. These properties have been found by non-linear
  least squares fitting of a linear velocity gradient to the first moment
  of the C$^{18}$O $J=3\to 2$ data. Only significant detections are shown
  with $\sigma_{\cal G}/{\cal G} \geq 3$. The full version of this table is available as Supporting
  Information to the online version of this article.} 
\begin{tabular}{l c c c c c c c c}
\hline
Sub-region~$^{a}$  & Clump~$^{b}$  &
Hatchell~$^{c}$  & Number~$^{d}$
& ${\cal G}\pm \sigma_{\cal G}$~$^{e}$  & $\theta_{\cal G}$~$^{f}$ & ${\cal
  G}/\sigma_{\cal G}$~$^{g}$ & $\beta_\mathrm{rot}$~$^{h}$ & $j$~$^{i}$  \\
 &  ID  & class & of points & (km\,s$^{-1}$\,pc$^{-1}$) & (deg E
of N) & & & (km\,s$^{-1}$\,pc)  \\
\hline
NGC1333 & 2 & I & 1117 & $3.16\pm0.76$ & 155.44 & 4 & 6.98E-04 & 1.05E-03\\
NGC1333 & 3 & 0 & 2106 & $4.85\pm0.35$ & 26.88 & 14 & 5.39E-03 & 3.51E-03\\
NGC1333 & 5 & I & 2229 & $1.54\pm0.41$ & 21.10 & 4 & 8.91E-04 & 1.63E-03\\
NGC1333 & 6 & 0 & 978 & $4.66\pm0.84$ & 125.79 & 6 & 3.35E-03 & 1.44E-03\\
NGC1333 & 8 & I & 330 & $10.63\pm2.67$ & -58.59 & 4 & 4.67E-03 &
9.83E-04\\
...\\
\hline
\end{tabular}\\
\begin{flushleft}
$^{a}$~Name of sub-region map. \\
$^{b}$~Clump identification from Curtis \& Richer (2010). \\
$^{c}$~H07 clump class, S=starless, 0=Class 0 protostar and
  I=Class I protostar. $^\mathrm{d}$~Number of points fitted in the
  map. \\
$^{e}$~Fitted velocity gradient and its error. \\
$^{f}$~Angle of the fitted velocity gradient, east of north. \\
$^{g}$~Level of significance of the fit, $\sigma_{\cal G}/{\cal
    G}$. \\
$^{h}$~Ratio of the kinetic to gravitational energy. \\
$^{i}$~Specific angular momentum.
\end{flushleft}
\label{table:rotationalproperties_clfind}
\end{table*}

\begin{table*}
\caption{Rotational properties of \scuba\ clumps located with
  \gclumps. These properties have been found by non-linear
  least squares fitting of a linear velocity gradient to the first moment
  of the C$^{18}$O $J=3\to 2$ data. Only significant detections are shown
  with $\sigma_{\cal G}/{\cal G} \geq 3$. The full version of this table is available as Supporting
  Information to the online version of this article.} 
\begin{tabular}{l c c c c c c c c}
\hline
Sub-region~$^{a}$  & Clump~$^{b}$  &
Hatchell~$^{c}$  & Number~$^{d}$
& ${\cal G}\pm \sigma_{\cal G}$~$^{e}$  & $\theta_{\cal G}$~$^{f}$ & ${\cal
  G}/\sigma_{\cal G}$~$^{g}$ & $\beta_\mathrm{rot}$~$^{h}$ & $j$~$^{i}$  \\
 &  ID  & class & of points & (km\,s$^{-1}$\,pc$^{-1}$) & (deg E
of N) & & & (km\,s$^{-1}$\,pc)  \\
\hline
NGC1333 & 1 & 0 & 800 & $5.33\pm1.33$ & 135.67 & 4 & 6.90E-04 & 7.86E-04\\
NGC1333 & 4 & I & 922 & $4.93\pm1.16$ & -79.61 & 4 & 1.22E-03 & 7.41E-04\\
NGC1333 & 7 & I & 1222 & $6.73\pm0.84$ & -77.95 & 8 & 4.27E-03 & 1.56E-03\\
NGC1333 & 8 & 0 & 812 & $4.25\pm1.11$ & 26.44 & 4 & 1.07E-03 & 5.54E-04\\
NGC1333 & 9 & 0 & 1374 & $5.37\pm0.65$ & -38.35 & 8 & 3.70E-03 &
1.63E-03\\
...\\
\hline
\end{tabular}\\
\begin{flushleft}
$^{a}$~Name of sub-region map. \\
$^{b}$~Clump identification from Curtis \& Richer (2010). \\
$^{c}$~H07 clump class, S=starless, 0=Class 0 protostar and
  I=Class I protostar. $^\mathrm{d}$~Number of points fitted in the
  map. \\
$^{e}$~Fitted velocity gradient and its error. \\
$^{f}$~Angle of the fitted velocity gradient, east of north. \\
$^{g}$~Level of significance of the fit, $\sigma_{\cal G}/{\cal
    G}$. \\
$^{h}$~Ratio of the kinetic to gravitational energy. \\
$^{i}$~Specific angular momentum.
\end{flushleft}
\label{table:rotationalproperties_gclumps}
\end{table*}

\begin{thebibliography}{}

\bibitem[\protect\citeauthoryear{{Alves}, {Lombardi} \& {Lada}}{{Alves}
  et~al.}{2007}]{alves07}
{Alves} J.,  {Lombardi} M.,    {Lada} C.~J.,  2007, \aap, 462, L17

\bibitem[\protect\citeauthoryear{{Andr{\'e}}, {Belloche}, {Motte} \&
  {Peretto}}{{Andr{\'e}} et~al.}{2007}]{andre07}
{Andr{\'e}} P.,  {Belloche} A.,  {Motte} F.,    {Peretto} N.,  2007, \aap, 472,
  519

\bibitem[\protect\citeauthoryear{{Arquilla} \& {Goldsmith}}{{Arquilla} \&
  {Goldsmith}}{1985}]{arquilla85}
{Arquilla} R.,  {Goldsmith} P.~F.,  1985, \apj, 297, 436

\bibitem[\protect\citeauthoryear{{Ballesteros-Paredes}, {Klessen} \&
  {V{\'a}zquez-Semadeni}}{{Ballesteros-Paredes}
  et~al.}{2003}]{ballesteros-paredes03}
{Ballesteros-Paredes} J.,  {Klessen} R.~S.,    {V{\'a}zquez-Semadeni} E.,
  2003, \apj, 592, 188

\bibitem[\protect\citeauthoryear{{Barranco} \& {Goodman}}{{Barranco} \&
  {Goodman}}{1998}]{barranco98}
{Barranco} J.~A.,  {Goodman} A.~A.,  1998, \apj, 504, 207

\bibitem[\protect\citeauthoryear{{Bate} \& {Bonnell}}{{Bate} \&
  {Bonnell}}{2005}]{bate05}
{Bate} M.~R.,  {Bonnell} I.~A.,  2005, \mnras, 356, 1201

\bibitem[\protect\citeauthoryear{{Belloche}, {Andr{\'e}}, {Despois} \&
  {Blinder}}{{Belloche} et~al.}{2002}]{belloche02}
{Belloche} A.,  {Andr{\'e}} P.,  {Despois} D.,    {Blinder} S.,  2002, \aap,
  393, 927

\bibitem[\protect\citeauthoryear{{Benson} \& {Myers}}{{Benson} \&
  {Myers}}{1989}]{benson89}
{Benson} P.~J.,  {Myers} P.~C.,  1989, \apjss, 71, 89

\bibitem[\protect\citeauthoryear{{Bertoldi} \& {McKee}}{{Bertoldi} \&
  {McKee}}{1992}]{bertoldi92}
{Bertoldi} F.,  {McKee} C.~F.,  1992, \apj, 395, 140

\bibitem[\protect\citeauthoryear{{Binney} \& {Tremaine}}{{Binney} \&
  {Tremaine}}{2008}]{binneytremaine}
{Binney} J.,  {Tremaine} S.,  2008, {Galactic Dynamics}.
Princeton Univ. Press, Princeton, NJ

\bibitem[\protect\citeauthoryear{{Bonnell}, {Bate}, {Clarke} \&
  {Pringle}}{{Bonnell} et~al.}{2001}]{bonnell01}
{Bonnell} I.~A.,  {Bate} M.~R.,  {Clarke} C.~J.,    {Pringle} J.~E.,  2001,
  \mnras, 323, 785

\bibitem[\protect\citeauthoryear{{Bontemps}, {Andre}, {Terebey} \&
  {Cabrit}}{{Bontemps} et~al.}{1996}]{bontemps96}
{Bontemps} S.,  {Andre} P.,  {Terebey} S.,    {Cabrit} S.,  1996, \aap, 311,
  858

\bibitem[\protect\citeauthoryear{{Buckle} et~al.,}{{Buckle}
  et~al.}{2009}]{harp_paper}
{Buckle} J.~V.,  et~al., 2009, \mnras, 399, 1026

\bibitem[\protect\citeauthoryear{{Buckle} et~al.,}{{Buckle} et~al.}{2010}]{buckle10}
{Buckle} J.~V.,  et~al., 2010, \mnras, 401, 204

\bibitem[\protect\citeauthoryear{{Burkert} \& {Bodenheimer}}{{Burkert} \&
  {Bodenheimer}}{2000}]{burkert00}
{Burkert} A.,  {Bodenheimer} P.,  2000, \apj, 543, 822

\bibitem[\protect\citeauthoryear{{Caselli} \& {Myers}}{{Caselli} \&
  {Myers}}{1995}]{caselli95}
{Caselli} P.,  {Myers} P.~C.,  1995, \apj, 446, 665

\bibitem[\protect\citeauthoryear{{Caselli}, {Benson}, {Myers} \&
  {Tafalla}}{{Caselli} et~al.}{2002}]{caselli02}
{Caselli} P.,  {Benson} P.~J.,  {Myers} P.~C.,    {Tafalla} M.,  2002, \apj,
  572, 238

\bibitem[\protect\citeauthoryear{{Clark}, {Klessen} \& {Bonnell}}{{Clark}
  et~al.}{2007}]{clark07}
{Clark} P.~C.,  {Klessen} R.~S.,    {Bonnell} I.~A.,  2007, \mnras, 379, 57

\bibitem[\protect\citeauthoryear{{Curtis}}{{Curtis}}{2009}]{curtisthesis}
{Curtis} E.~I.,  2009, PhD thesis, {Univ.\ of Cambridge}

\bibitem[\protect\citeauthoryear{{Curtis} \& {Richer}}{{Curtis} \&
  {Richer}}{2010}]{scubapaper}
{Curtis} E.~I.,  {Richer} J.~S.,  2010, \mnras, 402, 603

\bibitem[\protect\citeauthoryear{{Curtis}, {Richer} \&
    {Buckle}}{{Curtis} et~al.}{2010a}]{paper1} {Curtis} E.~I.,
  {Richer} J.~S.,    {Buckle} J.~V.,  2010a, \mnras, 401, 455 \citepalias{paper1}

\bibitem[\protect\citeauthoryear{{Curtis}, {Richer}, {Swift} \&
  {Williams}}{{Curtis} et~al.}{2010b}]{paper3}
{Curtis} E.~I.,  {Richer} J.~S.,  {Swift} J.~J.,    {Williams} J.~P.,  2010b,
  \mnras, in press \citepalias{paper3}

\bibitem[\protect\citeauthoryear{{Dib}, {Kim}, {V{\'a}zquez-Semadeni},
  {Burkert} \& {Shadmehri}}{{Dib} et~al.}{2007}]{dib07}
{Dib} S.,  {Kim} J.,  {V{\'a}zquez-Semadeni} E.,  {Burkert} A.,    {Shadmehri}
  M.,  2007, \apj, 661, 262

\bibitem[\protect\citeauthoryear{{Dib}, {Hennebelle}, {Pineda}, {Csengeri},
  {Bontemps}, {Audit} \& {Goodman}}{{Dib} et~al.}{2010}]{dib10}
{Dib} S.,  {Hennebelle} P.,  {Pineda} J.~E.,  {Csengeri} T.,  {Bontemps} S.,
  {Audit} E.,    {Goodman} A.~A.,  2010, \apj, in press

\bibitem[\protect\citeauthoryear{{Duquennoy} \& {Mayor}}{{Duquennoy} \&
  {Mayor}}{1991}]{duquennoy91}
{Duquennoy} A.,  {Mayor} M.,  1991, \aap, 248, 485

\bibitem[\protect\citeauthoryear{{Elmegreen}, {Efremov}, {Pudritz} \&
  {Zinnecker}}{{Elmegreen} et~al.}{2000}]{elmegreen00}
{Elmegreen} B.~G.,  {Efremov} Y.,  {Pudritz} R.~E.,    {Zinnecker} H.,  2000,
  in {Mannings} V.,  {Boss} A.~P.,   {Russell} S.~S.,  eds, Protostars and
  Planets IV. Univ. of Arizona Press, Tucson, p.~179

\bibitem[\protect\citeauthoryear{{Elmegreen} \& {Scalo}}{{Elmegreen} \&
  {Scalo}}{2004}]{elmegreen04}
{Elmegreen} B.~G.,  {Scalo} J.,  2004, \araa, 42, 211

\bibitem[\protect\citeauthoryear{{Enoch} et~al.,}{{Enoch}
  et~al.}{2006}]{enoch06}
{Enoch} M.~L.,  et~al., 2006, \apj, 638, 293

\bibitem[\protect\citeauthoryear{{Enoch}, {Evans} II, {Sargent}, {Glenn},
  {Rosolowsky} \& {Myers}}{{Enoch} et~al.}{2008}]{enoch08}
{Enoch} M.~L.,  {Evans} II N.~J.,  {Sargent} A.~I.,  {Glenn} J.,  {Rosolowsky}
  E.,    {Myers} P.,  2008, \apj, 684, 1240

\bibitem[\protect\citeauthoryear{{Galli}, {Walmsley} \& {Gon{\c c}alves}}{{Galli} et~al.}{2002}]{galli02}
{Galli} D.,  {Walmsley} M.,  {Gon{\c c}alves} J., 2002, \aap, 394, 275

\bibitem[\protect\citeauthoryear{{Gammie}, {Lin}, {Stone} \&
  {Ostriker}}{{Gammie} et~al.}{2003}]{gammie03}
{Gammie} C.~F.,  {Lin} Y.-T.,  {Stone} J.~M.,    {Ostriker} E.~C.,  2003, \apj,
  592, 203

\bibitem[\protect\citeauthoryear{{Goodman}, {Benson}, {Fuller} \&
  {Myers}}{{Goodman} et~al.}{1993}]{goodman93}
{Goodman} A.~A.,  {Benson} P.~J.,  {Fuller} G.~A.,    {Myers} P.~C.,  1993,
  \apj, 406, 528 \citepalias{goodman93}

\bibitem[\protect\citeauthoryear{{Goodman}, {Barranco}, {Wilner} \& {Heyer}}{{Goodman} et~al.}{1998}]{goodman98}
{Goodman} A.~A.,  {Barranco} J.~A.,  {Wilner} D.~J.,    {Heyer} M.~H.,  1998,
  \apj, 504, 223 

\bibitem[\protect\citeauthoryear{{Goodwin} \& {Kouwenhoven}}{{Goodwin} \&
  {Kouwenhoven}}{2009}]{goodwin09}
{Goodwin} S.~P.,  {Kouwenhoven} M.~B.~N.,  2009, \mnras, 397, L36

\bibitem[\protect\citeauthoryear{{Goodwin}, {Kroupa}, {Goodman} \&
  {Burkert}}{{Goodwin} et~al.}{2007}]{goodwin07}
{Goodwin} S.~P.,  {Kroupa} P.,  {Goodman} A.,    {Burkert} A.,  2007, in
  {Reipurth} B.,  {Jewitt} D.,   {Keil} K.,  eds, Protostars and
  Planets V. Univ. of Arizona Press, Tucson, p.~133

\bibitem[\protect\citeauthoryear{{Goodwin}, {Nutter}, {Kroupa}, {Ward-Thompson}
  \& {Whitworth}}{{Goodwin} et~al.}{2008}]{goodwin08}
{Goodwin} S.~P.,  {Nutter} D.,  {Kroupa} P.,  {Ward-Thompson} D.,
  {Whitworth} A.~P.,  2008, \aap, 477, 823

\bibitem[\protect\citeauthoryear{{Hartmann}}{{Hartmann}}{2001}]{hartmann01}
{Hartmann} L.,  2001, \aj, 121, 1030

\bibitem[\protect\citeauthoryear{{Hatchell} \& {Fuller}}{{Hatchell} \&
  {Fuller}}{2008}]{hatchell08}
{Hatchell} J.,  {Fuller} G.~A.,  2008, \aap, 482, 855

\bibitem[\protect\citeauthoryear{{Hatchell}, {Richer}, {Fuller}, {Qualtrough},
  {Ladd} \& {Chandler}}{{Hatchell} et~al.}{2005}]{hatchell05}
{Hatchell} J.,  {Richer} J.~S.,  {Fuller} G.~A.,  {Qualtrough} C.~J.,  {Ladd}
  E.~F.,    {Chandler} C.~J.,  2005, \aap, 440, 151

\bibitem[\protect\citeauthoryear{{Hatchell}, {Fuller}, {Richer}, {Harries} \& {Ladd}}{{Hatchell} et~al.}{2007}]{hatch07}
{Hatchell} J.,  {Fuller} G.~A.,  {Richer} J.~S.,  {Harries} T.~J.,    {Ladd}
  E.~F.,  2007, \aap, 468, 1009 \citepalias{hatch07}

\bibitem[\protect\citeauthoryear{{Jappsen} \& {Klessen}}{{Jappsen} \&
  {Klessen}}{2004}]{jappsen04}
{Jappsen} A.-K.,  {Klessen} R.~S.,  2004, \aap, 423, 1 \citepalias{jappsen04}

\bibitem[\protect\citeauthoryear{{Jijina}, {Myers} \& {Adams}}{{Jijina}
  et~al.}{1999}]{jijina99}
{Jijina} J.,  {Myers} P.~C.,    {Adams} F.~C.,  1999, \apjss, 125, 161

\bibitem[\protect\citeauthoryear{{Jones}, {Basu} \& {Dubinski}}{{Jones}
  et~al.}{2001}]{jones01}
{Jones} C.~E.,  {Basu} S.,    {Dubinski} J.,  2001, \apj, 551, 387

\bibitem[\protect\citeauthoryear{{Kirk}, {Johnstone} \& {Di Francesco}}{{Kirk}
  et~al.}{2006}]{kirk06}
{Kirk} H.,  {Johnstone} D.,    {Di Francesco} J.,  2006, \apj, 646, 1009

\bibitem[\protect\citeauthoryear{{Kirk}, {Johnstone} \&
    {Tafalla}}{{Kirk} et~al.}{2007}]{hkirk07}
{Kirk} H.,  {Johnstone} D.,    {Tafalla} M.,  2007, \apj, 668, 1042 \citepalias{hkirk07}

\bibitem[\protect\citeauthoryear{{Klessen}, {Ballesteros-Paredes},
  {V{\'a}zquez-Semadeni} \& {Dur{\'a}n-Rojas}}{{Klessen}
  et~al.}{2005}]{klessen05}
{Klessen} R.~S.,  {Ballesteros-Paredes} J.,  {V{\'a}zquez-Semadeni} E.,
  {Dur{\'a}n-Rojas} C.,  2005, \apj, 620, 786

\bibitem[\protect\citeauthoryear{{Krumholz} \& {Tan}}{{Krumholz} \&
  {Tan}}{2007}]{krumholz07}
{Krumholz} M.~R.,  {Tan} J.~C.,  2007, \apj, 654, 304

\bibitem[\protect\citeauthoryear{{Krumholz}, {McKee} \& {Klein}}{{Krumholz}
  et~al.}{2005}]{krumholz05}
{Krumholz} M.~R.,  {McKee} C.~F.,    {Klein} R.~I.,  2005, \nat, 438, 332

\bibitem[\protect\citeauthoryear{{Larson}}{{Larson}}{1981}]{larson81}
{Larson} R.~B.,  1981, \mnras, 194, 809

\bibitem[\protect\citeauthoryear{{Larson}}{{Larson}}{2003}]{larson03}
{Larson} R.~B.,  2003, Rep. of Progress in Phys., 66, 1651

\bibitem[\protect\citeauthoryear{{Li}, {Norman}, {Mac Low} \& {Heitsch}}{{Li}
  et~al.}{2004}]{li04}
{Li} P.~S.,  {Norman} M.~L.,  {Mac Low} M.-M.,    {Heitsch} F.,  2004, \apj,
  605, 800

\bibitem[\protect\citeauthoryear{{Lissauer}}{{Lissauer}}{1993}]{lissauer93}
{Lissauer} J.~J.,  1993, \araa, 31, 129

\bibitem[\protect\citeauthoryear{{Mac Low} \& {Klessen}}{{Mac Low} \&
  {Klessen}}{2004}]{maclow04}
{Mac Low} M.-M.,  {Klessen} R.~S.,  2004, Rev. of Modern Phys., 76, 125

\bibitem[\protect\citeauthoryear{{MacLaren}, {Richardson} \&
  {Wolfendale}}{{MacLaren} et~al.}{1988}]{maclaren88}
{MacLaren} I.,  {Richardson} K.~M.,    {Wolfendale} A.~W.,  1988, \apj, 333,
  821

\bibitem[\protect\citeauthoryear{{McKee}}{{McKee}}{1999}]{mckee99}
{McKee} C.~F.,  1999, in {Lada} C.~J.,  {Kylafis} N.~D.,  eds, The Origin of Stars and Planetary Systems. Kluwer, Dordrecht, p.~29

\bibitem[\protect\citeauthoryear{{Motte}, {Andre} \& {Neri}}{{Motte}
  et~al.}{1998}]{motte98}
{Motte} F.,  {Andre} P.,    {Neri} R.,  1998, \aap, 336, 150

\bibitem[\protect\citeauthoryear{{Mouschovias}}{{Mouschovias}}{1987}]{mouschov%
ias87}
{Mouschovias} T.~C.,  1987, in {Morfill} G.~E.,  {Scholer} M.,  eds, Physical Processes in Interstellar Clouds. Reidel,
  Dordrecht, p.~453

\bibitem[\protect\citeauthoryear{{Myers}}{{Myers}}{1983}]{myers83}
{Myers} P.~C.,  1983, \apj, 270, 105

\bibitem[\protect\citeauthoryear{{Nakamura} \& {Li}}{{Nakamura} \&
  {Li}}{2005}]{nakamura05}
{Nakamura} F.,  {Li} Z.-Y.,  2005, \apj, 631, 411

\bibitem[\protect\citeauthoryear{{Nakamura} \& {Li}}{{Nakamura} \&
  {Li}}{2007}]{nakamura07}
{Nakamura} F.,  {Li} Z.-Y.,  2007, \apj, 662, 395

\bibitem[\protect\citeauthoryear{{Offner}, {Klein} \& {McKee}}{{Offner}
  et~al.}{2008a}]{offner08a}
{Offner} S.~S.~R.,  {Klein} R.~I.,    {McKee} C.~F.,  2008a, \apj, 686, 1174

\bibitem[\protect\citeauthoryear{{Offner}, {Krumholz}, {Klein} \&
  {McKee}}{{Offner} et~al.}{2008b}]{offner08b}
{Offner} S.~S.~R.,  {Krumholz} M.~R.,  {Klein} R.~I.,    {McKee} C.~F.,  2008b,
  \aj, 136, 404

\bibitem[\protect\citeauthoryear{{Ohashi}, {Hayashi}, {Ho}, {Momose}, {Tamura},
  {Hirano} \& {Sargent}}{{Ohashi} et~al.}{1997}]{ohashi97}
{Ohashi} N.,  {Hayashi} M.,  {Ho} P.~T.~P.,  {Momose} M.,  {Tamura} M.,
  {Hirano} N.,    {Sargent} A.~I.,  1997, \apj, 488, 317

\bibitem[\protect\citeauthoryear{{Olmi}, {Testi} \& {Sargent}}{{Olmi}
  et~al.}{2005}]{olmi05}
{Olmi} L.,  {Testi} L.,    {Sargent} A.~I.,  2005, \aap, 431, 253

\bibitem[\protect\citeauthoryear{{Padoan}, {Juvela}, {Goodman} \&
  {Nordlund}}{{Padoan} et~al.}{2001}]{padoan01b}
{Padoan} P.,  {Juvela} M.,  {Goodman} A.~A.,    {Nordlund} {\AA}.,  2001, \apj,
  553, 227

\bibitem[\protect\citeauthoryear{{Pineda}, {Goodman}, {Arce}, {Caselli},
  {Foster}, {Myers} \& {Rosolowsky}}{{Pineda} et~al.}{2010}]{pineda10}
{Pineda} J.~E.,  {Goodman} A.~A.,  {Arce} H.~G.,  {Caselli} P.,  {Foster}
  J.~B.,  {Myers} P.~C.,    {Rosolowsky} E.~W.,  2010, \apjl, 712, L116

\bibitem[\protect\citeauthoryear{{Pirogov}, {Zinchenko}, {Caselli}, {Johansson}
  \& {Myers}}{{Pirogov} et~al.}{2003}]{pirogov03}
{Pirogov} L.,  {Zinchenko} I.,  {Caselli} P.,  {Johansson} L.~E.~B.,    {Myers}
  P.~C.,  2003, \aap, 405, 639

\bibitem[\protect\citeauthoryear{{Redman}, {Keto}, {Rawlings} \&
  {Williams}}{{Redman} et~al.}{2004}]{redman04}
{Redman} M.~P.,  {Keto} E.,  {Rawlings} J.~M.~C.,    {Williams} D.~A.,  2004,
  \mnras, 352, 1365

\bibitem[\protect\citeauthoryear{{Rohlfs} \& {Wilson}}{{Rohlfs} \&
  {Wilson}}{2004}]{rohlfswilson}
{Rohlfs} K.,  {Wilson} T.~L.,  2004, {Tools of Radio
  Astronomy}. Springer, Berlin

\bibitem[\protect\citeauthoryear{{Rosolowsky}, {Pineda}, {Foster}, {Borkin},
  {Kauffmann}, {Caselli}, {Myers} \& {Goodman}}{{Rosolowsky}
  et~al.}{2008}]{rosolowsky08}
{Rosolowsky} E.~W.,  {Pineda} J.~E.,  {Foster} J.~B.,  {Borkin} M.~A.,
  {Kauffmann} J.,  {Caselli} P.,  {Myers} P.~C.,    {Goodman} A.~A.,  2008,
  \apjss, 175, 509

\bibitem[\protect\citeauthoryear{{Ruden}}{{Ruden}}{1999}]{ruden99}
{Ruden} S.~P.,  1999, in {Lada} C.~J.,  {Kylafis} N.~D.,  eds, The
Origin of Stars and Planetary Systems. Kluwer, Dordrecht, p.~643

\bibitem[\protect\citeauthoryear{{Schnee}, {Caselli}, {Goodman}, {Arce},
  {Ballesteros-Paredes} \& {Kuchibhotla}}{{Schnee} et~al.}{2007}]{schnee07}
{Schnee} S.,  {Caselli} P.,  {Goodman} A.,  {Arce} H.~G.,
  {Ballesteros-Paredes} J.,    {Kuchibhotla} K.,  2007, \apj, 671, 1839

\bibitem[\protect\citeauthoryear{{Shu}, {Adams} \& {Lizano}}{{Shu}
  et~al.}{1987}]{shu87}
{Shu} F.~H.,  {Adams} F.~C.,    {Lizano} S.,  1987, \araa, 25, 23

\bibitem[\protect\citeauthoryear{{Simon}}{{Simon}}{1992}]{simon92}
{Simon} M.,  1992, in {McAlister} H.~A.,  {Hartkopf} W.~I.,  eds, IAU Colloq.
  135, Complementary Approaches to Double and Multiple Star
  Research. Astron. Soc. Pac., San Francisco, p.~41

\bibitem[\protect\citeauthoryear{{Solomon}, {Rivolo}, {Barrett} \&
  {Yahil}}{{Solomon} et~al.}{1987}]{solomon87}
{Solomon} P.~M.,  {Rivolo} A.~R.,  {Barrett} J.,    {Yahil} A.,  1987, \apj,
  319, 730

\bibitem[\protect\citeauthoryear{{Spitzer}}{{Spitzer}}{1978}]{spitzer78}
{Spitzer} L.,  1978, {Physical processes in the interstellar medium}.
Wiley-Interscience, New York, NY

\bibitem[\protect\citeauthoryear{{Stutzki} \& {G\"usten}}{{Stutzki} \&
  {G\"usten}}{1990}]{stutzki90}
{Stutzki} J.,  {G\"usten} R.,  1990, \apj, 356, 513

\bibitem[\protect\citeauthoryear{{Swift} \& {Williams}}{{Swift} \&
  {Williams}}{2008}]{swift08}
{Swift} J.~J.,  {Williams} J.~P.,  2008, \apj, 679, 552

\bibitem[\protect\citeauthoryear{{Tachihara}, {Onishi}, {Mizuno} \&
  {Fukui}}{{Tachihara} et~al.}{2002}]{tachihara02}
{Tachihara} K.,  {Onishi} T.,  {Mizuno} A.,    {Fukui} Y.,  2002, \aap, 385,
  909

\bibitem[\protect\citeauthoryear{{Tilley} \& {Pudritz}}{{Tilley} \&
  {Pudritz}}{2004}]{tilley04}
{Tilley} D.~A.,  {Pudritz} R.~E.,  2004, \mnras, 353, 769

\bibitem[\protect\citeauthoryear{{Walmsley}, {Flower} \& {Pineau des
  For{\^e}ts}}{{Walmsley} et~al.}{2004}]{walmsley04}
{Walmsley} C.~M.,  {Flower} D.~R.,    {Pineau des For{\^e}ts} G.,  2004, \aap,
  418, 1035

\bibitem[\protect\citeauthoryear{{Williams}, {de Geus} \& {Blitz}}{{Williams}
  et~al.}{1994}]{williams94}
{Williams} J.~P.,  {de Geus} E.~J.,    {Blitz} L.,  1994, \apj, 428, 693

\bibitem[\protect\citeauthoryear{{Williams}, {Blitz} \& {McKee}}{{Williams}
  et~al.}{2000}]{williams00}
{Williams} J.~P.,  {Blitz} L.,    {McKee} C.~F.,  2000, in {Mannings} V.,
  {Boss} A.~P.,   {Russell} S.~S.,  eds, Protostars and Planets
  IV. Univ. of Arizona Press, Tucson, p.~97

\bibitem[\protect\citeauthoryear{{Zuckerman} \& {Evans} II}{{Zuckerman} \&
  {Evans}}{1974}]{zuckerman74}
{Zuckerman} B.,  {Evans} II N.~J.,  1974, \apjl, 192, L149

\end{thebibliography}
\end{document}